\documentclass[%
 reprint,
 amsmath,amssymb,
 aps,
]{revtex4-2}

\usepackage{graphicx}% Include figure files
\usepackage{dcolumn}% Align table columns on decimal point
\usepackage{bm}% bold math
\usepackage{aas_macros}
\usepackage{hyperref}% add hypertext capabilities
\usepackage{xcolor}
\newcommand{\fnl}{f_{\rm NL}~}
\newcommand{\fnlloc}{f_{\rm NL}~}%}^{\mathrm{loc}}}
\newcommand{\M}[0]{\mathrm{Mpc}/h}

\newcommand{\bphi}{b_{\phi}}

\begin{document}

\preprint{APS/123-QED}

\title{Local Primordial non-Gaussian Bias from Time Evolution}% Force line breaks with \\

\author{James M. Sullivan}
 \email{jms3@mit.edu}
 \thanks{Brinson Prize Fellow}
 \affiliation{Department of Astronomy, University of California, Berkeley, CA 94720, USA\\ Berkeley Center for Cosmological Physics, University of California, Berkeley, CA 94720, USA \\
 Center for Theoretical Physics, Massachusetts Institute of Technology, Cambridge, MA 02139, USA}%Lines break automatically or can be forced with \\
 
\author{Uro\v{s} Seljak}%
\affiliation{%
Department of Astronomy, University of California, Berkeley, CA 94720, USA\\ Berkeley Center for Cosmological Physics, University of California, Berkeley, CA 94720, USA\\
Department of Physics, University of California, Berkeley, CA 94720, USA\\
Physics Division, Lawrence Berkeley National Laboratory, Berkeley, California 94720, USA
}%

\date{\today}% It is always \today, today,

\begin{abstract}
Primordial non-Gaussianity (PNG) is a signature of fundamental physics in the early universe that is probed by cosmological observations.
It is well known that the local type of PNG generates a strong signal in the two-point function of large-scale structure tracers, such as galaxies. 
This signal, often termed ``scale-dependent bias'' is a generic feature of modulation of gravitational structure formation by a large-scale mode.
It is less well-appreciated that the coefficient controlling this signal, $b_{\phi}$, is closely connected to the time evolution of the tracer number density.
This correspondence between time evolution and local PNG can be simply explained for a universal tracer whose mass function only depends on peak height, and more generally for non-universal tracers in the separate universe picture, which we validate in simulations.
We also describe how to recover the bias of tracers subject to a survey selection function, and perform a simple demonstration on simulated galaxies.
Since the local PNG amplitude in $n-$point statistics ($f_{\rm NL}$) is largely degenerate with the coefficient $b_{\phi}$, this proof of concept study demonstrates that galaxy survey data can
allow for more optimal and robust extraction of local PNG information from upcoming surveys.
\end{abstract}

\maketitle

%\tableofcontents

\section{\label{sec:intro}Introduction}

We are entering a golden age of large-scale structure (LSS) surveys.
DESI \footnote{\url{https://www.desi.lbl.gov/}} and Euclid \footnote{\url{https://www.cosmos.esa.int/web/euclid}} are currently operating, SPHEREx \footnote{\url{https://spherex.caltech.edu/}} has launched, LSST\footnote{\url{https://www.lsst.org/}} will begin taking data, and Spec-S5\footnote{\url{https://www.spec-s5.org/}} will spearhead future spectroscopic survey cosmology.
These surveys span a range of galaxy number densities and volumes, but on the whole are tending toward increased sky coverage and higher redshifts.
Primordial non-Gaussianity (PNG) is a prime target of these surveys, as any deviation from Gaussian initial conditions will be imprinted on tracers of LSS, such as galaxies, and will generally be easier to detect at the large scales these surveys will be sensitive to.

While there are several well-understood ways in which the detailed physics of the early universe can generate non-Gaussian statistics, galaxy surveys are particularly well-suited to probe \textit{local} PNG (with amplitude parameter $\fnl = f_{\rm NL}^{(\rm loc)}$).
A detection of the local PNG (LPNG) signal, under standard assumptions, would indicate the presence of multiple light fields during inflation. 
LSS surveys most sensitively detect LPNG through a large-scale signal in galaxy $n-$point statistics that cannot be generated by gravitational evolution \cite{consistency_kehagias_13,consistency_peloso_13,consistency_creminelli_13}. 
This signal is often termed ``scale-dependent bias''  and can be understood as the coupling of modes in the initial conditions of structure formation (e.g., generating is a non-zero squeezed bispectrum, where one mode is much larger than the size of any gravitational displacement).

LPNG bias has accumulated a long history of study \cite{Dalal08,Slosar08,verde_matarrese_pngbias,carbone_png_bas,desjacques_seljak_iliev_png_bias_halos,2014arXiv1412.4671A,jeong_komatsu_09_sdb,dePutterDore17,Giri23}, but it was understood immediately that for universal tracers \footnote{We will frequently use the word ``halo'' or ``galaxy'', since those are the cases of interest here, but all conclusions stated here generally apply to any LSS tracer.} \footnote{With a universal mass function (UMF) depending only on peak height $\nu(M,z) = \frac{\delta_c}{\sigma(M,z)}$ for critical overdensity $\delta_c$ and amplitude of smoothed matter fluctuations $\sigma$ } the LPNG bias coefficient $\bphi$ is closely related to the the tracer Lagrangian bias $\bphi(b_{L}) = 2\delta_{c} b_{L} = 2\delta_{c} (b_{E}-1)$ \cite{Dalal08, Slosar08}.
This relationship has been explored in dark-matter-only simulations and hydrodynamical simulations \cite{Biagettibphi,2020JCAP...12..013B,barreira_field_b1_b2,2022JCAP...04..057B,2022JCAP...01..033B}, where it generally holds for dark matter halos selected by mass.

A richer picture of $\bphi$ has emerged that extends the core connection to Lagrangian bias.
From nearly the outset, possible deviations from the UMF relation have been captured by a free parameter $p$, the presence of which was motivated by the possibility of recent halo mergers \cite{Slosar08,reid_ab_10}.
But, recently, there has been an ambitious rally to physically model the deviation of $\bphi$ from its UMF relation (i.e., accounting for $\bphi$ ``assembly bias'') \cite{sullivan_bphi_forecast_ab,Lazeyras22,marinucci_bphi_sam,barreira_krause_23,2023MNRAS.524.1746L,fondi_ab_forecast_24,boryana_AB_bphi_2,boryana_png_ab_1,pngunit_bphi_ab}, as samples with large differences in $\bphi$ (with similar $b_1$, as typical of strong assembly bias splits) lead to improved constraints on $\fnl$. 
However, even with the multi-tracer technique \cite{seljak_mt,SchmittfullSeljak} and machine learning tools trained on simulations, previous efforts require a muscular assumption - that one knows $\bphi$ with high confidence. 
Relaxing this assumption can significantly degrade constraints on $\fnl$ \cite{2022JCAP...11..013B,field_level_bias_stephen}.

The question we aim to answer in this paper is simple - can we wriggle out of strong assumptions about $\bphi$ by \textit{estimating $\bphi$ directly from the data?}.
We suggest that the answer to this question is in the affirmative, exploiting the close connection between $\bphi$ and the response of galaxy number density to the growth of structure, which can be related to the evolution bias $b_e$ \cite{challinor_lewis_gr, alonso_ultra_large_scale,rossiter_fnl_gr_bias,DJS}.
We first motivate the connection between time evolution and LPNG (Section~\ref{sec:theory}), before validating this effect in N-body and hydrodynamical  simulations (Section~\ref{sec:sims}),  showing that it generalizes beyond the universal case for galaxies in hydrodynamical simulations in addition to dark matter halos.
We then address the important issue of selection effects (Section~\ref{sec:selection}), which must be properly accounted for before attempting to inform the value of $b_\phi$ from tracer number density data.
Finally we discuss possible extensions of the method develop here (Section~\ref{sec:discussion}) before concluding with an outlook towards future more optimal and robust $\fnl$ analyses (Section~\ref{sec:conclusions}).

\section{\label{sec:theory} Theory}

We present a simple derivation of the approximate equivalence of LPNG bias and evolution bias for universal tracers in Section~\ref{subsec:theory_umf} and extend this to more general tracers in Section~\ref{subsec:theory_cfc}.

\subsection{\label{subsec:theory_umf} Bias from universal tracers}

For a tracer whose abundance $n_{t}(\nu)$ (in terms of comoving number density) is assumed to depend only on peak height $\nu(M,z,\sigma_8) \equiv \frac{\delta_{c}}{\sigma(M,z,\sigma_8)}$, it is simple to see the correspondence between Local PNG bias $\bphi$ and (time-) evolution bias $b_e$.

\begin{figure}[h!]
    \centering
    \includegraphics[width=0.45\textwidth]{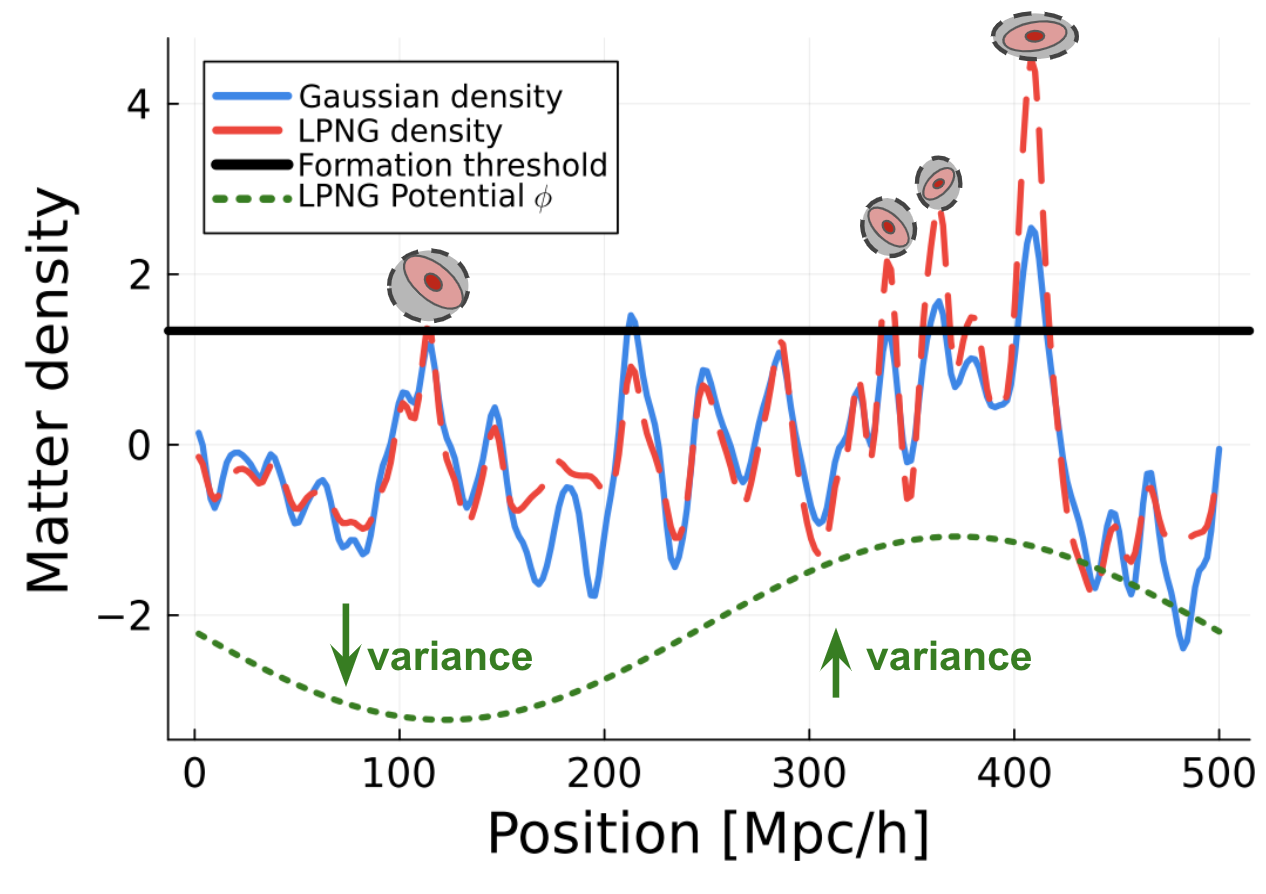}
    \caption{
    A cartoon illustrating the key intuition behind the impact of LPNG on tracer formation.
    Due to PNG, long-wavelength changes in the Bardeen potential $\phi$ (green)
    modulate the variance of the matter density field.
    In regions of high or low variance, the peaks collapse more or less readily into galaxy-hosting dark matter halos.
    The purpose of this work is to demonstrate that this modulation of variance is approximately equivalent to a modulation of the time evolution of large-scale structure.
    }
    \label{fig:bphi_ev_cartoon}
\end{figure}

Figure~\ref{fig:bphi_ev_cartoon} gives a heuristic depiction of why this might be the case using field realizations \footnote{Here using an unrealistically large $f_{NL}=5000$ for illustration.}.
In a simple picture of tracer formation, halos/tracers form when the (smoothed) matter density field exceeds a certain collapse threshold.
Local PNG (orange-dashed) acts to rescale the variance of the matter overdensity field via the large scale potential (green dotted), which affects tracer formation (some orange-dashed peaks cross the threshold when they did not in the Gaussian case [blue] and vice versa).
One can roughly imagine that the orange-dashed curve is actually a future version of the blue curve in regions where the variance is enhanced - some peaks have grown while nearby ``voids'' have emptied out.
We now show there is some merit to this story.

Following  Ref.~\cite{jsh_gr}, for universal tracers, we have
\begin{align}
    b_{e} &= \frac{\partial \log n_{t}(\nu(M,z,\sigma_8))}{\partial \log (1+z)},
    \label{eqn:UMF_be_bphi}
\end{align}
while the usual variance separate universe definition of $b_{\phi}$ is
\begin{align}
    b_{\phi} &= \frac{\partial \log n_{t}(\nu(M,z,\sigma_8))}{\partial \log \sigma_8}.
    \label{eqn:UMF_bphi}
\end{align}
A universal abundance function, regardless of its form, depends only on peak height $\nu$, and therefore, for fixed $\delta_c$ (as is usually assumed, with $\delta_c=1.686$), only on the linear variance $\sigma^2$.
We can immediately see the reason we expect the bias form time evolution in eqn.~\ref{eqn:UMF_be_bphi} to be connected to the standard variance separate universe (SU) LPNG bias in eqn.~\ref{eqn:UMF_bphi} by considering the usual expression for the variance:
\begin{align}
    \sigma^{2}(M,z,\sigma_8) &= \frac{1}{2\pi^{2}} \int_{0}^{\infty} k^{2} dk |W_{R(M,z)}^{2}|(k)D^{2}(z)P(k), 
    \label{eqn:UMF_sigma}
\end{align}
where, in an EdS universe, $D(z) \propto \frac{1}{(1+z)}$, and $P(k)$ is the $z=0$ power spectrum.
With the standard form of the power spectrum, $P(k) \propto \sigma_8^2$, so clearly $\frac{d \sigma}{d\sigma_8} = \frac{d \sigma}{d D(z)}$.
For EdS, the latter is just the negative derivative with respect to redshift, wile for a general growth factor we pick up a factor of $\frac{d D(z)}{d z}$, which we can rewrite as a factor of the growth rate $f$, in agreement with Ref.~\cite{jsh_gr}.
In matter domination, EdS is an excellent approximation for growth, which is reflected in the fact that the growth rate $f\approx 1$ at this time.
The computed expression of Ref.~\cite{jsh_gr} for $b_e$,
\begin{equation}
    \label{eqn:be_UMF}
    b_e = \delta_c f(z)(b_1-1),
\end{equation}
then reduces to the universality prediction for $\frac{b_{\phi}}{2}$, 
\begin{equation}
    \label{eqn:bphi_UMF}
    b_\phi = 2 \delta_c (b_1-1).
\end{equation}
The simple picture offered by these predictions eqns.~\ref{eqn:be_UMF} and \ref{eqn:bphi_UMF} is expected to hold for dark matter halos selected by mass, though not for general tracers that deviate from universality.
Accordingly, we now provide a more generally applicable argument for the correspondence between time evolution and LPNG. 

% FIGURE FOR CARTOON REALIZATION RESCALING (in 1D)

\subsection{\label{subsec:theory_cfc} Beyond universality - PNG bias from time shift Separate Universe}

We can extend the argument above to non-universal tracers through an extremely simple application of Conformal Fermi Coordinates (CFCs) \cite{baldauf_gr, pajer_3pt_cfc,liang_cfc_1,liang_su} and the Separate Universe (SU) picture \cite{salopek_bond, liang_su,baldauf_su,baldauf_su2,sirko_su,gnedin_su,wagner_su,yin_su_14}.
The key intuition is that the effect of a long-wavelength mode on small scale modes within some region can be absorbed into a redefinition of the spacetime coordinates and metric in a patch smaller than the spatial extent of the long-wavelength mode.
In this sense, local observers experiencing the effects of different long-wavelength modes live in ``separate universes'' and the long mode-absorbing coordinates are CFCs.

More specifically, for a given metric modified by long a wavelength perturbation, CFCs provide a modified coordinate system in which the metric is of the Friedmann–Lema\^{i}tre–Robertson–Walker (FLRW) form \footnote{In the presence of tidal long mode perturbations, the metric is of perturbed FLRW form instead.}.
We will only briefly review the construction of CFCs and then quote basic results on the form of the CFC metric.
For a detailed presentation of the construction of CFCs (and the assumptions going into them) we refer the reader to, e.g., Ref.~\cite{liang_cfc_1}.

For isotropic scalar perturbations $\phi,\psi$ with wavelength longer than a length scale $\Lambda^{-1}$, the ``global'' perturbed metric is
\begin{equation}
\label{eqn:cng_global_metric}
ds^2 = a^2(\tau)\left[-\left(1 + 2\psi \right)d\tau^2 + \left(1 - 2\phi \right) \delta_{ij} dx^i dx^j\right],
\end{equation}
where we work in the conformal Newtonian gauge convention of Ref.~\cite{ma_bertschinger}\footnote{Such that the lack of anisotropic stress implies $\phi=\psi$, which we will not assume.}.
As noted in Ref.~\cite{liang_su}, the corresponding CFC metric can be written in the form
\begin{equation}
\label{eqn:cfc_iso_metric}
ds^2 = a_F^2(\tau_F)\left[-d\tau_F^2 + \frac{\delta_{ij} dx_F^i dx_F^j}{\left(1+K_F \tilde{r}_{F}^{2}\right)^2} \right].%,
\end{equation}
In these coordinates, all the effects of the long mode perturbation are parameterized through the local scale factor $a_F$, altered local Hubble parameter $H_F$ and local curvature $K_F$:
\begin{align}
\label{eqn:cfc_aF_kF}
a_F(\tau_F) &= a(\tau)\left[1 + \frac23 \psi_{\mathrm{ini}} + \int_0^\tau d\tau'\left(-\phi' + \frac13 \partial_i V^i\right)\right]\\
H_{F}(\tau_F) &= H\left[1-\left(\psi + \frac{1}{\mathcal{H}}\phi'\right) + \frac{1}{3\mathcal{H}}\partial_i V^i\right]\\
K_F &= \frac23 \left(\partial^2 \phi - \mathcal{H} \partial_i V^i\right).
\end{align}

These results were developed with a long mode of the matter density in mind, which is an isotropic scalar perturbation. 
We will remain interested only in the isotropic scalar perturbation case.
In fact, we are concerned with something even simpler than this, which is simply to describe how the presence a long mode of \textit{potential} impacts time-evolution in a (large-scale) CFC patch of size $\Lambda^{-1}$.

For a scalar perturbation in the potential that is approximately constant in space 
and time,
the impact of a long wavelength mode of potential during matter domination can be absorbed entirely into a shift in the local scale factor \cite{liang_su}
\begin{equation}
\label{eqn:cfc_simple_aF} 
    a_{F}(\tau_{F}) = a(\tau)\left(1 + \frac23 \psi_{\mathrm{ini}}\right)
\end{equation}
or, 
\begin{equation*}
    \log\left(\frac{a_{F}(\tau_F)}{a(\tau)}\right) = \log(1+\frac23 \psi_{\mathrm{ini}}),
\end{equation*} 
while the local FLRW curvature parameter vanishes ($K_F = 0$), and the CFC Hubble parameter is the same as the global one ($H_F =H$).

In other words, the CFC metric eqn.~\ref{eqn:cfc_iso_metric} reduces simply to the Minkowski metric with the local scale factor $a_F$ instead of the global one $a$.
\begin{equation}
\label{eqn:cfc_simple_metric}
ds^2 = a_F^2(\tau_F)\left(-d\tau_F^2 + \delta_{ij} dx_F^i dx_F^j\right).
\end{equation}
This is the same form of the metric we would obtain from simply restricting attention to the spatial origin ($x_F^i \to 0$, e.g. for a freely falling observer), which should be thought of as restricting to only quantities that change over the whole spatial patch and so do not vary spatially \footnote{By pursuing this logic further, one can calculate finite wavelength corrections to $\phi_{l}=\mathrm{const.}$ at each order in $x_{F}^n$.}.

Eqn.~\ref{eqn:cfc_simple_aF} 
suggests the connection to the evolution bias.
The evolution bias multiplies the proper time-perturbation $\mathcal{T}$ \cite{cosmic_clocks,pajer_3pt_cfc,DJS}:
\begin{equation}
\label{eqn:or_time_linear_bias}
\delta_{g}^{\mathrm{or}}(\tilde{z}) =  \delta_{g}^{\mathrm{pt}}(\tilde{z}) + b_e \mathcal{T}(\tilde{z}),
\end{equation}
where, in the notation of Ref.~\cite{DJS} in which $\tilde{z}, \tilde{\tau}, \tilde{a}$ are the conformal time and scale factor in the observed redshift gauge,
\begin{equation}
\label{eqn:time_pert}
\mathcal{T}(\tilde{\tau}) = \log \left(\frac{a(t_F(\tilde{\tau}))}{\tilde{a}}\right)
\end{equation}
is the proper time perturbation.
From eqn.~\ref{eqn:time_pert}, we see that the connection between LPNG bias $b_\phi$ and evolution bias $b_e$ can be boiled down to the shared form of equations~\ref{eqn:cfc_simple_aF} and \ref{eqn:time_pert} \footnote{The time shift expression depends on the chosen gauge, but $\mathcal{T}$ is a gauge-invariant quantity \cite{cosmic_clocks}.}
The shift in the scale factor that absorbs the LPNG long mode and the shift in the scale factor between a constant proper time and constant observed redshift are due to distinct physical effects, but both have a time-dependent effect on the galaxy number density.
In other words, since the impact of LPNG on the galaxy number density can be recast as a shift in the scale factor $\log\left(\frac{a_{F}(t_{F})}{a}\right) \neq 0$ \footnote{More precisely, the shift from coordinate time $a$ to proper time $a_{F}(t_{F}(a))$, where $a$ is analogous to the choice of ``observed redshift'' coordinates.}, 
the LPNG bias (in its peak-background split form, assuming EdS) has exactly the form of the evolution bias.

Another way to see the correspondence between LPNG bias and evolution bias is the following.
Since a long-wavelength potential mode $\phi_l$ shifts the proper time relative to no perturbation (i.e. in a Gaussian universe), we can expose this perturbation explicitly when in global coordinates. 
Then in local CFC coordinates we can reuse the ``cosmic clocks'' picture \cite{cosmic_clocks}, where arbitrary perturbations (in the CFC patch) are allowed.
Following this picture through, the photon propagation effects that lead to evolution bias, can, as usual, be written in terms of the proper time (p.t.) - observed redshift (o.r.) difference in scale factor at any given spacetime point. 
However, since a LPNG long mode causes a scale factor shift independent of the usual photon propagation effects in a Gaussian universe, we can separate it out from the proper time (p.t.-o.r.) perturbation $\mathcal{T}$ as an additional shift.
Since the size of this shift will scale like $f_{\rm NL}$, it should be separated out from the Gaussian contribution.
We can then see by defining a new coefficient $\mathcal{T}_{NG}$ we arrive at $b_\phi$
\begin{align}
\label{eqn:pt_pert_split}
\delta_g^{or} &= b_1 \delta^{pt} + b_\phi f_{NL}\phi_{l} + b_e \mathcal{T}_G\\
\delta_{g}^{or} &= b_1 \delta^{pt} + b_e(\mathcal{T}_{G} + \mathcal{T}_{NG})
\end{align}
where $\mathcal{T}_{NG}$ describes the shift in proper time relative to a no $\phi_{l}$-perturbed universe - aka the change from $a_{F}(t_{F})$ vs $a_{G}(t_{G})$, $t_{G}$ also being proper time in the global unperturbed pre-CFC coordinates. 
At the end of the day, these two expressions are the same (in the SU/PBS approximation).
The discussion of the non-PBS picture where $\phi_{l}$ is not approximated as a constant merits further discussion, which we leave for future work.

One \textit{consequence} of this correspondence is the change in the matter fluctuation variance $\sigma^2 (R)$, which depends on the growth factor $D(z)$.
However, it is clear that the relationship between evolution bias and PNG bias is due to the common interpretation of a long mode of scalar potential and a general observed redshift perturbation as shifts between time coordinate and proper time.
Therefore, one can think of the ``variance'' separate universe \cite{baldauf_su, barreira_field_b1_b2} can be thought of as a ``time shift'' separate universe.

Such an interpretation naturally predicts that the impact of local PNG on galaxy number density should persist even for non-universal tracers - i.e. any tracer population that is defined according to a time dependent property \footnote{More precisely, any tracer that responds to the growth of structure.}.
For halos, such a property could of course include the peak height $\nu$ but could also include the halo concentration or assembly history features, while for galaxies such a property could be the star formation or color.
The presence of the recently much-explored assembly bias of $b_\phi$ \cite{Slosar08,reid_ab_10,lazeyras_AB_quadratic_halos_21,barreira_krause_23,sullivan_bphi_forecast_ab,fondi_ab_forecast_24} can be explained in this way.

\section{\label{sec:sims} Validation in simulations}

We illustrate that the simple picture sketched above  holds in cosmological simulations.

\subsection{\label{subsec:sims_dm} Dark-matter only}

\begin{figure}[h!]
    \centering
    \includegraphics[width=0.45\textwidth]{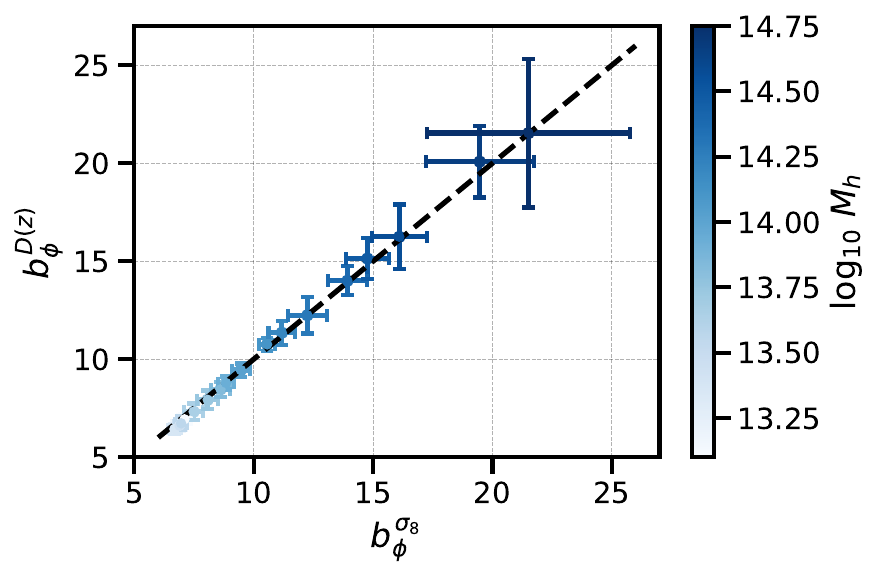}
    \caption{Equivalence of time-evolution LPNG bias $b_{\phi}^{D(z)}$ and bias computed from variance separate universe simulations $b_{\phi}^{\sigma_8}$.
    These biases are measured using finite difference at $z=1$ over several mass bins (of varying width, see text).
    The black dashed line indicates a perfect match between the two measurements.
    }
    \label{fig:nbody_halos}
\end{figure}

We first show that LPNG bias computed from the variance separate universe method (using $\sigma_8$ \cite{baldauf_su}) is approximately equivalent to that computed from time evolution for dark matter halos in Figure~\ref{fig:nbody_halos}.
To obtain the halo catalogs we use, we evolve the matter field from redshift $z_{\rm{ini}} = 99$ to $z=0$ using \texttt{FastPM} \cite{fastpm} in a box of size $L = 2000 ~ \M$, with $N_{p}=1024^3$ simulation particles, force resolution ``boost'' factor of $B = 2$, using 40 timesteps.
We generate initial conditions using \texttt{CAMB} \cite{camb}, with a $\Lambda$CDM cosmology using: $\Omega_m = 0.3175$, $\Omega_b = 0.049$, $h = 0.6711$, $A_s = 1.91 \times 10^{-9}$ ($\sigma_8 = 0.834$), $n_s = 0.9624$.
Halos are found using Friends-of-Friends (FoF) \cite{fof_davis} with linking length $\ell=0.2$.
We use simulations with snapshots at $z=0.95,1.00,1.05$.
These are the same simulations as used in Ref.~\cite{lpng_field_level}.

Figure~\ref{fig:nbody_halos} shows the values of LPNG bias $\bphi$ estimated from the variance separate universe ($b_{\phi}^{\sigma_8}$)  on the horizontal axis and from time evolution ($b_\phi^{D(z)}$) on the vertical axis.
Specifically, we take centered finite differences in $\sigma_8$ and growth factor\footnote{Though at $z=1$ just taking derivatives in redshift, as in Einstein-de Sitter $D(z) = (1+z)^{-1}$, is not a bad approximation.} $D(z)$ (at each redshift $z$), and estimate uncertainties on these quantities using ten realizations of the initial density field. 
We consider a range of halo masses (shades of blue), and agreement between $b_{\phi}^{\sigma_8}$ and $b_{\phi}^{D(z)}$ within the errorbars at all but the lowest mass halos (which may be affected by the simulation mass resolution).
These results clearly show that the agreement between these values is quite good, and is what we expect from the universality predictions (eqns.~\ref{eqn:be_UMF} and \ref{eqn:bphi_UMF}) for mass-selected halos.

\begin{figure}[h!]
    \centering
    \includegraphics[width=0.45\textwidth]{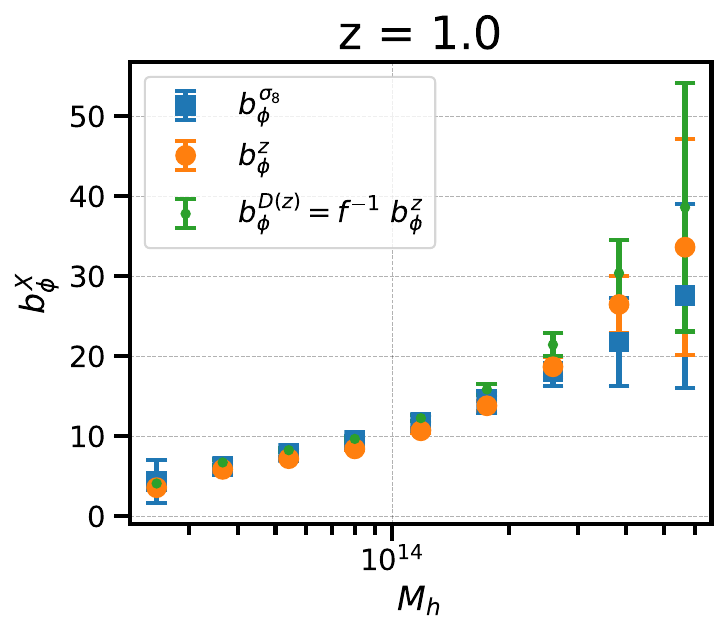}
    \includegraphics[width=0.45\textwidth]{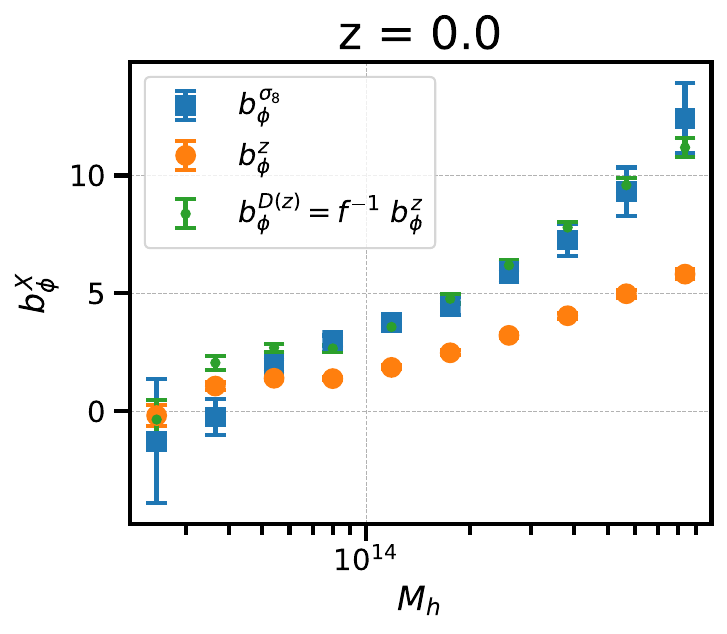}
    \caption{Similar to Figure~\ref{fig:nbody_halos}, but for the Quijote simulations at redshifts $z=1$ (\textit{top}) and $z=0$ (\textit{bottom}), and as a function of halo mass (in $M_\odot/h$) on the horizontal axis.
    The significant deviation from matter domination at $z\lesssim1$, as expected, demonstrates the difference between $b_e$ and $b_\phi^{D(z)}$.
    }
    \label{fig:nbody_halos_quijote}
\end{figure}

\begin{figure}[h!]
    \centering
    \includegraphics[width=0.45\textwidth]{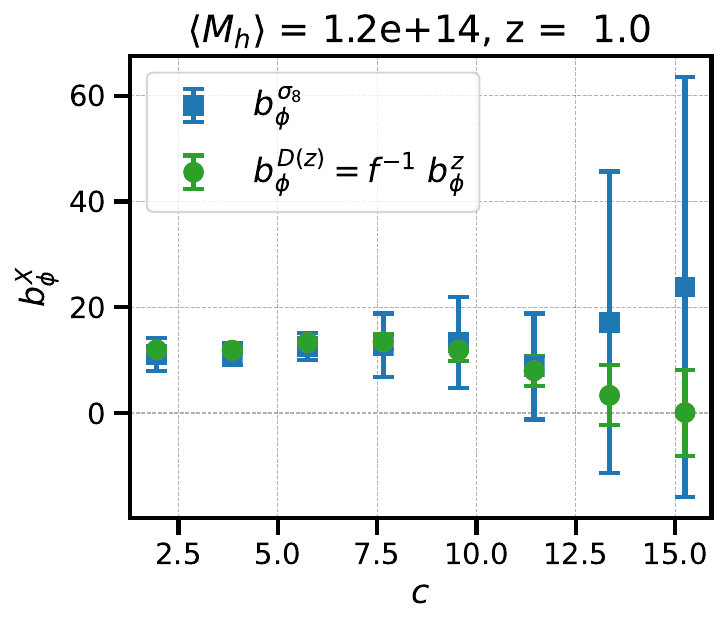}
    \includegraphics[width=0.45\textwidth]{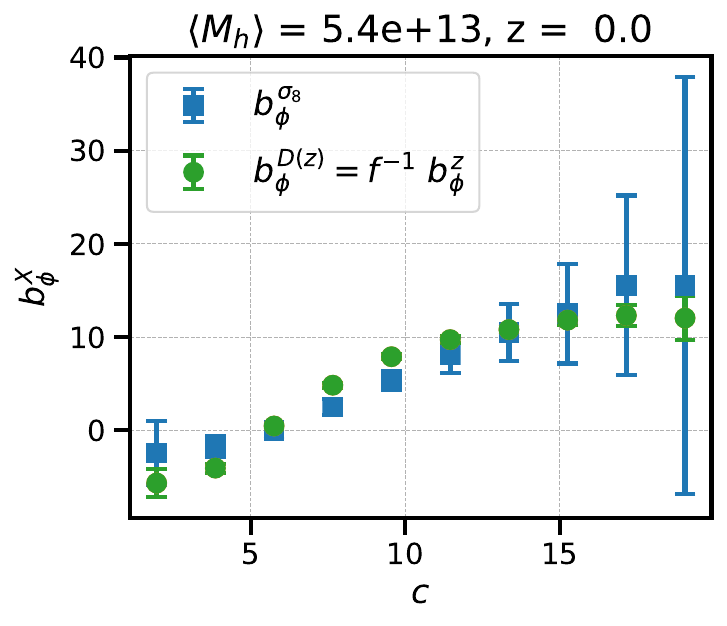}
    \caption{Similar to Figure~\ref{fig:nbody_halos_quijote}, but for non-universal tracers.
    $b_\phi$ is computed as a function of halo concentration at $z=1$ for a halo mass bin centered at $M_{h} = 1.8 \times 10^{14} h^{-1}~M_\odot$ (\textit{top}) and $z=0$ for a halo mass bin centered at $M_{h} = 5.4 \times 10^{13} ~ h^{-1}~M_\odot$ (\textit{bottom}).
    }
    \label{fig:nbody_halos_conc_quijote}
\end{figure}

We also demonstrate the equivalence of LPNG bias from time evolution and from using the Quijote simulations both for universal tracers (mass-binned halos)  in Figure~\ref{fig:nbody_halos_quijote} and non-universal tracers (halo concentration and mass-binned halos) in Figure~\ref{fig:nbody_halos_conc_quijote}.
In contrast to the FastPM simulations used in Fig.~\ref{fig:nbody_halos}, we use the Quijote Rockstar \cite{rockstar_behroozi} halos rather than Friends-of-Friends.
This serves as check that is independent of $N$-body simulation code or halo finder.
As for the FastPM simulations, the number density derivatives as a function of $\sigma_8$ are defined via variance separate universe finite differences.
However, due to the low frequency of Quijote redshift output, time (and growth factor) finite difference derivatives are estimated using the derivative of a spline fit at the nodes (using the \texttt{scipy} implementation of fitpack \cite{2020SciPy-NMeth}).
Uncertainties are computed from 30 simulation realizations.
For mass-binned halos in Fig.~\ref{fig:nbody_halos_quijote}, we see that, as in the FastPM simulations, the agreement at $z=1$ is excellent between the variance SU derivative (blue squares), the number density derivative with respect to redshift (orange circles), and the number density derivative with respect to the growth factor (green points).
At $z=0$, since the response of the number density to the growth of structure (as realized by the growth factor) no longer has EdS as a good approximation, the redshift derivative points (orange circles) deviate significantly from the SU derivatives.
However, the finite difference derivative with respect to $D(z)$ (green points) remains in excellent agreement with the variance SU derivatives (blue squares).
The correspondence between $b_\phi^{(\mathrm{SU})}$ and $b_\phi^{D(z)}$ also holds for non-universal tracers in the Quijote simulations \cite{quijote_sims}.
When binning by both mass and (virial) concentration, we find that the match between these two quantities in Fig.~\ref{fig:nbody_halos_conc_quijote} generally remains excellent.

\subsection{\label{subsec:sims_hydro} Hydrodynamical simulations}
\begin{figure}[h!]
    \centering
    \includegraphics[width=0.45\textwidth]{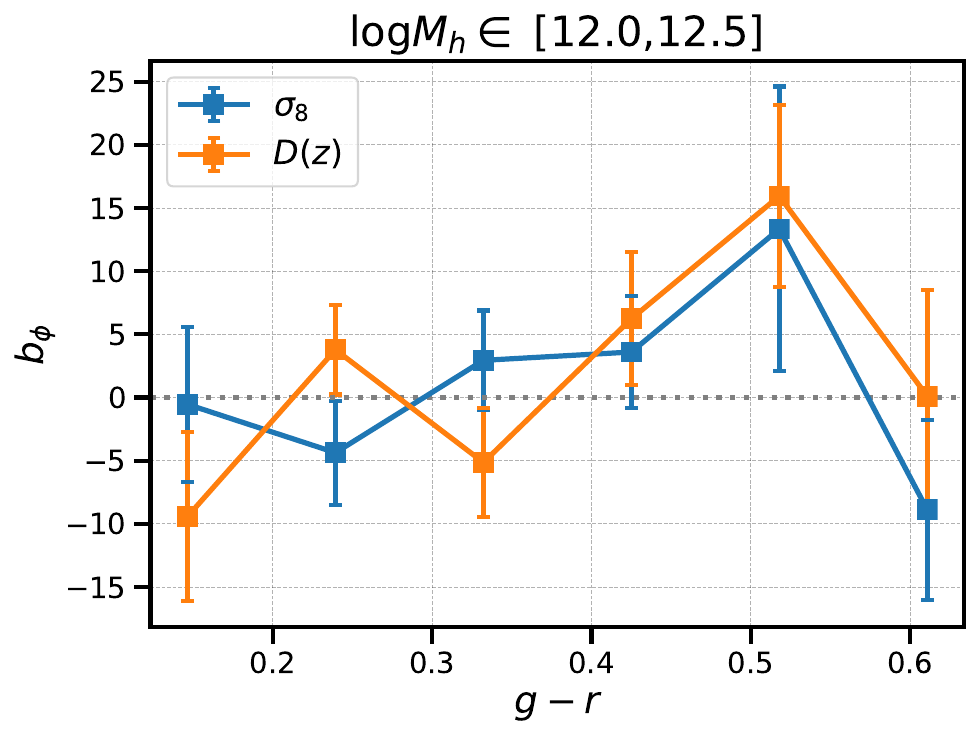}
    \caption{
    Approximate equivalence of LPNG bias from time evolution (blue) and from variance separate universe simulations (orange) in hydrodynamical simulations.
    Values are again provided at $z=1$, in a halo mass bin, but here for a variety of simulated $g-r$ colors (with the TNG model in the CAMELS simulations).
    Errorbars are estimated via jackknife resampling.
    }
    \label{fig:hydro_gmr_camels}
\end{figure}

The correspondence between time evolution and LPNG bias also appears to hold for more complex tracers than dark matter halos.
Figure~\ref{fig:hydro_gmr_camels} shows results similar to those presented in Figure~\ref{fig:nbody_halos_quijote} at $z=1$, but for simulated galaxies from the \texttt{CAMELS} suite of hydrodynamical simulations \cite{camels} with the IllustrisTNG galaxy formation model \cite{illustristng}.
To estimate SU LPNG bias (blue points), we use the CAMELS $\sigma_8$ ``\texttt{1P}'' simulations in the role of a variance separate universe. 
For the redshift evolution (orange points) we take simulations at $z=0.95,1.05$ compute a central finite difference estimate.
The plotted uncertainties are obtained from jackknife resampling of sub-volumes of the data defined by 1/5 of the linear size of the box side.

The vertical axis of Figure~\ref{fig:hydro_gmr_camels} shows the estimates for $\bphi$ from the different finite difference estimation methods for different bins of (rest-frame) SDSS color $g-r$ \footnote{We refer the reader to Refs.~\cite{camels,illustristng} for details about how these colors are assigned to simulated galaxies.} in a single halo mass bin.
This choice is inspired by recent work (in Ref.~\cite{barreira_krause_23} and later followed up by in Ref.~\cite{marinucci_bphi_sam}) showing that splitting high host halo mass simulated galaxies by $g-r$ leads to an extreme change in $\bphi$ as estimated by the variance separate universe method \footnote{This is a manifestation of the ``galaxy assembly bias'' of $\bphi$ that has also been seen for halos (e.g. in Refs.~\cite{reid_ab_10,lazeyras_AB_quadratic_halos_21}}.
We also see there is a trend in the estimated value of $\bphi$ with $g-r$ even given the significant errorbars in Fig.~\ref{fig:hydro_gmr_camels}.
However, the most interesting aspect of Fig.~\ref{fig:hydro_gmr_camels} is that the two methods of estimating $\bphi$ track each other quite well even in the presence of $g-r$ assembly bias.
While we only show one halo mass bin, this approximate equivalence (within the errorbars) persists for all but the smallest halo masses in CAMELS $M_{h}\gtrsim 3 \times 10^{10}~h^{-1} M_\odot$.
It would be interesting to explore these results further with larger volume simulations, though this would require hydrodynamical simulations with both multiple values of $\sigma_8$, as are available for CAMELS.

\begin{figure}[h!]
    \centering
    \includegraphics[width=0.45\textwidth]{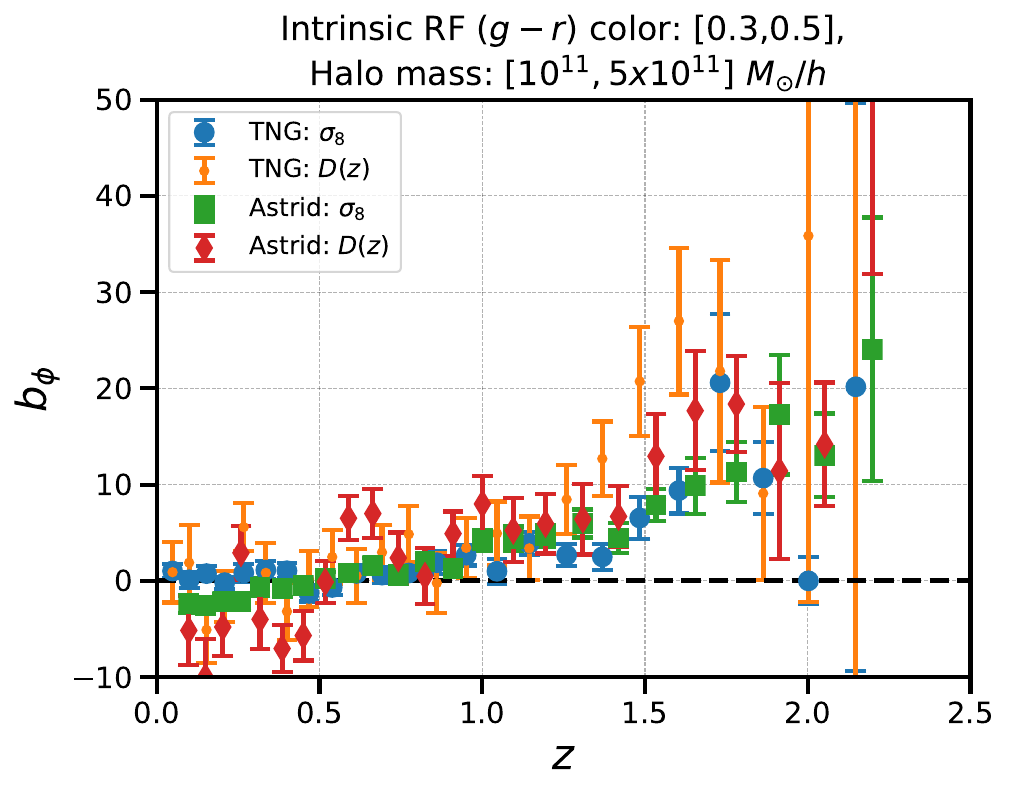}
    \includegraphics[width=0.45\textwidth]{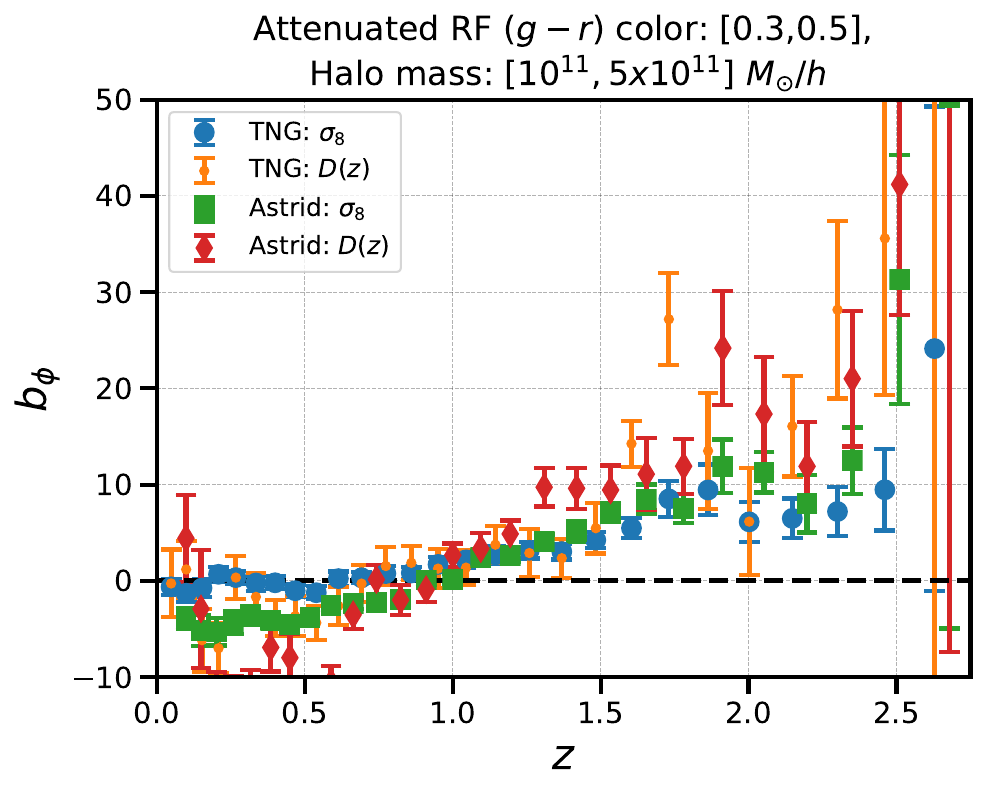}
    \caption{
    Comparison of variance separate universe and time evolution $\bphi$ with (\textit{top}) and without \textit{(bottom)} dust attenuation in hydrodynamical simulations with two different galaxy formation models as a function of redshift for a single halo mass bin ($\log\left(\frac{M_h}{M_\odot/h} \right) = [10^{11},10^{11.5}]$ and a rest-frame $g-r$ color $(g-r)_{r} = [0.3,0.5]$
    Blue circles (green squares) show the IllustrisTNG (Astrid) variance SU estimates of $\bphi$, while the orange points (red diamonds) show the estimates from time evolution.
    Uncertainties are determined using the same jackknife procedure as in Figure~\ref{fig:hydro_gmr_camels}.
    }
    \label{fig:hydro_gmr_camels_dust}
\end{figure}

Figure~\ref{fig:hydro_gmr_camels_dust} shows the correspondence between variance separate universe $\bphi$ and time evolution $\bphi$, as well as a view of the correspondence between time evolution and separate universe bias in hydrodynamical simulations, now as a function of redshift, for a fixed halo mass and (intrinsic) color bin with and without dust attenuation.
The CAMELS photometry and dust attenuation model is is described in Ref.~\cite{lovell_camels_photometry} \footnote{As a technical note, in Figure~\ref{fig:hydro_gmr_camels}, we used the older CAMELS SDSS band photometry, while in Figure~\ref{fig:hydro_gmr_camels_dust}, we use the more recent simulated photometry from Ref.~\cite{lovell_camels_photometry}}.
While it remains somewhat challenging especially at high redshift to determine agreement between the two estimates due to noise, it is clear they are relatively consistent with and without dust attenuation applied to the galaxies, though there is some visible impact of dust attenuation especially at low redshifts.
The agreement between Astrid and TNG for the variance SU in Fig.~\ref{fig:hydro_gmr_camels_dust} is consistent with the findings of Ref.~\cite{marinucci_bphi_sam}, who found robustness of the response of variance SU $\bphi^{\sigma_8}$ to changes in SAM galaxy formation models.
We additionally find a similar level of agreement between the time evolution derivatives between the two hydrodynamical galaxy formation models.

\section{\label{sec:selection} Tracer Selection}

We have established the correspondence between evolution bias and LPNG bias for simulated galaxies as a function of physical galaxy properties (e.g., intrinsic color, host halo mass).
However, for observed galaxy samples, galaxies are selected based on observed properties, such as apparent magnitude (or observed flux) and observed-frame colors.
This complicates the interpretation of the observed number density as a function of redshift \cite{challinor_lewis_gr}.
We will now discuss the impact of this reality on estimating LPNG bias from observed number densities as a function of redshift.
Here we will focus only on the qualitative impact of color and flux cuts in this case, as these that are the dominant effects for existing galaxy surveys, and lessons from these will be generally applicable to different surveys.
There are, of course, other survey-specific effects due to telescope/instrument properties (e.g. redshift failures, star/galaxy separation, fiber collisions, etc.), all of which we will neglect here and will characterize appropriately in future work applying our methodology to LPNG analysis of a specific survey.

\subsubsection{\label{subssec:selfunc_formalism} Selection function}

A general LSS tracer population $\bar{n}_t^{\mathrm{obs}}$ can depend explicitly on redshift $z$, on (redshift-independent) cosmology parameters $\theta = \{\Omega_m, \sigma_8,...\}$, physical tracer properties $\phi(z,\theta)$  
and on observable selection parameters $\phi^{(s)}$ (such as observed colors, magnitudes, isophotes,...), which are themselves functions of intrinsic tracer properties $\phi(z)$ \cite{DJS}, $\bar{n}^{\mathrm{obs}}_t(z,\phi(z),\theta,\phi^{(s)}(\phi(z),z)) = f^{\mathrm{sel}}\left[\bar{n}^{\mathrm{phys}}(z,\phi(z),\theta),\phi^{(s)}(z,\phi(z)),z \right]$.
In this general case, the selected galaxy number density can be written as
\begin{align}
    \label{eqn:general_selection_derivative_1}
    \frac{d\log\bar{n}_t^{\mathrm{obs}}}{d \log (1+z)} &= \frac{\partial \log\bar{n}_t^{\mathrm{obs}}}{\partial \log (1+z)}\bigg \rvert_{\phi^{(s)}} \\
    & + \frac{\partial \log\bar{n}_t^{\mathrm{obs}}}{\partial \log \phi^{(s)}(\phi,z)} \frac{\partial \log \phi^{(s)}(\phi,z)}{\partial \log (1+z)}\bigg\rvert_{z}
\end{align}
where several of these terms can be further expanded (e.g. if $\phi^{(s)}(z) = \phi^{(s)}(z,\phi(z),...)$, as is often the case)
For the special case of simulated selections, or where the observed selection parameter is an intrinsic property of the LSS tracer, then the expression simplifies and $\phi^{(s)} = \phi$
\begin{align}
    \label{eqn:general_selection_derivative}
    \frac{d\log\bar{n}_t^{\mathrm{obs}}}{d \log (1+z)} &= \frac{\partial \log\bar{n}_t^{\mathrm{obs}}}{\partial \log (1+z)}\bigg \rvert_{\phi(z)} \\
    & + \frac{\partial \log\bar{n}_t^{\mathrm{obs}}}{\partial \log \phi(z)} \frac{\partial \log \phi(z)}{\partial \log (1+z)}\bigg\rvert_{z}.
\end{align}
To obtain the evolution bias, one must of account for all contributions arising from the terms following the first on the RHS of eqn.~\ref{eqn:general_selection_derivative} (see, e.g., Ref.~\cite{maartens_bev}).

One example of a selection-based correction comes from the flux/magnitude limit of a survey.
Real cosmological redshift surveys have a flux (or magnitude) limit, at minimum due to limited telescope imaging depth.
The flux limit combined with the fact that photons travel on perturbed geodesics impacts the number of galaxies observed most obviously through magnification due to lensing.
This is true even with perfect knowledge of the tracer population model as a function of redshift and perfect knowledge of the flux cut \footnote{In the fictitious case where we observe all galaxies in the universe, the number of objects would not be altered by this magnification, only the observed flux.}.
Additionally, number counts of galaxies are obtained from measured redshifts, which are not the same as the cosmological redshift of a source moving with the background expansion of the universe \cite{challinor_lewis_gr, jsh_gr, yoo_plus_09_gr,baldauf_gr}.
(see also Appendix~\ref{app:mag}). 
The magnification effect \cite{bartelmann_schneider_wl_review} is crucial to model properly for $n$-point function analyses of the observed tracer density (``magnification bias''). 
The flux cut also alters the redshift evolution of the observed galaxy sample number density in an angle-integrated sense \cite{challinor_lewis_gr,maartens_bev} in addition to the contribution from magnification.

For the simple case of a magnitude/flux-limited survey, it is straightforward to include this well-known effect.
The constant (idealized as bolometric) flux cut is unrelated to the intrinsic galaxy properties, but induces a redshift-dependent cut in the intrinsic luminosity.
The resulting luminosity limit then serves as the physical selection parameter $\phi(z)$  in eqn.~\ref{eqn:general_selection_derivative} in this simple example.
Following Ref.~\cite{challinor_lewis_gr}, for a flux limit $F_{\mathrm{min}}$, the corresponding redshift-dependence luminosity is $L_{\mathrm{min}}(z) = 4\pi d_L^2(z) F_{\mathrm{min}}$, and so the total redshift  derivative is
\begin{widetext}
    \label{eqn:maglim_sample}
    \begin{align}
    \frac{d\log\bar{n}^{(\mathrm{obs})}_{t}(z,L_{\mathrm{min}}(z))}{d\log(1+z)} &= \frac{\partial\log\bar{n}^{(\mathrm{obs})}_{t}(z,L_{\mathrm{min}}(z))}{\partial\log(1+z)}\bigg\rvert_{L_{\mathrm{min}}(z)} + \frac{\partial\log\bar{n}^{(\mathrm{obs})}_{t}(m_c)}{\partial \log L_{\mathrm{min}}(z)}\bigg\rvert_{z} \frac{\partial \log L_{\mathrm{min}}(z) }{\partial \log(1+z)}\bigg\rvert_{z} \\
        % &= b_{e} + \left(\frac52 s\right)\left(2 + \frac{2}{\mathcal{H}(z)\chi(z)} \right)\\
         &= -b_{e} + \left(\frac{\partial\log\bar{n}^{(\mathrm{obs})}_{t}(m_c)}{\partial \log L_{\mathrm{min}}(z)}\right)\left(2 + \frac{2}{\mathcal{H}(z)\chi(z)} \right)
        % &= b_{e} + 5 s\left(1 + \frac{1}{\mathcal{H}(z)\chi(z)} \right),
    \end{align}
\end{widetext}
where replacing the first term in parentheses with $-\frac52 s= -\frac52s(z,m_{c})$ gives the familiar magnification contribution, which can be interpreted in terms of the faint end slope of the sample luminosity function.
While the magnification affect must be accounted for in the observed number density \cite{maartens_bev,challinor_lewis_gr}
(See Appendix~\ref{app:mag} for an application of magnification contributions to observed number density, and discussion of other related contributions.), any magnitude limit also impacts the time evolution of number density independent of any magnification.
We now turn to a simulated galaxy sample to show the impact of this effect in a simplified context. 

\subsubsection{\label{subsec:selfunc_2} Simulated photometric selections}

Realistic surveys are not as simple as the flux-limited picture above, and almost always involve other selection criteria, such as with apparent magnitude cuts in different bands, as well as color cuts, to obtain a tracer source population suitable for LSS analysis (e.g. that approaches uniform spatial number density).
For example, color combinations (such as the $c_\parallel, ~d_\perp$ used for BOSS LRGs \cite{reid_target_selection}) are based in astrophysical models of galaxy formation, often informed by previous observations, that tie the observed colors (and other properties) to the stellar mass or age of galaxies of interest.
Such a sequence of cuts is often necessarily complicated, and strongly influences the redshift dependence of the observed tracer number density. 
Rather than attempt to model a realistic galaxy sample in its full complexity, we instead provide a qualitative example of the kinds of effects these types of selection criteria can produce.

\begin{figure}[h!]
    \includegraphics[width=0.5\textwidth]{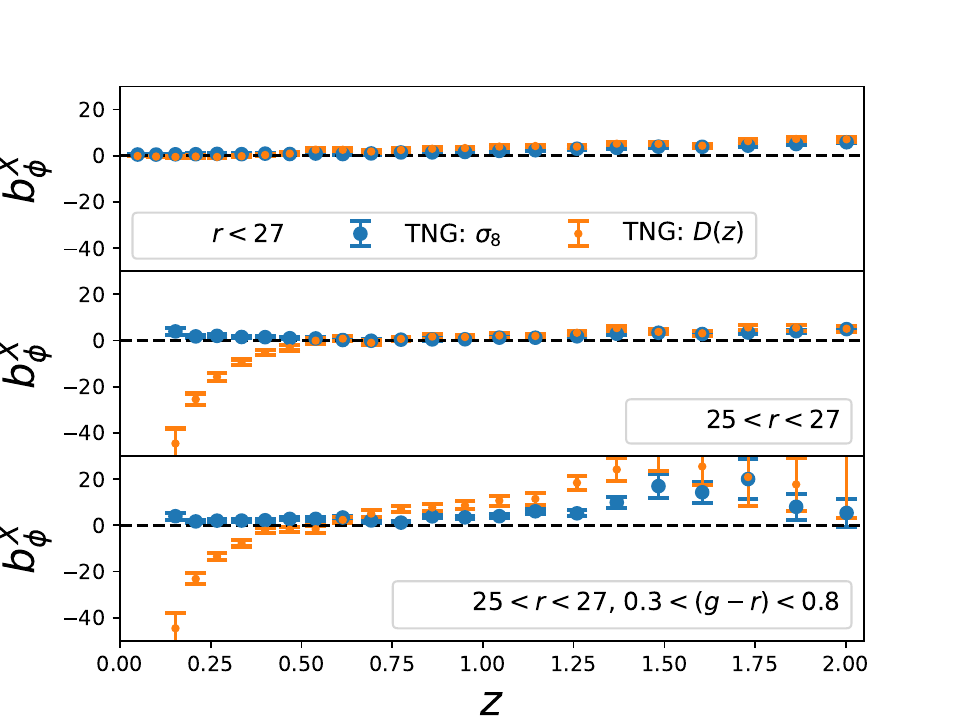}
    \caption{
    The effect of simple apparent magnitude and color selection on simulated galaxies. 
    The application of observed-frame magnitude cuts and color cuts spoils the equivalence between variance SU estimates of $\bphi$ and estimates from the growth of structure.
    \textit{Top:} The impact of a $r$ maximum magnitude cut in apparent $r$, which is rather limited.
    \textit{Center:} The impact of both a minimum and maximum cut in $r$, which leads to a strong disagreement between the two estimates at low redshift.
    \textit{Bottom:} The impact of the cuts described above as well as cuts in $(g-r)$ color - significant disagreement between the estimates is now present at most redshifts.
        }
    \label{fig:amag_dependence}
\end{figure}

Figure~\ref{fig:amag_dependence} shows the estimates of $\bphi$ from the variance SU and time evolution for a sample defined according to a simple selection.
The selection is performed with apparent magnitudes (computed simply via distance modulus while neglecting k corrections \cite{hogg_kcorr,maartens_bev,rossiter_fnl_gr_bias,guo_13_sdss_kpluse,Sobral-Blanco:2024qlb}) for the SDSS bands and in the (unattenuated) colors using the CAMELS photometry and TNG galaxy formation model.
In Fig.~\ref{fig:amag_dependence}, we consider a minimum $r$ magnitude cut (\textit{top panel}), a bin in $r$ magnitude (\textit{middle panel}), and a simultaneous cut bin in $r$ magnitude and $(g-r)$ color (\textit{bottom panel}).
There we see the impact of selection on the agreement between $\bphi$ computed via the variance SU (blue circles) and from time evolution (orange points).
Errorbars are computed from jackknife realizations as discussed previously.
The impact is limited for a simple maximum magnitude cut (\textit{top panel}), but the transition to quite negative $\bphi^{(D(z))}$ at $z\lesssim 0.5$ as a result of the bright end magnitude cut is striking.
When additionally adding a color cut, we see that at high $z$ especially, both methods predict higher $\bphi$.
This expected due to the physical response of the color distribution to LPNG, as discussed in the previous Section.
However, we also see that at higher $z$, before the errorbars get too large, there is significant disagreement between the two methods, indicating the impact of the color selection on the time evolution estimate.

Roughly, these choices of selection are meant to give a sense of the impact of basic selection on bright, red galaxies.
In particular, a more complicated composition of such cuts is used in, e.g., BOSS \cite{reid_target_selection} and DESI \cite{zhou_lrg_target} LRG target selections.
While out of scope of the current work, it would be especially interesting to extend the simple discussion here to both a realistic selection model for LRGs in these surveys.
This is not possible at the moment with CAMELS, principally due to the limited volume leading to insufficiently massive halos to host LRGs (hence low values of $\bphi$ from either method), and, to the best of our knowledge, there is not another suite of publicly-available simulations run at different values of $\sigma_8$ that facilitate a comparison with the variance separate universe $\bphi$.
Using a larger volume would also allow for more complicated selections in this demonstration thanks to the higher number of galaxies.

The qualitative features seen in Figure~\ref{fig:amag_dependence} are of the type that appear when attempting to take the number density derivative of measured galaxy simulations in BOSS (see Appendix~\ref{app:mag}), though are less quantitatively dramatic.

\section{\label{sec:discussion} Discussion}

Obtaining an estimate of $b_\phi$ from time evolution of LSS tracers has immediate application to $n$-point function analysis in searches for multi-field inflation via $\fnlloc$.
While finding a detection of the amplitude of local PNG in the galaxy distribution, $\bphi \fnlloc$ does not require knowledge of $\bphi$, obtaining an upper bound on $\fnlloc$, or, determining the value of $\fnlloc$ if $\bphi \fnlloc$ is detected, does require knowledge of $\bphi$.
For this reason, developing trustworthy priors on $\bphi$ remains urgent, especially as PNG analyses beyond DESI, such as SPHEREx \cite{spherex14}, or the planned Stage 5 spectroscopic experiment \cite{spec_s5} are forecasted to exceed the precision of Planck on $\fnlloc$.

Given the hurdles alluded to in the previous Section, it may be difficult to exactly estimate $\bphi$ from the measured number density evolution with redshift for a general LSS tracer.
However, even partial information can inform a $\bphi$ prior that can aid in robust analysis of $\fnlloc$, and after a LSS survey is completed, further follow-up study of the galaxy population in a galaxy evolution analysis may improve knowledge of the underlying galaxy population, which can provide more confident $\bphi$ priors with this method \footnote{E.g., such priors may be obtained from simulations \cite{ivanov_sfpng_priors}}. 
It is also clear that the galaxy halo connection itself, contrary to what is often assumed in HOD modeling, depends on cosmology, at least through $\sigma_8$ \cite{lovell_camels_photometry,2020JCAP...12..013B,marinucci_bphi_sam,voivodic_response_phi}.
Further exploration of how to best separate the contribution to halo occupation that is physical from that induced by the selection, e.g. when using HOD models, requires further study.

On the other hand, with sufficient understanding of the survey selection function, a subsample of galaxies may be chosen \textit{specifically for the purpose} of maximizing the value of $\bphi$, which, depending on the sample number density, can give the tightest constraints on $\fnlloc$.
This can be done via educated guess, or even at random, and can be pushed even further using a multi-tracer analysis applied to these $\Delta \bphi$-maximizing samples \cite{sullivan_bphi_forecast_ab,barreira_krause_23,fondi_ab_forecast_24}\footnote{For the case of strong assembly bias, the difference in $\Delta \bphi$ is most important, but the sensitivity of multitracer in the sample variance limit involves the product of $\bphi$ and $b_1$ across samples. }.
Though we note care must be taken to properly treat systematic effects on the redshift-dependence of the galaxy number density (as discussed here and beyond) when splitting real galaxy catalogs using such multi-tracer methods \footnote{For example, while argued not to be an issue for LRGs, for emission line galaxies (ELGs), it may be necessary to correct for the interaction between galaxy velocities and color selection \cite{doppler_bias}.}

Partial information on $b_\phi$ can also come from other measurements of the evolution bias.
In the absence of PNG, the evolution bias (and magnification bias) is critical to accurate modeling of the large-scale tracer power spectrum \cite{challinor_lewis_gr}.
In principle, these parameters can be estimated by performing multi-tracer splits and measuring the dipole and octopole of the large-scale tracer power spectrum \cite{Sobral-Blanco:2024qlb}, which could be folded into an estimate of $\bphi$ from time evolution.
This is true especially at higher redshifts where the number density redshift derivative is a good approximation to the response to the growth of structure.
Incorporating knowledge of $b_{e}$ (and magnification) obtained through such estimation, or jointly with the methods here, is worth further study.

Through perhaps most readily applicable to the case of LRGs, it would also be interesting to apply this method to a variety tracers that will dominate future LSS surveys (such as ELGs or higher-redshift star-forming galaxies), especially since the redshift evolution of such tracers are quite challenging to model.
Another example is that of quasars, which trace high redshifts where the LPNG signal is stronger \cite{edmonnd_qso_lrg_desi_lpng,cagliari_qso_eboss_24} \footnote{Quasars potentially also have a less-complicated sample definition, at least at the high-redshift end \cite{palanque-delabrouille_qso_lf_sdss,palanque-delabrouille_qso_lf_eboss, wang_quasar_bev}} or even more general tracers, such as the Lyman-alpha forest or line intensity.

\section{\label{sec:conclusions} Conclusions}
Local Primordial non-Gaussianity is a primary target of large-scale structure surveys for the next decade.
Leading efforts to constrain the amplitude of LPNG, $\fnlloc$, are driven by the prior on the LPNG bias parameter $\bphi$.
Here we introduced a new way to estimate $\bphi$ directly from the time evolution of the number density of tracers.
We motivated the connection between time-evolution and LPNG, which boils down to the physical impact of long-wavelength metric (``potential'') fluctuation as a rescaling of the local time evolution.
As a result, by measuring the time dependence of a modeled population of tracers, one can simulate the impact of adding a long-wavelength potential mode as is relevant for LPNG.

We showed that this correspondence persists in dark-matter-only and hydrodynamical simulations, and is \textit{not} limited to universal tracers.
In fact, it applies to tracers defined by any properties that themselves respond to the growth of structure (including halo concentration and intrinsic galaxy colors).
We also discussed, using simulated galaxies, the care that must be taken to correctly model the selection function used to define a real galaxy population to perform this measurement in practice.
This natural next step following this work is to obtain constraints on $\fnlloc$ from these galaxy samples in a $n-$point function analysis using priors informed by time evolution in the data, which will be the subject of future work.

\acknowledgments
JMS thanks Liang Dai, Fabian Schmidt, Martin White, and Matias Zaldarriaga for helpful conversations about CFCs, evolution bias, BOSS LRGs, and bias universality respectively, as well as Neal Dalal, Olivier Dor\'e, Elisabeth Krause, and Salman Habib for useful discussion, and Francisco Villaescusa-Navarro, Chris Lovell, and the CAMELS and Quijote teams for making their simulations publicly accessible and assisting with their usage.
JMS was partially supported by a U.S. Dept. of Energy SCGSR award during the completion of this work. 
JMS also acknowledges that, in part, support for this work was provided by The Brinson Foundation through a Brinson Prize Fellowship grant.
This research used resources of the National Energy Research Scientific Computing Center (NERSC), a Department of Energy Office of Science User Facility using NERSC award HEP-ERCAP0028635, as well as the MIT Engaging system.
This research has made use of NASA's Astrophysics Data System.

\appendix
\section{Magnification and BOSS galaxies \label{app:mag}}

As a demonstration of the type of selection corrections that must be applied to real galaxy number densities when estimating $\bphi$ from time evolution, here we show the result of applying the magnification correction to BOSS DR12 Luminous Red Galaxies (LRGs).
We use galaxy catalog data from the DR12 data release of the completed Baryon Acoustic Oscillation Spectroscopic Survey (BOSS) \cite{alam_boss_data}.
The BOSS LRG selection \cite{reid_target_selection} was informed by the ``passively evolving'' LRG model of Ref.~\cite{maraston_passive_lrg_sps_model,maraston_cmass_stellar_mass}.
We use the LOWZ and CMASS LSS catalog samples and their responses of number density to foreground magnification computed in Ref.~\cite{reid_target_selection,wenzl_mag_bias_boss} to correct for the impact of magnification on comoving number density time evolution.
We compare both the case of $s = 1$ to the mean effective values of $s$ obtained by directly simulating the impact of magnification of the parameters used to define the tracer population from Ref.~\cite{wenzl_mag_bias_boss}.

%%%% FIGURE
\begin{figure}[h!]
    \includegraphics[width=0.5\textwidth]{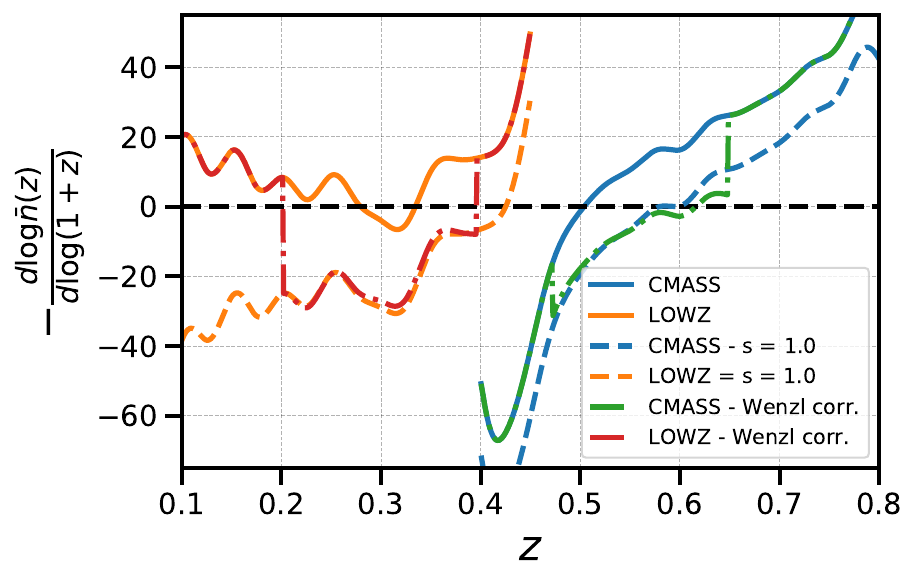}
    \caption{
    Observed number densities of the LOWZ and CMASS BOSS samples with and without redshift magnification corrections. 
    The solid curves show the ``raw'' uncorrected number density, dashed shows the $s=1$ magnification correction, and dash-dotted shows the color-aware magnification correction of Ref.~\cite{wenzl_mag_bias_boss}.
    }
    \label{fig:boss_nbar_bev}
\end{figure}

Figure~\ref{fig:boss_nbar_bev} shows the redshift finite difference derivatives of BOSS galaxies with (dashed lines for $s=1$ and dash-dotted for the mean correction of Ref.~\cite{wenzl_mag_bias_boss}) and without (solid lines) the magnification corrections applied.
The impact of the magnification corrections on the number density evolution, as well as the impact of using a selection-aware magnification correction to the number density evolution, while significant, is subdominant to the other selection criteria used to define the samples.
These results appear to be roughly consistent with recent work studying other contributions to the redshift dependence of DESI galaxies \cite{doppler_bias}.

\bibliography{apssamp}% Produces the bibliography via BibTeX.

\begin{thebibliography}{105}%
\makeatletter
\providecommand \@ifxundefined [1]{%
 \@ifx{#1\undefined}
}%
\providecommand \@ifnum [1]{%
 \ifnum #1\expandafter \@firstoftwo
 \else \expandafter \@secondoftwo
 \fi
}%
\providecommand \@ifx [1]{%
 \ifx #1\expandafter \@firstoftwo
 \else \expandafter \@secondoftwo
 \fi
}%
\providecommand \natexlab [1]{#1}%
\providecommand \enquote  [1]{``#1''}%
\providecommand \bibnamefont  [1]{#1}%
\providecommand \bibfnamefont [1]{#1}%
\providecommand \citenamefont [1]{#1}%
\providecommand \href@noop [0]{\@secondoftwo}%
\providecommand \href [0]{\begingroup \@sanitize@url \@href}%
\providecommand \@href[1]{\@@startlink{#1}\@@href}%
\providecommand \@@href[1]{\endgroup#1\@@endlink}%
\providecommand \@sanitize@url [0]{\catcode `\\12\catcode `\$12\catcode `\&12\catcode `\#12\catcode `\^12\catcode `\_12\catcode `\%12\relax}%
\providecommand \@@startlink[1]{}%
\providecommand \@@endlink[0]{}%
\providecommand \url  [0]{\begingroup\@sanitize@url \@url }%
\providecommand \@url [1]{\endgroup\@href {#1}{\urlprefix }}%
\providecommand \urlprefix  [0]{URL }%
\providecommand \Eprint [0]{\href }%
\providecommand \doibase [0]{https://doi.org/}%
\providecommand \selectlanguage [0]{\@gobble}%
\providecommand \bibinfo  [0]{\@secondoftwo}%
\providecommand \bibfield  [0]{\@secondoftwo}%
\providecommand \translation [1]{[#1]}%
\providecommand \BibitemOpen [0]{}%
\providecommand \bibitemStop [0]{}%
\providecommand \bibitemNoStop [0]{.\EOS\space}%
\providecommand \EOS [0]{\spacefactor3000\relax}%
\providecommand \BibitemShut  [1]{\csname bibitem#1\endcsname}%
\let\auto@bib@innerbib\@empty
%</preamble>
\bibitem [{Note1()}]{Note1}%
  \BibitemOpen
  \bibinfo {note} {\protect \url {https://www.desi.lbl.gov/}}\BibitemShut {NoStop}%
\bibitem [{Note2()}]{Note2}%
  \BibitemOpen
  \bibinfo {note} {\protect \url {https://www.cosmos.esa.int/web/euclid}}\BibitemShut {NoStop}%
\bibitem [{Note3()}]{Note3}%
  \BibitemOpen
  \bibinfo {note} {\protect \url {https://spherex.caltech.edu/}}\BibitemShut {NoStop}%
\bibitem [{Note4()}]{Note4}%
  \BibitemOpen
  \bibinfo {note} {\protect \url {https://www.lsst.org/}}\BibitemShut {NoStop}%
\bibitem [{Note5()}]{Note5}%
  \BibitemOpen
  \bibinfo {note} {\protect \url {https://www.spec-s5.org/}}\BibitemShut {NoStop}%
\bibitem [{\citenamefont {{Kehagias}}\ and\ \citenamefont {{Riotto}}(2013)}]{consistency_kehagias_13}%
  \BibitemOpen
  \bibfield  {author} {\bibinfo {author} {\bibfnamefont {A.}~\bibnamefont {{Kehagias}}}\ and\ \bibinfo {author} {\bibfnamefont {A.}~\bibnamefont {{Riotto}}},\ }\bibfield  {title} {\bibinfo {title} {{Symmetries and consistency relations in the large scale structure of the universe}},\ }\href {https://doi.org/10.1016/j.nuclphysb.2013.05.009} {\bibfield  {journal} {\bibinfo  {journal} {Nuclear Physics B}\ }\textbf {\bibinfo {volume} {873}},\ \bibinfo {pages} {514} (\bibinfo {year} {2013})},\ \Eprint {https://arxiv.org/abs/1302.0130} {arXiv:1302.0130 [astro-ph.CO]} \BibitemShut {NoStop}%
\bibitem [{\citenamefont {{Peloso}}\ and\ \citenamefont {{Pietroni}}(2013)}]{consistency_peloso_13}%
  \BibitemOpen
  \bibfield  {author} {\bibinfo {author} {\bibfnamefont {M.}~\bibnamefont {{Peloso}}}\ and\ \bibinfo {author} {\bibfnamefont {M.}~\bibnamefont {{Pietroni}}},\ }\bibfield  {title} {\bibinfo {title} {{Galilean invariance and the consistency relation for the nonlinear squeezed bispectrum of large scale structure}},\ }\href {https://doi.org/10.1088/1475-7516/2013/05/031} {\bibfield  {journal} {\bibinfo  {journal} {\jcap}\ }\textbf {\bibinfo {volume} {2013}},\ \bibinfo {eid} {031} (\bibinfo {year} {2013})},\ \Eprint {https://arxiv.org/abs/1302.0223} {arXiv:1302.0223 [astro-ph.CO]} \BibitemShut {NoStop}%
\bibitem [{\citenamefont {{Creminelli}}\ \emph {et~al.}(2013)\citenamefont {{Creminelli}}, \citenamefont {{Nore{\~n}a}}, \citenamefont {{Simonovi{\'c}}},\ and\ \citenamefont {{Vernizzi}}}]{consistency_creminelli_13}%
  \BibitemOpen
  \bibfield  {author} {\bibinfo {author} {\bibfnamefont {P.}~\bibnamefont {{Creminelli}}}, \bibinfo {author} {\bibfnamefont {J.}~\bibnamefont {{Nore{\~n}a}}}, \bibinfo {author} {\bibfnamefont {M.}~\bibnamefont {{Simonovi{\'c}}}},\ and\ \bibinfo {author} {\bibfnamefont {F.}~\bibnamefont {{Vernizzi}}},\ }\bibfield  {title} {\bibinfo {title} {{Single-field consistency relations of large scale structure}},\ }\href {https://doi.org/10.1088/1475-7516/2013/12/025} {\bibfield  {journal} {\bibinfo  {journal} {\jcap}\ }\textbf {\bibinfo {volume} {2013}},\ \bibinfo {eid} {025} (\bibinfo {year} {2013})},\ \Eprint {https://arxiv.org/abs/1309.3557} {arXiv:1309.3557 [astro-ph.CO]} \BibitemShut {NoStop}%
\bibitem [{\citenamefont {{Dalal}}\ \emph {et~al.}(2008)\citenamefont {{Dalal}}, \citenamefont {{Dor{\'e}}}, \citenamefont {{Huterer}},\ and\ \citenamefont {{Shirokov}}}]{Dalal08}%
  \BibitemOpen
  \bibfield  {author} {\bibinfo {author} {\bibfnamefont {N.}~\bibnamefont {{Dalal}}}, \bibinfo {author} {\bibfnamefont {O.}~\bibnamefont {{Dor{\'e}}}}, \bibinfo {author} {\bibfnamefont {D.}~\bibnamefont {{Huterer}}},\ and\ \bibinfo {author} {\bibfnamefont {A.}~\bibnamefont {{Shirokov}}},\ }\bibfield  {title} {\bibinfo {title} {{Imprints of primordial non-Gaussianities on large-scale structure: Scale-dependent bias and abundance of virialized objects}},\ }\href {https://doi.org/10.1103/PhysRevD.77.123514} {\bibfield  {journal} {\bibinfo  {journal} {\prd}\ }\textbf {\bibinfo {volume} {77}},\ \bibinfo {eid} {123514} (\bibinfo {year} {2008})},\ \Eprint {https://arxiv.org/abs/0710.4560} {arXiv:0710.4560 [astro-ph]} \BibitemShut {NoStop}%
\bibitem [{\citenamefont {{Slosar}}\ \emph {et~al.}(2008)\citenamefont {{Slosar}}, \citenamefont {{Hirata}}, \citenamefont {{Seljak}}, \citenamefont {{Ho}},\ and\ \citenamefont {{Padmanabhan}}}]{Slosar08}%
  \BibitemOpen
  \bibfield  {author} {\bibinfo {author} {\bibfnamefont {A.}~\bibnamefont {{Slosar}}}, \bibinfo {author} {\bibfnamefont {C.}~\bibnamefont {{Hirata}}}, \bibinfo {author} {\bibfnamefont {U.}~\bibnamefont {{Seljak}}}, \bibinfo {author} {\bibfnamefont {S.}~\bibnamefont {{Ho}}},\ and\ \bibinfo {author} {\bibfnamefont {N.}~\bibnamefont {{Padmanabhan}}},\ }\bibfield  {title} {\bibinfo {title} {{Constraints on local primordial non-Gaussianity from large scale structure}},\ }\href {https://doi.org/10.1088/1475-7516/2008/08/031} {\bibfield  {journal} {\bibinfo  {journal} {\jcap}\ }\textbf {\bibinfo {volume} {2008}},\ \bibinfo {eid} {031} (\bibinfo {year} {2008})},\ \Eprint {https://arxiv.org/abs/0805.3580} {arXiv:0805.3580 [astro-ph]} \BibitemShut {NoStop}%
\bibitem [{\citenamefont {{Verde}}\ and\ \citenamefont {{Matarrese}}(2009)}]{verde_matarrese_pngbias}%
  \BibitemOpen
  \bibfield  {author} {\bibinfo {author} {\bibfnamefont {L.}~\bibnamefont {{Verde}}}\ and\ \bibinfo {author} {\bibfnamefont {S.}~\bibnamefont {{Matarrese}}},\ }\bibfield  {title} {\bibinfo {title} {{Detectability of the Effect of Inflationary Non-Gaussianity on Halo Bias}},\ }\href {https://doi.org/10.1088/0004-637X/706/1/L91} {\bibfield  {journal} {\bibinfo  {journal} {\apjl}\ }\textbf {\bibinfo {volume} {706}},\ \bibinfo {pages} {L91} (\bibinfo {year} {2009})},\ \Eprint {https://arxiv.org/abs/0909.3224} {arXiv:0909.3224 [astro-ph.CO]} \BibitemShut {NoStop}%
\bibitem [{\citenamefont {{Carbone}}\ \emph {et~al.}(2008)\citenamefont {{Carbone}}, \citenamefont {{Verde}},\ and\ \citenamefont {{Matarrese}}}]{carbone_png_bas}%
  \BibitemOpen
  \bibfield  {author} {\bibinfo {author} {\bibfnamefont {C.}~\bibnamefont {{Carbone}}}, \bibinfo {author} {\bibfnamefont {L.}~\bibnamefont {{Verde}}},\ and\ \bibinfo {author} {\bibfnamefont {S.}~\bibnamefont {{Matarrese}}},\ }\bibfield  {title} {\bibinfo {title} {{Non-Gaussian Halo Bias and Future Galaxy Surveys}},\ }\href {https://doi.org/10.1086/592020} {\bibfield  {journal} {\bibinfo  {journal} {\apjl}\ }\textbf {\bibinfo {volume} {684}},\ \bibinfo {pages} {L1} (\bibinfo {year} {2008})},\ \Eprint {https://arxiv.org/abs/0806.1950} {arXiv:0806.1950 [astro-ph]} \BibitemShut {NoStop}%
\bibitem [{\citenamefont {{Desjacques}}\ \emph {et~al.}(2009)\citenamefont {{Desjacques}}, \citenamefont {{Seljak}},\ and\ \citenamefont {{Iliev}}}]{desjacques_seljak_iliev_png_bias_halos}%
  \BibitemOpen
  \bibfield  {author} {\bibinfo {author} {\bibfnamefont {V.}~\bibnamefont {{Desjacques}}}, \bibinfo {author} {\bibfnamefont {U.}~\bibnamefont {{Seljak}}},\ and\ \bibinfo {author} {\bibfnamefont {I.~T.}\ \bibnamefont {{Iliev}}},\ }\bibfield  {title} {\bibinfo {title} {{Scale-dependent bias induced by local non-Gaussianity: a comparison to N-body simulations}},\ }\href {https://doi.org/10.1111/j.1365-2966.2009.14721.x} {\bibfield  {journal} {\bibinfo  {journal} {\mnras}\ }\textbf {\bibinfo {volume} {396}},\ \bibinfo {pages} {85} (\bibinfo {year} {2009})},\ \Eprint {https://arxiv.org/abs/0811.2748} {arXiv:0811.2748 [astro-ph]} \BibitemShut {NoStop}%
\bibitem [{\citenamefont {{Alvarez}}\ \emph {et~al.}(2014)\citenamefont {{Alvarez}}, \citenamefont {{Baldauf}}, \citenamefont {{Bond}}, \citenamefont {{Dalal}}, \citenamefont {{de Putter}}, \citenamefont {{Dor{\'e}}}, \citenamefont {{Green}}, \citenamefont {{Hirata}}, \citenamefont {{Huang}}, \citenamefont {{Huterer}}, \citenamefont {{Jeong}}, \citenamefont {{Johnson}}, \citenamefont {{Krause}}, \citenamefont {{Loverde}}, \citenamefont {{Meyers}}, \citenamefont {{Meerburg}}, \citenamefont {{Senatore}}, \citenamefont {{Shandera}}, \citenamefont {{Silverstein}}, \citenamefont {{Slosar}}, \citenamefont {{Smith}}, \citenamefont {{Zaldarriaga}}, \citenamefont {{Assassi}}, \citenamefont {{Braden}}, \citenamefont {{Hajian}}, \citenamefont {{Kobayashi}}, \citenamefont {{Stein}},\ and\ \citenamefont {{van Engelen}}}]{2014arXiv1412.4671A}%
  \BibitemOpen
  \bibfield  {author} {\bibinfo {author} {\bibfnamefont {M.}~\bibnamefont {{Alvarez}}}, \bibinfo {author} {\bibfnamefont {T.}~\bibnamefont {{Baldauf}}}, \bibinfo {author} {\bibfnamefont {J.~R.}\ \bibnamefont {{Bond}}}, \bibinfo {author} {\bibfnamefont {N.}~\bibnamefont {{Dalal}}}, \bibinfo {author} {\bibfnamefont {R.}~\bibnamefont {{de Putter}}}, \bibinfo {author} {\bibfnamefont {O.}~\bibnamefont {{Dor{\'e}}}}, \bibinfo {author} {\bibfnamefont {D.}~\bibnamefont {{Green}}}, \bibinfo {author} {\bibfnamefont {C.}~\bibnamefont {{Hirata}}}, \bibinfo {author} {\bibfnamefont {Z.}~\bibnamefont {{Huang}}}, \bibinfo {author} {\bibfnamefont {D.}~\bibnamefont {{Huterer}}}, \bibinfo {author} {\bibfnamefont {D.}~\bibnamefont {{Jeong}}}, \bibinfo {author} {\bibfnamefont {M.~C.}\ \bibnamefont {{Johnson}}}, \bibinfo {author} {\bibfnamefont {E.}~\bibnamefont {{Krause}}}, \bibinfo {author} {\bibfnamefont {M.}~\bibnamefont {{Loverde}}}, \bibinfo {author} {\bibfnamefont {J.}~\bibnamefont {{Meyers}}}, \bibinfo {author}
  {\bibfnamefont {P.~D.}\ \bibnamefont {{Meerburg}}}, \bibinfo {author} {\bibfnamefont {L.}~\bibnamefont {{Senatore}}}, \bibinfo {author} {\bibfnamefont {S.}~\bibnamefont {{Shandera}}}, \bibinfo {author} {\bibfnamefont {E.}~\bibnamefont {{Silverstein}}}, \bibinfo {author} {\bibfnamefont {A.}~\bibnamefont {{Slosar}}}, \bibinfo {author} {\bibfnamefont {K.}~\bibnamefont {{Smith}}}, \bibinfo {author} {\bibfnamefont {M.}~\bibnamefont {{Zaldarriaga}}}, \bibinfo {author} {\bibfnamefont {V.}~\bibnamefont {{Assassi}}}, \bibinfo {author} {\bibfnamefont {J.}~\bibnamefont {{Braden}}}, \bibinfo {author} {\bibfnamefont {A.}~\bibnamefont {{Hajian}}}, \bibinfo {author} {\bibfnamefont {T.}~\bibnamefont {{Kobayashi}}}, \bibinfo {author} {\bibfnamefont {G.}~\bibnamefont {{Stein}}},\ and\ \bibinfo {author} {\bibfnamefont {A.}~\bibnamefont {{van Engelen}}},\ }\bibfield  {title} {\bibinfo {title} {{Testing Inflation with Large Scale Structure: Connecting Hopes with Reality}},\ }\href {https://doi.org/10.48550/arXiv.1412.4671}
  {\bibfield  {journal} {\bibinfo  {journal} {arXiv e-prints}\ ,\ \bibinfo {eid} {arXiv:1412.4671}} (\bibinfo {year} {2014})},\ \Eprint {https://arxiv.org/abs/1412.4671} {arXiv:1412.4671 [astro-ph.CO]} \BibitemShut {NoStop}%
\bibitem [{\citenamefont {{Jeong}}\ and\ \citenamefont {{Komatsu}}(2009)}]{jeong_komatsu_09_sdb}%
  \BibitemOpen
  \bibfield  {author} {\bibinfo {author} {\bibfnamefont {D.}~\bibnamefont {{Jeong}}}\ and\ \bibinfo {author} {\bibfnamefont {E.}~\bibnamefont {{Komatsu}}},\ }\bibfield  {title} {\bibinfo {title} {{Primordial Non-Gaussianity, Scale-dependent Bias, and the Bispectrum of Galaxies}},\ }\href {https://doi.org/10.1088/0004-637X/703/2/1230} {\bibfield  {journal} {\bibinfo  {journal} {\apj}\ }\textbf {\bibinfo {volume} {703}},\ \bibinfo {pages} {1230} (\bibinfo {year} {2009})},\ \Eprint {https://arxiv.org/abs/0904.0497} {arXiv:0904.0497 [astro-ph.CO]} \BibitemShut {NoStop}%
\bibitem [{\citenamefont {{de Putter}}\ and\ \citenamefont {{Dor{\'e}}}(2017)}]{dePutterDore17}%
  \BibitemOpen
  \bibfield  {author} {\bibinfo {author} {\bibfnamefont {R.}~\bibnamefont {{de Putter}}}\ and\ \bibinfo {author} {\bibfnamefont {O.}~\bibnamefont {{Dor{\'e}}}},\ }\bibfield  {title} {\bibinfo {title} {{Designing an inflation galaxy survey: How to measure {\ensuremath{\sigma}} (f$_{NL}$){\ensuremath{\sim}}1 using scale-dependent galaxy bias}},\ }\href {https://doi.org/10.1103/PhysRevD.95.123513} {\bibfield  {journal} {\bibinfo  {journal} {\prd}\ }\textbf {\bibinfo {volume} {95}},\ \bibinfo {eid} {123513} (\bibinfo {year} {2017})},\ \Eprint {https://arxiv.org/abs/1412.3854} {arXiv:1412.3854 [astro-ph.CO]} \BibitemShut {NoStop}%
\bibitem [{\citenamefont {{Giri}}\ \emph {et~al.}(2023)\citenamefont {{Giri}}, \citenamefont {{M{\"u}nchmeyer}},\ and\ \citenamefont {{Smith}}}]{Giri23}%
  \BibitemOpen
  \bibfield  {author} {\bibinfo {author} {\bibfnamefont {U.}~\bibnamefont {{Giri}}}, \bibinfo {author} {\bibfnamefont {M.}~\bibnamefont {{M{\"u}nchmeyer}}},\ and\ \bibinfo {author} {\bibfnamefont {K.~M.}\ \bibnamefont {{Smith}}},\ }\bibfield  {title} {\bibinfo {title} {{Constraining $f_{NL}$ using the Large-Scale Modulation of Small-Scale Statistics}},\ }\href {https://doi.org/10.48550/arXiv.2305.03070} {\bibfield  {journal} {\bibinfo  {journal} {arXiv e-prints}\ ,\ \bibinfo {eid} {arXiv:2305.03070}} (\bibinfo {year} {2023})},\ \Eprint {https://arxiv.org/abs/2305.03070} {arXiv:2305.03070 [astro-ph.CO]} \BibitemShut {NoStop}%
\bibitem [{Note6()}]{Note6}%
  \BibitemOpen
  \bibinfo {note} {We will frequently use the word ``halo'' or ``galaxy'', since those are the cases of interest here, but all conclusions stated here generally apply to any LSS tracer.}\BibitemShut {Stop}%
\bibitem [{Note7()}]{Note7}%
  \BibitemOpen
  \bibinfo {note} {With a universal mass function (UMF) depending only on peak height $\nu (M,z) = \protect \frac {\delta _c}{\sigma (M,z)}$ for critical overdensity $\delta _c$ and amplitude of smoothed matter fluctuations $\sigma $}\BibitemShut {NoStop}%
\bibitem [{\citenamefont {{Biagetti}}\ \emph {et~al.}(2017)\citenamefont {{Biagetti}}, \citenamefont {{Lazeyras}}, \citenamefont {{Baldauf}}, \citenamefont {{Desjacques}},\ and\ \citenamefont {{Schmidt}}}]{Biagettibphi}%
  \BibitemOpen
  \bibfield  {author} {\bibinfo {author} {\bibfnamefont {M.}~\bibnamefont {{Biagetti}}}, \bibinfo {author} {\bibfnamefont {T.}~\bibnamefont {{Lazeyras}}}, \bibinfo {author} {\bibfnamefont {T.}~\bibnamefont {{Baldauf}}}, \bibinfo {author} {\bibfnamefont {V.}~\bibnamefont {{Desjacques}}},\ and\ \bibinfo {author} {\bibfnamefont {F.}~\bibnamefont {{Schmidt}}},\ }\bibfield  {title} {\bibinfo {title} {{Verifying the consistency relation for the scale-dependent bias from local primordial non-Gaussianity}},\ }\href {https://doi.org/10.1093/mnras/stx714} {\bibfield  {journal} {\bibinfo  {journal} {\mnras}\ }\textbf {\bibinfo {volume} {468}},\ \bibinfo {pages} {3277} (\bibinfo {year} {2017})},\ \Eprint {https://arxiv.org/abs/1611.04901} {arXiv:1611.04901 [astro-ph.CO]} \BibitemShut {NoStop}%
\bibitem [{\citenamefont {{Barreira}}\ \emph {et~al.}(2020)\citenamefont {{Barreira}}, \citenamefont {{Cabass}}, \citenamefont {{Schmidt}}, \citenamefont {{Pillepich}},\ and\ \citenamefont {{Nelson}}}]{2020JCAP...12..013B}%
  \BibitemOpen
  \bibfield  {author} {\bibinfo {author} {\bibfnamefont {A.}~\bibnamefont {{Barreira}}}, \bibinfo {author} {\bibfnamefont {G.}~\bibnamefont {{Cabass}}}, \bibinfo {author} {\bibfnamefont {F.}~\bibnamefont {{Schmidt}}}, \bibinfo {author} {\bibfnamefont {A.}~\bibnamefont {{Pillepich}}},\ and\ \bibinfo {author} {\bibfnamefont {D.}~\bibnamefont {{Nelson}}},\ }\bibfield  {title} {\bibinfo {title} {{Galaxy bias and primordial non-Gaussianity: insights from galaxy formation simulations with IllustrisTNG}},\ }\href {https://doi.org/10.1088/1475-7516/2020/12/013} {\bibfield  {journal} {\bibinfo  {journal} {\jcap}\ }\textbf {\bibinfo {volume} {2020}},\ \bibinfo {eid} {013} (\bibinfo {year} {2020})},\ \Eprint {https://arxiv.org/abs/2006.09368} {arXiv:2006.09368 [astro-ph.CO]} \BibitemShut {NoStop}%
\bibitem [{\citenamefont {{Barreira}}\ \emph {et~al.}(2021)\citenamefont {{Barreira}}, \citenamefont {{Lazeyras}},\ and\ \citenamefont {{Schmidt}}}]{barreira_field_b1_b2}%
  \BibitemOpen
  \bibfield  {author} {\bibinfo {author} {\bibfnamefont {A.}~\bibnamefont {{Barreira}}}, \bibinfo {author} {\bibfnamefont {T.}~\bibnamefont {{Lazeyras}}},\ and\ \bibinfo {author} {\bibfnamefont {F.}~\bibnamefont {{Schmidt}}},\ }\bibfield  {title} {\bibinfo {title} {{Galaxy bias from forward models: linear and second-order bias of IllustrisTNG galaxies}},\ }\href {https://doi.org/10.1088/1475-7516/2021/08/029} {\bibfield  {journal} {\bibinfo  {journal} {\jcap}\ }\textbf {\bibinfo {volume} {2021}},\ \bibinfo {eid} {029} (\bibinfo {year} {2021})},\ \Eprint {https://arxiv.org/abs/2105.02876} {arXiv:2105.02876 [astro-ph.CO]} \BibitemShut {NoStop}%
\bibitem [{\citenamefont {{Barreira}}(2022{\natexlab{a}})}]{2022JCAP...04..057B}%
  \BibitemOpen
  \bibfield  {author} {\bibinfo {author} {\bibfnamefont {A.}~\bibnamefont {{Barreira}}},\ }\bibfield  {title} {\bibinfo {title} {{The local PNG bias of neutral Hydrogen, H$_{I}$}},\ }\href {https://doi.org/10.1088/1475-7516/2022/04/057} {\bibfield  {journal} {\bibinfo  {journal} {\jcap}\ }\textbf {\bibinfo {volume} {2022}},\ \bibinfo {eid} {057} (\bibinfo {year} {2022}{\natexlab{a}})},\ \Eprint {https://arxiv.org/abs/2112.03253} {arXiv:2112.03253 [astro-ph.CO]} \BibitemShut {NoStop}%
\bibitem [{\citenamefont {{Barreira}}(2022{\natexlab{b}})}]{2022JCAP...01..033B}%
  \BibitemOpen
  \bibfield  {author} {\bibinfo {author} {\bibfnamefont {A.}~\bibnamefont {{Barreira}}},\ }\bibfield  {title} {\bibinfo {title} {{Predictions for local PNG bias in the galaxy power spectrum and bispectrum and the consequences for f $_{NL}$ constraints}},\ }\href {https://doi.org/10.1088/1475-7516/2022/01/033} {\bibfield  {journal} {\bibinfo  {journal} {\jcap}\ }\textbf {\bibinfo {volume} {2022}},\ \bibinfo {eid} {033} (\bibinfo {year} {2022}{\natexlab{b}})},\ \Eprint {https://arxiv.org/abs/2107.06887} {arXiv:2107.06887 [astro-ph.CO]} \BibitemShut {NoStop}%
\bibitem [{\citenamefont {{Reid}}\ \emph {et~al.}(2010)\citenamefont {{Reid}}, \citenamefont {{Verde}}, \citenamefont {{Dolag}}, \citenamefont {{Matarrese}},\ and\ \citenamefont {{Moscardini}}}]{reid_ab_10}%
  \BibitemOpen
  \bibfield  {author} {\bibinfo {author} {\bibfnamefont {B.~A.}\ \bibnamefont {{Reid}}}, \bibinfo {author} {\bibfnamefont {L.}~\bibnamefont {{Verde}}}, \bibinfo {author} {\bibfnamefont {K.}~\bibnamefont {{Dolag}}}, \bibinfo {author} {\bibfnamefont {S.}~\bibnamefont {{Matarrese}}},\ and\ \bibinfo {author} {\bibfnamefont {L.}~\bibnamefont {{Moscardini}}},\ }\bibfield  {title} {\bibinfo {title} {{Non-Gaussian halo assembly bias}},\ }\href {https://doi.org/10.1088/1475-7516/2010/07/013} {\bibfield  {journal} {\bibinfo  {journal} {\jcap}\ }\textbf {\bibinfo {volume} {2010}},\ \bibinfo {eid} {013} (\bibinfo {year} {2010})},\ \Eprint {https://arxiv.org/abs/1004.1637} {arXiv:1004.1637 [astro-ph.CO]} \BibitemShut {NoStop}%
\bibitem [{\citenamefont {{Sullivan}}\ \emph {et~al.}(2023)\citenamefont {{Sullivan}}, \citenamefont {{Prijon}},\ and\ \citenamefont {{Seljak}}}]{sullivan_bphi_forecast_ab}%
  \BibitemOpen
  \bibfield  {author} {\bibinfo {author} {\bibfnamefont {J.~M.}\ \bibnamefont {{Sullivan}}}, \bibinfo {author} {\bibfnamefont {T.}~\bibnamefont {{Prijon}}},\ and\ \bibinfo {author} {\bibfnamefont {U.}~\bibnamefont {{Seljak}}},\ }\bibfield  {title} {\bibinfo {title} {{Learning to concentrate: multi-tracer forecasts on local primordial non-Gaussianity with machine-learned bias}},\ }\href {https://doi.org/10.1088/1475-7516/2023/08/004} {\bibfield  {journal} {\bibinfo  {journal} {\jcap}\ }\textbf {\bibinfo {volume} {2023}},\ \bibinfo {eid} {004} (\bibinfo {year} {2023})},\ \Eprint {https://arxiv.org/abs/2303.08901} {arXiv:2303.08901 [astro-ph.CO]} \BibitemShut {NoStop}%
\bibitem [{\citenamefont {{Lazeyras}}\ \emph {et~al.}(2022)\citenamefont {{Lazeyras}}, \citenamefont {{Barreira}}, \citenamefont {{Schmidt}},\ and\ \citenamefont {{Desjacques}}}]{Lazeyras22}%
  \BibitemOpen
  \bibfield  {author} {\bibinfo {author} {\bibfnamefont {T.}~\bibnamefont {{Lazeyras}}}, \bibinfo {author} {\bibfnamefont {A.}~\bibnamefont {{Barreira}}}, \bibinfo {author} {\bibfnamefont {F.}~\bibnamefont {{Schmidt}}},\ and\ \bibinfo {author} {\bibfnamefont {V.}~\bibnamefont {{Desjacques}}},\ }\bibfield  {title} {\bibinfo {title} {{Assembly bias in the local PNG halo bias and its implication for $f_{\rm NL}$ constraints}},\ }\href@noop {} {\bibfield  {journal} {\bibinfo  {journal} {arXiv e-prints}\ ,\ \bibinfo {eid} {arXiv:2209.07251}} (\bibinfo {year} {2022})},\ \Eprint {https://arxiv.org/abs/2209.07251} {arXiv:2209.07251 [astro-ph.CO]} \BibitemShut {NoStop}%
\bibitem [{\citenamefont {{Marinucci}}\ \emph {et~al.}(2023)\citenamefont {{Marinucci}}, \citenamefont {{Desjacques}},\ and\ \citenamefont {{Benson}}}]{marinucci_bphi_sam}%
  \BibitemOpen
  \bibfield  {author} {\bibinfo {author} {\bibfnamefont {M.}~\bibnamefont {{Marinucci}}}, \bibinfo {author} {\bibfnamefont {V.}~\bibnamefont {{Desjacques}}},\ and\ \bibinfo {author} {\bibfnamefont {A.}~\bibnamefont {{Benson}}},\ }\bibfield  {title} {\bibinfo {title} {{Non-Gaussian assembly bias from a semi-analytic galaxy formation model}},\ }\href {https://doi.org/10.1093/mnras/stad1884} {\bibfield  {journal} {\bibinfo  {journal} {\mnras}\ }\textbf {\bibinfo {volume} {524}},\ \bibinfo {pages} {325} (\bibinfo {year} {2023})},\ \Eprint {https://arxiv.org/abs/2303.10337} {arXiv:2303.10337 [astro-ph.CO]} \BibitemShut {NoStop}%
\bibitem [{\citenamefont {{Barreira}}\ and\ \citenamefont {{Krause}}(2023)}]{barreira_krause_23}%
  \BibitemOpen
  \bibfield  {author} {\bibinfo {author} {\bibfnamefont {A.}~\bibnamefont {{Barreira}}}\ and\ \bibinfo {author} {\bibfnamefont {E.}~\bibnamefont {{Krause}}},\ }\bibfield  {title} {\bibinfo {title} {{Towards optimal and robust f\_NL constraints with multi-tracer analyses}},\ }\href {https://doi.org/10.1088/1475-7516/2023/10/044} {\bibfield  {journal} {\bibinfo  {journal} {\jcap}\ }\textbf {\bibinfo {volume} {2023}},\ \bibinfo {eid} {044} (\bibinfo {year} {2023})},\ \Eprint {https://arxiv.org/abs/2302.09066} {arXiv:2302.09066 [astro-ph.CO]} \BibitemShut {NoStop}%
\bibitem [{\citenamefont {{Lucie-Smith}}\ \emph {et~al.}(2023)\citenamefont {{Lucie-Smith}}, \citenamefont {{Barreira}},\ and\ \citenamefont {{Schmidt}}}]{2023MNRAS.524.1746L}%
  \BibitemOpen
  \bibfield  {author} {\bibinfo {author} {\bibfnamefont {L.}~\bibnamefont {{Lucie-Smith}}}, \bibinfo {author} {\bibfnamefont {A.}~\bibnamefont {{Barreira}}},\ and\ \bibinfo {author} {\bibfnamefont {F.}~\bibnamefont {{Schmidt}}},\ }\bibfield  {title} {\bibinfo {title} {{Halo assembly bias from a deep learning model of halo formation}},\ }\href {https://doi.org/10.1093/mnras/stad2003} {\bibfield  {journal} {\bibinfo  {journal} {\mnras}\ }\textbf {\bibinfo {volume} {524}},\ \bibinfo {pages} {1746} (\bibinfo {year} {2023})},\ \Eprint {https://arxiv.org/abs/2304.09880} {arXiv:2304.09880 [astro-ph.CO]} \BibitemShut {NoStop}%
\bibitem [{\citenamefont {{Fondi}}\ \emph {et~al.}(2024)\citenamefont {{Fondi}}, \citenamefont {{Verde}}, \citenamefont {{Villaescusa-Navarro}}, \citenamefont {{Baldi}}, \citenamefont {{Coulton}}, \citenamefont {{Jung}}, \citenamefont {{Karagiannis}}, \citenamefont {{Liguori}}, \citenamefont {{Ravenni}},\ and\ \citenamefont {{Wandelt}}}]{fondi_ab_forecast_24}%
  \BibitemOpen
  \bibfield  {author} {\bibinfo {author} {\bibfnamefont {E.}~\bibnamefont {{Fondi}}}, \bibinfo {author} {\bibfnamefont {L.}~\bibnamefont {{Verde}}}, \bibinfo {author} {\bibfnamefont {F.}~\bibnamefont {{Villaescusa-Navarro}}}, \bibinfo {author} {\bibfnamefont {M.}~\bibnamefont {{Baldi}}}, \bibinfo {author} {\bibfnamefont {W.~R.}\ \bibnamefont {{Coulton}}}, \bibinfo {author} {\bibfnamefont {G.}~\bibnamefont {{Jung}}}, \bibinfo {author} {\bibfnamefont {D.}~\bibnamefont {{Karagiannis}}}, \bibinfo {author} {\bibfnamefont {M.}~\bibnamefont {{Liguori}}}, \bibinfo {author} {\bibfnamefont {A.}~\bibnamefont {{Ravenni}}},\ and\ \bibinfo {author} {\bibfnamefont {B.~D.}\ \bibnamefont {{Wandelt}}},\ }\bibfield  {title} {\bibinfo {title} {{Taming assembly bias for primordial non-Gaussianity}},\ }\href {https://doi.org/10.1088/1475-7516/2024/02/048} {\bibfield  {journal} {\bibinfo  {journal} {\jcap}\ }\textbf {\bibinfo {volume} {2024}},\ \bibinfo {eid} {048} (\bibinfo {year} {2024})},\ \Eprint
  {https://arxiv.org/abs/2311.10088} {arXiv:2311.10088 [astro-ph.CO]} \BibitemShut {NoStop}%
\bibitem [{\citenamefont {{Hadzhiyska}}\ and\ \citenamefont {{Ferraro}}(2025)}]{boryana_AB_bphi_2}%
  \BibitemOpen
  \bibfield  {author} {\bibinfo {author} {\bibfnamefont {B.}~\bibnamefont {{Hadzhiyska}}}\ and\ \bibinfo {author} {\bibfnamefont {S.}~\bibnamefont {{Ferraro}}},\ }\bibfield  {title} {\bibinfo {title} {{Refining local-type primordial non-Gaussianity: Sharpened $b_\phi$ constraints through bias expansion}},\ }\href {https://doi.org/10.48550/arXiv.2501.14873} {\bibfield  {journal} {\bibinfo  {journal} {arXiv e-prints}\ ,\ \bibinfo {eid} {arXiv:2501.14873}} (\bibinfo {year} {2025})},\ \Eprint {https://arxiv.org/abs/2501.14873} {arXiv:2501.14873 [astro-ph.CO]} \BibitemShut {NoStop}%
\bibitem [{\citenamefont {{Hadzhiyska}}\ \emph {et~al.}(2024)\citenamefont {{Hadzhiyska}}, \citenamefont {{Garrison}}, \citenamefont {{Eisenstein}},\ and\ \citenamefont {{Ferraro}}}]{boryana_png_ab_1}%
  \BibitemOpen
  \bibfield  {author} {\bibinfo {author} {\bibfnamefont {B.}~\bibnamefont {{Hadzhiyska}}}, \bibinfo {author} {\bibfnamefont {L.~H.}\ \bibnamefont {{Garrison}}}, \bibinfo {author} {\bibfnamefont {D.~J.}\ \bibnamefont {{Eisenstein}}},\ and\ \bibinfo {author} {\bibfnamefont {S.}~\bibnamefont {{Ferraro}}},\ }\bibfield  {title} {\bibinfo {title} {{Modest set of simulations of local-type primordial non-Gaussianity in the DESI era}},\ }\href {https://doi.org/10.1103/PhysRevD.109.103530} {\bibfield  {journal} {\bibinfo  {journal} {\prd}\ }\textbf {\bibinfo {volume} {109}},\ \bibinfo {eid} {103530} (\bibinfo {year} {2024})},\ \Eprint {https://arxiv.org/abs/2402.10881} {arXiv:2402.10881 [astro-ph.CO]} \BibitemShut {NoStop}%
\bibitem [{\citenamefont {{Guti{\'e}rrez Adame}}\ \emph {et~al.}(2024)\citenamefont {{Guti{\'e}rrez Adame}}, \citenamefont {{Avila}}, \citenamefont {{Gonzalez-Perez}}, \citenamefont {{Yepes}}, \citenamefont {{Pellejero}}, \citenamefont {{Wang}}, \citenamefont {{Chuang}}, \citenamefont {{Feng}}, \citenamefont {{Garcia-Bellido}},\ and\ \citenamefont {{Knebe}}}]{pngunit_bphi_ab}%
  \BibitemOpen
  \bibfield  {author} {\bibinfo {author} {\bibfnamefont {A.}~\bibnamefont {{Guti{\'e}rrez Adame}}}, \bibinfo {author} {\bibfnamefont {S.}~\bibnamefont {{Avila}}}, \bibinfo {author} {\bibfnamefont {V.}~\bibnamefont {{Gonzalez-Perez}}}, \bibinfo {author} {\bibfnamefont {G.}~\bibnamefont {{Yepes}}}, \bibinfo {author} {\bibfnamefont {M.}~\bibnamefont {{Pellejero}}}, \bibinfo {author} {\bibfnamefont {M.~S.}\ \bibnamefont {{Wang}}}, \bibinfo {author} {\bibfnamefont {C.-H.}\ \bibnamefont {{Chuang}}}, \bibinfo {author} {\bibfnamefont {Y.}~\bibnamefont {{Feng}}}, \bibinfo {author} {\bibfnamefont {J.}~\bibnamefont {{Garcia-Bellido}}},\ and\ \bibinfo {author} {\bibfnamefont {A.}~\bibnamefont {{Knebe}}},\ }\bibfield  {title} {\bibinfo {title} {{PNG-UNITsims: Halo clustering response to primordial non-Gaussianities as a function of mass}},\ }\href {https://doi.org/10.1051/0004-6361/202349037} {\bibfield  {journal} {\bibinfo  {journal} {\aap}\ }\textbf {\bibinfo {volume} {689}},\ \bibinfo {eid} {A69} (\bibinfo {year}
  {2024})},\ \Eprint {https://arxiv.org/abs/2312.12405} {arXiv:2312.12405 [astro-ph.CO]} \BibitemShut {NoStop}%
\bibitem [{\citenamefont {{Seljak}}(2009)}]{seljak_mt}%
  \BibitemOpen
  \bibfield  {author} {\bibinfo {author} {\bibfnamefont {U.}~\bibnamefont {{Seljak}}},\ }\bibfield  {title} {\bibinfo {title} {{Extracting Primordial Non-Gaussianity without Cosmic Variance}},\ }\href {https://doi.org/10.1103/PhysRevLett.102.021302} {\bibfield  {journal} {\bibinfo  {journal} {\prl}\ }\textbf {\bibinfo {volume} {102}},\ \bibinfo {eid} {021302} (\bibinfo {year} {2009})},\ \Eprint {https://arxiv.org/abs/0807.1770} {arXiv:0807.1770 [astro-ph]} \BibitemShut {NoStop}%
\bibitem [{\citenamefont {{Schmittfull}}\ and\ \citenamefont {{Seljak}}(2018)}]{SchmittfullSeljak}%
  \BibitemOpen
  \bibfield  {author} {\bibinfo {author} {\bibfnamefont {M.}~\bibnamefont {{Schmittfull}}}\ and\ \bibinfo {author} {\bibfnamefont {U.}~\bibnamefont {{Seljak}}},\ }\bibfield  {title} {\bibinfo {title} {{Parameter constraints from cross-correlation of CMB lensing with galaxy clustering}},\ }\href {https://doi.org/10.1103/PhysRevD.97.123540} {\bibfield  {journal} {\bibinfo  {journal} {\prd}\ }\textbf {\bibinfo {volume} {97}},\ \bibinfo {eid} {123540} (\bibinfo {year} {2018})},\ \Eprint {https://arxiv.org/abs/1710.09465} {arXiv:1710.09465 [astro-ph.CO]} \BibitemShut {NoStop}%
\bibitem [{\citenamefont {{Barreira}}(2022{\natexlab{c}})}]{2022JCAP...11..013B}%
  \BibitemOpen
  \bibfield  {author} {\bibinfo {author} {\bibfnamefont {A.}~\bibnamefont {{Barreira}}},\ }\bibfield  {title} {\bibinfo {title} {{Can we actually constrain f$_{NL}$ using the scale-dependent bias effect? An illustration of the impact of galaxy bias uncertainties using the BOSS DR12 galaxy power spectrum}},\ }\href {https://doi.org/10.1088/1475-7516/2022/11/013} {\bibfield  {journal} {\bibinfo  {journal} {\jcap}\ }\textbf {\bibinfo {volume} {2022}},\ \bibinfo {eid} {013} (\bibinfo {year} {2022}{\natexlab{c}})},\ \Eprint {https://arxiv.org/abs/2205.05673} {arXiv:2205.05673 [astro-ph.CO]} \BibitemShut {NoStop}%
\bibitem [{\citenamefont {{Sullivan}}\ and\ \citenamefont {{Chen}}(2025)}]{field_level_bias_stephen}%
  \BibitemOpen
  \bibfield  {author} {\bibinfo {author} {\bibfnamefont {J.~M.}\ \bibnamefont {{Sullivan}}}\ and\ \bibinfo {author} {\bibfnamefont {S.-F.}\ \bibnamefont {{Chen}}},\ }\bibfield  {title} {\bibinfo {title} {{Local primordial non-Gaussian bias at the field level}},\ }\href {https://doi.org/10.1088/1475-7516/2025/03/016} {\bibfield  {journal} {\bibinfo  {journal} {\jcap}\ }\textbf {\bibinfo {volume} {2025}},\ \bibinfo {eid} {016} (\bibinfo {year} {2025})},\ \Eprint {https://arxiv.org/abs/2410.18039} {arXiv:2410.18039 [astro-ph.CO]} \BibitemShut {NoStop}%
\bibitem [{\citenamefont {{Challinor}}\ and\ \citenamefont {{Lewis}}(2011)}]{challinor_lewis_gr}%
  \BibitemOpen
  \bibfield  {author} {\bibinfo {author} {\bibfnamefont {A.}~\bibnamefont {{Challinor}}}\ and\ \bibinfo {author} {\bibfnamefont {A.}~\bibnamefont {{Lewis}}},\ }\bibfield  {title} {\bibinfo {title} {{Linear power spectrum of observed source number counts}},\ }\href {https://doi.org/10.1103/PhysRevD.84.043516} {\bibfield  {journal} {\bibinfo  {journal} {\prd}\ }\textbf {\bibinfo {volume} {84}},\ \bibinfo {eid} {043516} (\bibinfo {year} {2011})},\ \Eprint {https://arxiv.org/abs/1105.5292} {arXiv:1105.5292 [astro-ph.CO]} \BibitemShut {NoStop}%
\bibitem [{\citenamefont {{Alonso}}\ \emph {et~al.}(2015)\citenamefont {{Alonso}}, \citenamefont {{Bull}}, \citenamefont {{Ferreira}}, \citenamefont {{Maartens}},\ and\ \citenamefont {{Santos}}}]{alonso_ultra_large_scale}%
  \BibitemOpen
  \bibfield  {author} {\bibinfo {author} {\bibfnamefont {D.}~\bibnamefont {{Alonso}}}, \bibinfo {author} {\bibfnamefont {P.}~\bibnamefont {{Bull}}}, \bibinfo {author} {\bibfnamefont {P.~G.}\ \bibnamefont {{Ferreira}}}, \bibinfo {author} {\bibfnamefont {R.}~\bibnamefont {{Maartens}}},\ and\ \bibinfo {author} {\bibfnamefont {M.~G.}\ \bibnamefont {{Santos}}},\ }\bibfield  {title} {\bibinfo {title} {{Ultra-large-scale Cosmology in Next-generation Experiments with Single Tracers}},\ }\href {https://doi.org/10.1088/0004-637X/814/2/145} {\bibfield  {journal} {\bibinfo  {journal} {\apj}\ }\textbf {\bibinfo {volume} {814}},\ \bibinfo {eid} {145} (\bibinfo {year} {2015})},\ \Eprint {https://arxiv.org/abs/1505.07596} {arXiv:1505.07596 [astro-ph.CO]} \BibitemShut {NoStop}%
\bibitem [{\citenamefont {{Rossiter}}\ \emph {et~al.}(2024)\citenamefont {{Rossiter}}, \citenamefont {{Camera}}, \citenamefont {{Clarkson}},\ and\ \citenamefont {{Maartens}}}]{rossiter_fnl_gr_bias}%
  \BibitemOpen
  \bibfield  {author} {\bibinfo {author} {\bibfnamefont {S.}~\bibnamefont {{Rossiter}}}, \bibinfo {author} {\bibfnamefont {S.}~\bibnamefont {{Camera}}}, \bibinfo {author} {\bibfnamefont {C.}~\bibnamefont {{Clarkson}}},\ and\ \bibinfo {author} {\bibfnamefont {R.}~\bibnamefont {{Maartens}}},\ }\bibfield  {title} {\bibinfo {title} {{Decoupling Local Primordial non-Gaussianity from Relativistic Effects in the Galaxy Bispectrum}},\ }\href {https://doi.org/10.48550/arXiv.2407.06301} {\bibfield  {journal} {\bibinfo  {journal} {arXiv e-prints}\ ,\ \bibinfo {eid} {arXiv:2407.06301}} (\bibinfo {year} {2024})},\ \Eprint {https://arxiv.org/abs/2407.06301} {arXiv:2407.06301 [astro-ph.CO]} \BibitemShut {NoStop}%
\bibitem [{\citenamefont {{Desjacques}}\ \emph {et~al.}(2018)\citenamefont {{Desjacques}}, \citenamefont {{Jeong}},\ and\ \citenamefont {{Schmidt}}}]{DJS}%
  \BibitemOpen
  \bibfield  {author} {\bibinfo {author} {\bibfnamefont {V.}~\bibnamefont {{Desjacques}}}, \bibinfo {author} {\bibfnamefont {D.}~\bibnamefont {{Jeong}}},\ and\ \bibinfo {author} {\bibfnamefont {F.}~\bibnamefont {{Schmidt}}},\ }\bibfield  {title} {\bibinfo {title} {{Large-scale galaxy bias}},\ }\href {https://doi.org/10.1016/j.physrep.2017.12.002} {\bibfield  {journal} {\bibinfo  {journal} {\physrep}\ }\textbf {\bibinfo {volume} {733}},\ \bibinfo {pages} {1} (\bibinfo {year} {2018})},\ \Eprint {https://arxiv.org/abs/1611.09787} {arXiv:1611.09787 [astro-ph.CO]} \BibitemShut {NoStop}%
\bibitem [{Note8()}]{Note8}%
  \BibitemOpen
  \bibinfo {note} {Here using an unrealistically large $f_{NL}=5000$ for illustration.}\BibitemShut {Stop}%
\bibitem [{\citenamefont {{Jeong}}\ \emph {et~al.}(2012)\citenamefont {{Jeong}}, \citenamefont {{Schmidt}},\ and\ \citenamefont {{Hirata}}}]{jsh_gr}%
  \BibitemOpen
  \bibfield  {author} {\bibinfo {author} {\bibfnamefont {D.}~\bibnamefont {{Jeong}}}, \bibinfo {author} {\bibfnamefont {F.}~\bibnamefont {{Schmidt}}},\ and\ \bibinfo {author} {\bibfnamefont {C.~M.}\ \bibnamefont {{Hirata}}},\ }\bibfield  {title} {\bibinfo {title} {{Large-scale clustering of galaxies in general relativity}},\ }\href {https://doi.org/10.1103/PhysRevD.85.023504} {\bibfield  {journal} {\bibinfo  {journal} {\prd}\ }\textbf {\bibinfo {volume} {85}},\ \bibinfo {eid} {023504} (\bibinfo {year} {2012})},\ \Eprint {https://arxiv.org/abs/1107.5427} {arXiv:1107.5427 [astro-ph.CO]} \BibitemShut {NoStop}%
\bibitem [{\citenamefont {{Baldauf}}\ \emph {et~al.}(2011)\citenamefont {{Baldauf}}, \citenamefont {{Seljak}}, \citenamefont {{Senatore}},\ and\ \citenamefont {{Zaldarriaga}}}]{baldauf_gr}%
  \BibitemOpen
  \bibfield  {author} {\bibinfo {author} {\bibfnamefont {T.}~\bibnamefont {{Baldauf}}}, \bibinfo {author} {\bibfnamefont {U.}~\bibnamefont {{Seljak}}}, \bibinfo {author} {\bibfnamefont {L.}~\bibnamefont {{Senatore}}},\ and\ \bibinfo {author} {\bibfnamefont {M.}~\bibnamefont {{Zaldarriaga}}},\ }\bibfield  {title} {\bibinfo {title} {{Galaxy bias and non-linear structure formation in general relativity}},\ }\href {https://doi.org/10.1088/1475-7516/2011/10/031} {\bibfield  {journal} {\bibinfo  {journal} {\jcap}\ }\textbf {\bibinfo {volume} {2011}},\ \bibinfo {eid} {031} (\bibinfo {year} {2011})},\ \Eprint {https://arxiv.org/abs/1106.5507} {arXiv:1106.5507 [astro-ph.CO]} \BibitemShut {NoStop}%
\bibitem [{\citenamefont {{Pajer}}\ \emph {et~al.}(2013)\citenamefont {{Pajer}}, \citenamefont {{Schmidt}},\ and\ \citenamefont {{Zaldarriaga}}}]{pajer_3pt_cfc}%
  \BibitemOpen
  \bibfield  {author} {\bibinfo {author} {\bibfnamefont {E.}~\bibnamefont {{Pajer}}}, \bibinfo {author} {\bibfnamefont {F.}~\bibnamefont {{Schmidt}}},\ and\ \bibinfo {author} {\bibfnamefont {M.}~\bibnamefont {{Zaldarriaga}}},\ }\bibfield  {title} {\bibinfo {title} {{The Observed squeezed limit of cosmological three-point functions}},\ }\href {https://doi.org/10.1103/PhysRevD.88.083502} {\bibfield  {journal} {\bibinfo  {journal} {\prd}\ }\textbf {\bibinfo {volume} {88}},\ \bibinfo {eid} {083502} (\bibinfo {year} {2013})},\ \Eprint {https://arxiv.org/abs/1305.0824} {arXiv:1305.0824 [astro-ph.CO]} \BibitemShut {NoStop}%
\bibitem [{\citenamefont {{Dai}}\ \emph {et~al.}(2015{\natexlab{a}})\citenamefont {{Dai}}, \citenamefont {{Pajer}},\ and\ \citenamefont {{Schmidt}}}]{liang_cfc_1}%
  \BibitemOpen
  \bibfield  {author} {\bibinfo {author} {\bibfnamefont {L.}~\bibnamefont {{Dai}}}, \bibinfo {author} {\bibfnamefont {E.}~\bibnamefont {{Pajer}}},\ and\ \bibinfo {author} {\bibfnamefont {F.}~\bibnamefont {{Schmidt}}},\ }\bibfield  {title} {\bibinfo {title} {{Conformal Fermi Coordinates}},\ }\href {https://doi.org/10.1088/1475-7516/2015/11/043} {\bibfield  {journal} {\bibinfo  {journal} {\jcap}\ }\textbf {\bibinfo {volume} {2015}},\ \bibinfo {pages} {043} (\bibinfo {year} {2015}{\natexlab{a}})},\ \Eprint {https://arxiv.org/abs/1502.02011} {arXiv:1502.02011 [gr-qc]} \BibitemShut {NoStop}%
\bibitem [{\citenamefont {{Dai}}\ \emph {et~al.}(2015{\natexlab{b}})\citenamefont {{Dai}}, \citenamefont {{Pajer}},\ and\ \citenamefont {{Schmidt}}}]{liang_su}%
  \BibitemOpen
  \bibfield  {author} {\bibinfo {author} {\bibfnamefont {L.}~\bibnamefont {{Dai}}}, \bibinfo {author} {\bibfnamefont {E.}~\bibnamefont {{Pajer}}},\ and\ \bibinfo {author} {\bibfnamefont {F.}~\bibnamefont {{Schmidt}}},\ }\bibfield  {title} {\bibinfo {title} {{On separate universes}},\ }\href {https://doi.org/10.1088/1475-7516/2015/10/059} {\bibfield  {journal} {\bibinfo  {journal} {\jcap}\ }\textbf {\bibinfo {volume} {2015}},\ \bibinfo {pages} {059} (\bibinfo {year} {2015}{\natexlab{b}})},\ \Eprint {https://arxiv.org/abs/1504.00351} {arXiv:1504.00351 [astro-ph.CO]} \BibitemShut {NoStop}%
\bibitem [{\citenamefont {{Salopek}}\ and\ \citenamefont {{Bond}}(1990)}]{salopek_bond}%
  \BibitemOpen
  \bibfield  {author} {\bibinfo {author} {\bibfnamefont {D.~S.}\ \bibnamefont {{Salopek}}}\ and\ \bibinfo {author} {\bibfnamefont {J.~R.}\ \bibnamefont {{Bond}}},\ }\bibfield  {title} {\bibinfo {title} {{Nonlinear evolution of long-wavelength metric fluctuations in inflationary models}},\ }\href {https://doi.org/10.1103/PhysRevD.42.3936} {\bibfield  {journal} {\bibinfo  {journal} {\prd}\ }\textbf {\bibinfo {volume} {42}},\ \bibinfo {pages} {3936} (\bibinfo {year} {1990})}\BibitemShut {NoStop}%
\bibitem [{\citenamefont {{Baldauf}}\ \emph {et~al.}(2016{\natexlab{a}})\citenamefont {{Baldauf}}, \citenamefont {{Seljak}}, \citenamefont {{Senatore}},\ and\ \citenamefont {{Zaldarriaga}}}]{baldauf_su}%
  \BibitemOpen
  \bibfield  {author} {\bibinfo {author} {\bibfnamefont {T.}~\bibnamefont {{Baldauf}}}, \bibinfo {author} {\bibfnamefont {U.}~\bibnamefont {{Seljak}}}, \bibinfo {author} {\bibfnamefont {L.}~\bibnamefont {{Senatore}}},\ and\ \bibinfo {author} {\bibfnamefont {M.}~\bibnamefont {{Zaldarriaga}}},\ }\bibfield  {title} {\bibinfo {title} {{Linear response to long wavelength fluctuations using curvature simulations}},\ }\href {https://doi.org/10.1088/1475-7516/2016/09/007} {\bibfield  {journal} {\bibinfo  {journal} {\jcap}\ }\textbf {\bibinfo {volume} {2016}},\ \bibinfo {eid} {007} (\bibinfo {year} {2016}{\natexlab{a}})},\ \Eprint {https://arxiv.org/abs/1511.01465} {arXiv:1511.01465 [astro-ph.CO]} \BibitemShut {NoStop}%
\bibitem [{\citenamefont {{Baldauf}}\ \emph {et~al.}(2016{\natexlab{b}})\citenamefont {{Baldauf}}, \citenamefont {{Seljak}}, \citenamefont {{Senatore}},\ and\ \citenamefont {{Zaldarriaga}}}]{baldauf_su2}%
  \BibitemOpen
  \bibfield  {author} {\bibinfo {author} {\bibfnamefont {T.}~\bibnamefont {{Baldauf}}}, \bibinfo {author} {\bibfnamefont {U.}~\bibnamefont {{Seljak}}}, \bibinfo {author} {\bibfnamefont {L.}~\bibnamefont {{Senatore}}},\ and\ \bibinfo {author} {\bibfnamefont {M.}~\bibnamefont {{Zaldarriaga}}},\ }\bibfield  {title} {\bibinfo {title} {{Linear response to long wavelength fluctuations using curvature simulations}},\ }\href {https://doi.org/10.1088/1475-7516/2016/09/007} {\bibfield  {journal} {\bibinfo  {journal} {\jcap}\ }\textbf {\bibinfo {volume} {2016}},\ \bibinfo {eid} {007} (\bibinfo {year} {2016}{\natexlab{b}})},\ \Eprint {https://arxiv.org/abs/1511.01465} {arXiv:1511.01465 [astro-ph.CO]} \BibitemShut {NoStop}%
\bibitem [{\citenamefont {{Sirko}}(2005)}]{sirko_su}%
  \BibitemOpen
  \bibfield  {author} {\bibinfo {author} {\bibfnamefont {E.}~\bibnamefont {{Sirko}}},\ }\bibfield  {title} {\bibinfo {title} {{Initial Conditions to Cosmological N-Body Simulations, or, How to Run an Ensemble of Simulations}},\ }\href {https://doi.org/10.1086/497090} {\bibfield  {journal} {\bibinfo  {journal} {\apj}\ }\textbf {\bibinfo {volume} {634}},\ \bibinfo {pages} {728} (\bibinfo {year} {2005})},\ \Eprint {https://arxiv.org/abs/astro-ph/0503106} {arXiv:astro-ph/0503106 [astro-ph]} \BibitemShut {NoStop}%
\bibitem [{\citenamefont {{Gnedin}}\ \emph {et~al.}(2011)\citenamefont {{Gnedin}}, \citenamefont {{Kravtsov}},\ and\ \citenamefont {{Rudd}}}]{gnedin_su}%
  \BibitemOpen
  \bibfield  {author} {\bibinfo {author} {\bibfnamefont {N.~Y.}\ \bibnamefont {{Gnedin}}}, \bibinfo {author} {\bibfnamefont {A.~V.}\ \bibnamefont {{Kravtsov}}},\ and\ \bibinfo {author} {\bibfnamefont {D.~H.}\ \bibnamefont {{Rudd}}},\ }\bibfield  {title} {\bibinfo {title} {{Implementing the DC Mode in Cosmological Simulations with Supercomoving Variables}},\ }\href {https://doi.org/10.1088/0067-0049/194/2/46} {\bibfield  {journal} {\bibinfo  {journal} {\apjs}\ }\textbf {\bibinfo {volume} {194}},\ \bibinfo {eid} {46} (\bibinfo {year} {2011})},\ \Eprint {https://arxiv.org/abs/1104.1428} {arXiv:1104.1428 [astro-ph.CO]} \BibitemShut {NoStop}%
\bibitem [{\citenamefont {{Wagner}}\ \emph {et~al.}(2015)\citenamefont {{Wagner}}, \citenamefont {{Schmidt}}, \citenamefont {{Chiang}},\ and\ \citenamefont {{Komatsu}}}]{wagner_su}%
  \BibitemOpen
  \bibfield  {author} {\bibinfo {author} {\bibfnamefont {C.}~\bibnamefont {{Wagner}}}, \bibinfo {author} {\bibfnamefont {F.}~\bibnamefont {{Schmidt}}}, \bibinfo {author} {\bibfnamefont {C.~T.}\ \bibnamefont {{Chiang}}},\ and\ \bibinfo {author} {\bibfnamefont {E.}~\bibnamefont {{Komatsu}}},\ }\bibfield  {title} {\bibinfo {title} {{Separate universe simulations.}},\ }\href {https://doi.org/10.1093/mnrasl/slu187} {\bibfield  {journal} {\bibinfo  {journal} {\mnras}\ }\textbf {\bibinfo {volume} {448}},\ \bibinfo {pages} {L11} (\bibinfo {year} {2015})},\ \Eprint {https://arxiv.org/abs/1409.6294} {arXiv:1409.6294 [astro-ph.CO]} \BibitemShut {NoStop}%
\bibitem [{\citenamefont {{Li}}\ \emph {et~al.}(2014)\citenamefont {{Li}}, \citenamefont {{Hu}},\ and\ \citenamefont {{Takada}}}]{yin_su_14}%
  \BibitemOpen
  \bibfield  {author} {\bibinfo {author} {\bibfnamefont {Y.}~\bibnamefont {{Li}}}, \bibinfo {author} {\bibfnamefont {W.}~\bibnamefont {{Hu}}},\ and\ \bibinfo {author} {\bibfnamefont {M.}~\bibnamefont {{Takada}}},\ }\bibfield  {title} {\bibinfo {title} {{Super-sample covariance in simulations}},\ }\href {https://doi.org/10.1103/PhysRevD.89.083519} {\bibfield  {journal} {\bibinfo  {journal} {\prd}\ }\textbf {\bibinfo {volume} {89}},\ \bibinfo {eid} {083519} (\bibinfo {year} {2014})},\ \Eprint {https://arxiv.org/abs/1401.0385} {arXiv:1401.0385 [astro-ph.CO]} \BibitemShut {NoStop}%
\bibitem [{Note9()}]{Note9}%
  \BibitemOpen
  \bibinfo {note} {In the presence of tidal long mode perturbations, the metric is of perturbed FLRW form instead.}\BibitemShut {Stop}%
\bibitem [{\citenamefont {{Ma}}\ and\ \citenamefont {{Bertschinger}}(1995)}]{ma_bertschinger}%
  \BibitemOpen
  \bibfield  {author} {\bibinfo {author} {\bibfnamefont {C.-P.}\ \bibnamefont {{Ma}}}\ and\ \bibinfo {author} {\bibfnamefont {E.}~\bibnamefont {{Bertschinger}}},\ }\bibfield  {title} {\bibinfo {title} {{Cosmological Perturbation Theory in the Synchronous and Conformal Newtonian Gauges}},\ }\href {https://doi.org/10.1086/176550} {\bibfield  {journal} {\bibinfo  {journal} {\apj}\ }\textbf {\bibinfo {volume} {455}},\ \bibinfo {pages} {7} (\bibinfo {year} {1995})},\ \Eprint {https://arxiv.org/abs/astro-ph/9506072} {arXiv:astro-ph/9506072 [astro-ph]} \BibitemShut {NoStop}%
\bibitem [{Note10()}]{Note10}%
  \BibitemOpen
  \bibinfo {note} {Such that the lack of anisotropic stress implies $\phi =\psi $, which we will not assume.}\BibitemShut {Stop}%
\bibitem [{Note11()}]{Note11}%
  \BibitemOpen
  \bibinfo {note} {By pursuing this logic further, one can calculate finite wavelength corrections to $\phi _{l}=\protect \mathrm {const.}$ at each order in $x_{F}^n$.}\BibitemShut {Stop}%
\bibitem [{\citenamefont {{Jeong}}\ and\ \citenamefont {{Schmidt}}(2014)}]{cosmic_clocks}%
  \BibitemOpen
  \bibfield  {author} {\bibinfo {author} {\bibfnamefont {D.}~\bibnamefont {{Jeong}}}\ and\ \bibinfo {author} {\bibfnamefont {F.}~\bibnamefont {{Schmidt}}},\ }\bibfield  {title} {\bibinfo {title} {{Cosmic clocks}},\ }\href {https://doi.org/10.1103/PhysRevD.89.043519} {\bibfield  {journal} {\bibinfo  {journal} {\prd}\ }\textbf {\bibinfo {volume} {89}},\ \bibinfo {eid} {043519} (\bibinfo {year} {2014})},\ \Eprint {https://arxiv.org/abs/1305.1299} {arXiv:1305.1299 [astro-ph.CO]} \BibitemShut {NoStop}%
\bibitem [{Note12()}]{Note12}%
  \BibitemOpen
  \bibinfo {note} {The time shift expression depends on the chosen gauge, but $\protect \mathcal {T}$ is a gauge-invariant quantity \cite {cosmic_clocks}.}\BibitemShut {Stop}%
\bibitem [{Note13()}]{Note13}%
  \BibitemOpen
  \bibinfo {note} {More precisely, the shift from coordinate time $a$ to proper time $a_{F}(t_{F}(a))$, where $a$ is analogous to the choice of ``observed redshift'' coordinates.}\BibitemShut {Stop}%
\bibitem [{Note14()}]{Note14}%
  \BibitemOpen
  \bibinfo {note} {More precisely, any tracer that responds to the growth of structure.}\BibitemShut {Stop}%
\bibitem [{\citenamefont {{Lazeyras}}\ \emph {et~al.}(2021)\citenamefont {{Lazeyras}}, \citenamefont {{Barreira}},\ and\ \citenamefont {{Schmidt}}}]{lazeyras_AB_quadratic_halos_21}%
  \BibitemOpen
  \bibfield  {author} {\bibinfo {author} {\bibfnamefont {T.}~\bibnamefont {{Lazeyras}}}, \bibinfo {author} {\bibfnamefont {A.}~\bibnamefont {{Barreira}}},\ and\ \bibinfo {author} {\bibfnamefont {F.}~\bibnamefont {{Schmidt}}},\ }\bibfield  {title} {\bibinfo {title} {{Assembly bias in quadratic bias parameters of dark matter halos from forward modeling}},\ }\href {https://doi.org/10.1088/1475-7516/2021/10/063} {\bibfield  {journal} {\bibinfo  {journal} {\jcap}\ }\textbf {\bibinfo {volume} {2021}},\ \bibinfo {eid} {063} (\bibinfo {year} {2021})},\ \Eprint {https://arxiv.org/abs/2106.14713} {arXiv:2106.14713 [astro-ph.CO]} \BibitemShut {NoStop}%
\bibitem [{\citenamefont {{Feng}}\ \emph {et~al.}(2016)\citenamefont {{Feng}}, \citenamefont {{Chu}}, \citenamefont {{Seljak}},\ and\ \citenamefont {{McDonald}}}]{fastpm}%
  \BibitemOpen
  \bibfield  {author} {\bibinfo {author} {\bibfnamefont {Y.}~\bibnamefont {{Feng}}}, \bibinfo {author} {\bibfnamefont {M.-Y.}\ \bibnamefont {{Chu}}}, \bibinfo {author} {\bibfnamefont {U.}~\bibnamefont {{Seljak}}},\ and\ \bibinfo {author} {\bibfnamefont {P.}~\bibnamefont {{McDonald}}},\ }\bibfield  {title} {\bibinfo {title} {{FASTPM: a new scheme for fast simulations of dark matter and haloes}},\ }\href {https://doi.org/10.1093/mnras/stw2123} {\bibfield  {journal} {\bibinfo  {journal} {\mnras}\ }\textbf {\bibinfo {volume} {463}},\ \bibinfo {pages} {2273} (\bibinfo {year} {2016})},\ \Eprint {https://arxiv.org/abs/1603.00476} {arXiv:1603.00476 [astro-ph.CO]} \BibitemShut {NoStop}%
\bibitem [{\citenamefont {{Lewis}}\ \emph {et~al.}(2000)\citenamefont {{Lewis}}, \citenamefont {{Challinor}},\ and\ \citenamefont {{Lasenby}}}]{camb}%
  \BibitemOpen
  \bibfield  {author} {\bibinfo {author} {\bibfnamefont {A.}~\bibnamefont {{Lewis}}}, \bibinfo {author} {\bibfnamefont {A.}~\bibnamefont {{Challinor}}},\ and\ \bibinfo {author} {\bibfnamefont {A.}~\bibnamefont {{Lasenby}}},\ }\bibfield  {title} {\bibinfo {title} {{Efficient Computation of Cosmic Microwave Background Anisotropies in Closed Friedmann-Robertson-Walker Models}},\ }\href {https://doi.org/10.1086/309179} {\bibfield  {journal} {\bibinfo  {journal} {\apj}\ }\textbf {\bibinfo {volume} {538}},\ \bibinfo {pages} {473} (\bibinfo {year} {2000})},\ \Eprint {https://arxiv.org/abs/astro-ph/9911177} {arXiv:astro-ph/9911177 [astro-ph]} \BibitemShut {NoStop}%
\bibitem [{\citenamefont {{Davis}}\ \emph {et~al.}(1985)\citenamefont {{Davis}}, \citenamefont {{Efstathiou}}, \citenamefont {{Frenk}},\ and\ \citenamefont {{White}}}]{fof_davis}%
  \BibitemOpen
  \bibfield  {author} {\bibinfo {author} {\bibfnamefont {M.}~\bibnamefont {{Davis}}}, \bibinfo {author} {\bibfnamefont {G.}~\bibnamefont {{Efstathiou}}}, \bibinfo {author} {\bibfnamefont {C.~S.}\ \bibnamefont {{Frenk}}},\ and\ \bibinfo {author} {\bibfnamefont {S.~D.~M.}\ \bibnamefont {{White}}},\ }\bibfield  {title} {\bibinfo {title} {{The evolution of large-scale structure in a universe dominated by cold dark matter}},\ }\href {https://doi.org/10.1086/163168} {\bibfield  {journal} {\bibinfo  {journal} {\apj}\ }\textbf {\bibinfo {volume} {292}},\ \bibinfo {pages} {371} (\bibinfo {year} {1985})}\BibitemShut {NoStop}%
\bibitem [{\citenamefont {{Sullivan}}\ and\ \citenamefont {{Chen}}(2024)}]{lpng_field_level}%
  \BibitemOpen
  \bibfield  {author} {\bibinfo {author} {\bibfnamefont {J.~M.}\ \bibnamefont {{Sullivan}}}\ and\ \bibinfo {author} {\bibfnamefont {S.-F.}\ \bibnamefont {{Chen}}},\ }\bibfield  {title} {\bibinfo {title} {{Local Primordial Non-Gaussian Bias at the Field Level}},\ }\href@noop {} {\bibfield  {journal} {\bibinfo  {journal} {arXiv e-prints}\ ,\ \bibinfo {eid} {arXiv:2410.18039}} (\bibinfo {year} {2024})},\ \Eprint {https://arxiv.org/abs/2410.18039} {arXiv:2410.18039 [astro-ph.CO]} \BibitemShut {NoStop}%
\bibitem [{Note15()}]{Note15}%
  \BibitemOpen
  \bibinfo {note} {Though at $z=1$ just taking derivatives in redshift, as in Einstein-de Sitter $D(z) = (1+z)^{-1}$, is not a bad approximation.}\BibitemShut {Stop}%
\bibitem [{\citenamefont {{Behroozi}}\ \emph {et~al.}(2013)\citenamefont {{Behroozi}}, \citenamefont {{Wechsler}},\ and\ \citenamefont {{Wu}}}]{rockstar_behroozi}%
  \BibitemOpen
  \bibfield  {author} {\bibinfo {author} {\bibfnamefont {P.~S.}\ \bibnamefont {{Behroozi}}}, \bibinfo {author} {\bibfnamefont {R.~H.}\ \bibnamefont {{Wechsler}}},\ and\ \bibinfo {author} {\bibfnamefont {H.-Y.}\ \bibnamefont {{Wu}}},\ }\bibfield  {title} {\bibinfo {title} {{The ROCKSTAR Phase-space Temporal Halo Finder and the Velocity Offsets of Cluster Cores}},\ }\href {https://doi.org/10.1088/0004-637X/762/2/109} {\bibfield  {journal} {\bibinfo  {journal} {\apj}\ }\textbf {\bibinfo {volume} {762}},\ \bibinfo {eid} {109} (\bibinfo {year} {2013})},\ \Eprint {https://arxiv.org/abs/1110.4372} {arXiv:1110.4372 [astro-ph.CO]} \BibitemShut {NoStop}%
\bibitem [{\citenamefont {Virtanen}\ \emph {et~al.}(2020)\citenamefont {Virtanen}, \citenamefont {Gommers}, \citenamefont {Oliphant}, \citenamefont {Haberland}, \citenamefont {Reddy}, \citenamefont {Cournapeau}, \citenamefont {Burovski}, \citenamefont {Peterson}, \citenamefont {Weckesser}, \citenamefont {Bright}, \citenamefont {{van der Walt}}, \citenamefont {Brett}, \citenamefont {Wilson}, \citenamefont {Millman}, \citenamefont {Mayorov}, \citenamefont {Nelson}, \citenamefont {Jones}, \citenamefont {Kern}, \citenamefont {Larson}, \citenamefont {Carey}, \citenamefont {Polat}, \citenamefont {Feng}, \citenamefont {Moore}, \citenamefont {{VanderPlas}}, \citenamefont {Laxalde}, \citenamefont {Perktold}, \citenamefont {Cimrman}, \citenamefont {Henriksen}, \citenamefont {Quintero}, \citenamefont {Harris}, \citenamefont {Archibald}, \citenamefont {Ribeiro}, \citenamefont {Pedregosa}, \citenamefont {{van Mulbregt}},\ and\ \citenamefont {{SciPy 1.0 Contributors}}}]{2020SciPy-NMeth}%
  \BibitemOpen
  \bibfield  {author} {\bibinfo {author} {\bibfnamefont {P.}~\bibnamefont {Virtanen}}, \bibinfo {author} {\bibfnamefont {R.}~\bibnamefont {Gommers}}, \bibinfo {author} {\bibfnamefont {T.~E.}\ \bibnamefont {Oliphant}}, \bibinfo {author} {\bibfnamefont {M.}~\bibnamefont {Haberland}}, \bibinfo {author} {\bibfnamefont {T.}~\bibnamefont {Reddy}}, \bibinfo {author} {\bibfnamefont {D.}~\bibnamefont {Cournapeau}}, \bibinfo {author} {\bibfnamefont {E.}~\bibnamefont {Burovski}}, \bibinfo {author} {\bibfnamefont {P.}~\bibnamefont {Peterson}}, \bibinfo {author} {\bibfnamefont {W.}~\bibnamefont {Weckesser}}, \bibinfo {author} {\bibfnamefont {J.}~\bibnamefont {Bright}}, \bibinfo {author} {\bibfnamefont {S.~J.}\ \bibnamefont {{van der Walt}}}, \bibinfo {author} {\bibfnamefont {M.}~\bibnamefont {Brett}}, \bibinfo {author} {\bibfnamefont {J.}~\bibnamefont {Wilson}}, \bibinfo {author} {\bibfnamefont {K.~J.}\ \bibnamefont {Millman}}, \bibinfo {author} {\bibfnamefont {N.}~\bibnamefont {Mayorov}}, \bibinfo {author} {\bibfnamefont
  {A.~R.~J.}\ \bibnamefont {Nelson}}, \bibinfo {author} {\bibfnamefont {E.}~\bibnamefont {Jones}}, \bibinfo {author} {\bibfnamefont {R.}~\bibnamefont {Kern}}, \bibinfo {author} {\bibfnamefont {E.}~\bibnamefont {Larson}}, \bibinfo {author} {\bibfnamefont {C.~J.}\ \bibnamefont {Carey}}, \bibinfo {author} {\bibfnamefont {{\.I}.}~\bibnamefont {Polat}}, \bibinfo {author} {\bibfnamefont {Y.}~\bibnamefont {Feng}}, \bibinfo {author} {\bibfnamefont {E.~W.}\ \bibnamefont {Moore}}, \bibinfo {author} {\bibfnamefont {J.}~\bibnamefont {{VanderPlas}}}, \bibinfo {author} {\bibfnamefont {D.}~\bibnamefont {Laxalde}}, \bibinfo {author} {\bibfnamefont {J.}~\bibnamefont {Perktold}}, \bibinfo {author} {\bibfnamefont {R.}~\bibnamefont {Cimrman}}, \bibinfo {author} {\bibfnamefont {I.}~\bibnamefont {Henriksen}}, \bibinfo {author} {\bibfnamefont {E.~A.}\ \bibnamefont {Quintero}}, \bibinfo {author} {\bibfnamefont {C.~R.}\ \bibnamefont {Harris}}, \bibinfo {author} {\bibfnamefont {A.~M.}\ \bibnamefont {Archibald}}, \bibinfo {author}
  {\bibfnamefont {A.~H.}\ \bibnamefont {Ribeiro}}, \bibinfo {author} {\bibfnamefont {F.}~\bibnamefont {Pedregosa}}, \bibinfo {author} {\bibfnamefont {P.}~\bibnamefont {{van Mulbregt}}},\ and\ \bibinfo {author} {\bibnamefont {{SciPy 1.0 Contributors}}},\ }\bibfield  {title} {\bibinfo {title} {{{SciPy} 1.0: Fundamental Algorithms for Scientific Computing in Python}},\ }\href {https://doi.org/10.1038/s41592-019-0686-2} {\bibfield  {journal} {\bibinfo  {journal} {Nature Methods}\ }\textbf {\bibinfo {volume} {17}},\ \bibinfo {pages} {261} (\bibinfo {year} {2020})}\BibitemShut {NoStop}%
\bibitem [{\citenamefont {{Villaescusa-Navarro}}\ \emph {et~al.}(2020)\citenamefont {{Villaescusa-Navarro}}, \citenamefont {{Hahn}}, \citenamefont {{Massara}}, \citenamefont {{Banerjee}}, \citenamefont {{Delgado}}, \citenamefont {{Ramanah}}, \citenamefont {{Charnock}}, \citenamefont {{Giusarma}}, \citenamefont {{Li}}, \citenamefont {{Allys}}, \citenamefont {{Brochard}}, \citenamefont {{Uhlemann}}, \citenamefont {{Chiang}}, \citenamefont {{He}}, \citenamefont {{Pisani}}, \citenamefont {{Obuljen}}, \citenamefont {{Feng}}, \citenamefont {{Castorina}}, \citenamefont {{Contardo}}, \citenamefont {{Kreisch}}, \citenamefont {{Nicola}}, \citenamefont {{Alsing}}, \citenamefont {{Scoccimarro}}, \citenamefont {{Verde}}, \citenamefont {{Viel}}, \citenamefont {{Ho}}, \citenamefont {{Mallat}}, \citenamefont {{Wandelt}},\ and\ \citenamefont {{Spergel}}}]{quijote_sims}%
  \BibitemOpen
  \bibfield  {author} {\bibinfo {author} {\bibfnamefont {F.}~\bibnamefont {{Villaescusa-Navarro}}}, \bibinfo {author} {\bibfnamefont {C.}~\bibnamefont {{Hahn}}}, \bibinfo {author} {\bibfnamefont {E.}~\bibnamefont {{Massara}}}, \bibinfo {author} {\bibfnamefont {A.}~\bibnamefont {{Banerjee}}}, \bibinfo {author} {\bibfnamefont {A.~M.}\ \bibnamefont {{Delgado}}}, \bibinfo {author} {\bibfnamefont {D.~K.}\ \bibnamefont {{Ramanah}}}, \bibinfo {author} {\bibfnamefont {T.}~\bibnamefont {{Charnock}}}, \bibinfo {author} {\bibfnamefont {E.}~\bibnamefont {{Giusarma}}}, \bibinfo {author} {\bibfnamefont {Y.}~\bibnamefont {{Li}}}, \bibinfo {author} {\bibfnamefont {E.}~\bibnamefont {{Allys}}}, \bibinfo {author} {\bibfnamefont {A.}~\bibnamefont {{Brochard}}}, \bibinfo {author} {\bibfnamefont {C.}~\bibnamefont {{Uhlemann}}}, \bibinfo {author} {\bibfnamefont {C.-T.}\ \bibnamefont {{Chiang}}}, \bibinfo {author} {\bibfnamefont {S.}~\bibnamefont {{He}}}, \bibinfo {author} {\bibfnamefont {A.}~\bibnamefont {{Pisani}}}, \bibinfo
  {author} {\bibfnamefont {A.}~\bibnamefont {{Obuljen}}}, \bibinfo {author} {\bibfnamefont {Y.}~\bibnamefont {{Feng}}}, \bibinfo {author} {\bibfnamefont {E.}~\bibnamefont {{Castorina}}}, \bibinfo {author} {\bibfnamefont {G.}~\bibnamefont {{Contardo}}}, \bibinfo {author} {\bibfnamefont {C.~D.}\ \bibnamefont {{Kreisch}}}, \bibinfo {author} {\bibfnamefont {A.}~\bibnamefont {{Nicola}}}, \bibinfo {author} {\bibfnamefont {J.}~\bibnamefont {{Alsing}}}, \bibinfo {author} {\bibfnamefont {R.}~\bibnamefont {{Scoccimarro}}}, \bibinfo {author} {\bibfnamefont {L.}~\bibnamefont {{Verde}}}, \bibinfo {author} {\bibfnamefont {M.}~\bibnamefont {{Viel}}}, \bibinfo {author} {\bibfnamefont {S.}~\bibnamefont {{Ho}}}, \bibinfo {author} {\bibfnamefont {S.}~\bibnamefont {{Mallat}}}, \bibinfo {author} {\bibfnamefont {B.}~\bibnamefont {{Wandelt}}},\ and\ \bibinfo {author} {\bibfnamefont {D.~N.}\ \bibnamefont {{Spergel}}},\ }\bibfield  {title} {\bibinfo {title} {{The Quijote Simulations}},\ }\href
  {https://doi.org/10.3847/1538-4365/ab9d82} {\bibfield  {journal} {\bibinfo  {journal} {\apjs}\ }\textbf {\bibinfo {volume} {250}},\ \bibinfo {eid} {2} (\bibinfo {year} {2020})},\ \Eprint {https://arxiv.org/abs/1909.05273} {arXiv:1909.05273 [astro-ph.CO]} \BibitemShut {NoStop}%
\bibitem [{\citenamefont {{Villaescusa-Navarro}}\ \emph {et~al.}(2021)\citenamefont {{Villaescusa-Navarro}}, \citenamefont {{Angl{\'e}s-Alc{\'a}zar}}, \citenamefont {{Genel}}, \citenamefont {{Spergel}}, \citenamefont {{Somerville}}, \citenamefont {{Dave}}, \citenamefont {{Pillepich}}, \citenamefont {{Hernquist}}, \citenamefont {{Nelson}}, \citenamefont {{Torrey}}, \citenamefont {{Narayanan}}, \citenamefont {{Li}}, \citenamefont {{Philcox}}, \citenamefont {{La Torre}}, \citenamefont {{Maria Delgado}}, \citenamefont {{Ho}}, \citenamefont {{Hassan}}, \citenamefont {{Burkhart}}, \citenamefont {{Wadekar}}, \citenamefont {{Battaglia}}, \citenamefont {{Contardo}},\ and\ \citenamefont {{Bryan}}}]{camels}%
  \BibitemOpen
  \bibfield  {author} {\bibinfo {author} {\bibfnamefont {F.}~\bibnamefont {{Villaescusa-Navarro}}}, \bibinfo {author} {\bibfnamefont {D.}~\bibnamefont {{Angl{\'e}s-Alc{\'a}zar}}}, \bibinfo {author} {\bibfnamefont {S.}~\bibnamefont {{Genel}}}, \bibinfo {author} {\bibfnamefont {D.~N.}\ \bibnamefont {{Spergel}}}, \bibinfo {author} {\bibfnamefont {R.~S.}\ \bibnamefont {{Somerville}}}, \bibinfo {author} {\bibfnamefont {R.}~\bibnamefont {{Dave}}}, \bibinfo {author} {\bibfnamefont {A.}~\bibnamefont {{Pillepich}}}, \bibinfo {author} {\bibfnamefont {L.}~\bibnamefont {{Hernquist}}}, \bibinfo {author} {\bibfnamefont {D.}~\bibnamefont {{Nelson}}}, \bibinfo {author} {\bibfnamefont {P.}~\bibnamefont {{Torrey}}}, \bibinfo {author} {\bibfnamefont {D.}~\bibnamefont {{Narayanan}}}, \bibinfo {author} {\bibfnamefont {Y.}~\bibnamefont {{Li}}}, \bibinfo {author} {\bibfnamefont {O.}~\bibnamefont {{Philcox}}}, \bibinfo {author} {\bibfnamefont {V.}~\bibnamefont {{La Torre}}}, \bibinfo {author} {\bibfnamefont {A.}~\bibnamefont {{Maria
  Delgado}}}, \bibinfo {author} {\bibfnamefont {S.}~\bibnamefont {{Ho}}}, \bibinfo {author} {\bibfnamefont {S.}~\bibnamefont {{Hassan}}}, \bibinfo {author} {\bibfnamefont {B.}~\bibnamefont {{Burkhart}}}, \bibinfo {author} {\bibfnamefont {D.}~\bibnamefont {{Wadekar}}}, \bibinfo {author} {\bibfnamefont {N.}~\bibnamefont {{Battaglia}}}, \bibinfo {author} {\bibfnamefont {G.}~\bibnamefont {{Contardo}}},\ and\ \bibinfo {author} {\bibfnamefont {G.~L.}\ \bibnamefont {{Bryan}}},\ }\bibfield  {title} {\bibinfo {title} {{The CAMELS Project: Cosmology and Astrophysics with Machine-learning Simulations}},\ }\href {https://doi.org/10.3847/1538-4357/abf7ba} {\bibfield  {journal} {\bibinfo  {journal} {\apj}\ }\textbf {\bibinfo {volume} {915}},\ \bibinfo {eid} {71} (\bibinfo {year} {2021})},\ \Eprint {https://arxiv.org/abs/2010.00619} {arXiv:2010.00619 [astro-ph.CO]} \BibitemShut {NoStop}%
\bibitem [{\citenamefont {{Nelson}}\ \emph {et~al.}(2019)\citenamefont {{Nelson}}, \citenamefont {{Springel}}, \citenamefont {{Pillepich}}, \citenamefont {{Rodriguez-Gomez}}, \citenamefont {{Torrey}}, \citenamefont {{Genel}}, \citenamefont {{Vogelsberger}}, \citenamefont {{Pakmor}}, \citenamefont {{Marinacci}}, \citenamefont {{Weinberger}}, \citenamefont {{Kelley}}, \citenamefont {{Lovell}}, \citenamefont {{Diemer}},\ and\ \citenamefont {{Hernquist}}}]{illustristng}%
  \BibitemOpen
  \bibfield  {author} {\bibinfo {author} {\bibfnamefont {D.}~\bibnamefont {{Nelson}}}, \bibinfo {author} {\bibfnamefont {V.}~\bibnamefont {{Springel}}}, \bibinfo {author} {\bibfnamefont {A.}~\bibnamefont {{Pillepich}}}, \bibinfo {author} {\bibfnamefont {V.}~\bibnamefont {{Rodriguez-Gomez}}}, \bibinfo {author} {\bibfnamefont {P.}~\bibnamefont {{Torrey}}}, \bibinfo {author} {\bibfnamefont {S.}~\bibnamefont {{Genel}}}, \bibinfo {author} {\bibfnamefont {M.}~\bibnamefont {{Vogelsberger}}}, \bibinfo {author} {\bibfnamefont {R.}~\bibnamefont {{Pakmor}}}, \bibinfo {author} {\bibfnamefont {F.}~\bibnamefont {{Marinacci}}}, \bibinfo {author} {\bibfnamefont {R.}~\bibnamefont {{Weinberger}}}, \bibinfo {author} {\bibfnamefont {L.}~\bibnamefont {{Kelley}}}, \bibinfo {author} {\bibfnamefont {M.}~\bibnamefont {{Lovell}}}, \bibinfo {author} {\bibfnamefont {B.}~\bibnamefont {{Diemer}}},\ and\ \bibinfo {author} {\bibfnamefont {L.}~\bibnamefont {{Hernquist}}},\ }\bibfield  {title} {\bibinfo {title} {{The IllustrisTNG simulations:
  public data release}},\ }\href {https://doi.org/10.1186/s40668-019-0028-x} {\bibfield  {journal} {\bibinfo  {journal} {Computational Astrophysics and Cosmology}\ }\textbf {\bibinfo {volume} {6}},\ \bibinfo {eid} {2} (\bibinfo {year} {2019})},\ \Eprint {https://arxiv.org/abs/1812.05609} {arXiv:1812.05609 [astro-ph.GA]} \BibitemShut {NoStop}%
\bibitem [{Note16()}]{Note16}%
  \BibitemOpen
  \bibinfo {note} {We refer the reader to Refs.~\cite {camels,illustristng} for details about how these colors are assigned to simulated galaxies.}\BibitemShut {Stop}%
\bibitem [{Note17()}]{Note17}%
  \BibitemOpen
  \bibinfo {note} {This is a manifestation of the ``galaxy assembly bias'' of $b_{\phi }$ that has also been seen for halos (e.g. in Refs.~\cite {reid_ab_10,lazeyras_AB_quadratic_halos_21}}\BibitemShut {NoStop}%
\bibitem [{\citenamefont {{Lovell}}\ \emph {et~al.}(2024)\citenamefont {{Lovell}}, \citenamefont {{Starkenburg}}, \citenamefont {{Ho}}, \citenamefont {{Angl{\'e}s-Alc{\'a}zar}}, \citenamefont {{Dav{\'e}}}, \citenamefont {{Gabrielpillai}}, \citenamefont {{Iyer}}, \citenamefont {{Matthews}}, \citenamefont {{Roper}}, \citenamefont {{Somerville}}, \citenamefont {{Sommovigo}},\ and\ \citenamefont {{Villaescusa-Navarro}}}]{lovell_camels_photometry}%
  \BibitemOpen
  \bibfield  {author} {\bibinfo {author} {\bibfnamefont {C.~C.}\ \bibnamefont {{Lovell}}}, \bibinfo {author} {\bibfnamefont {T.}~\bibnamefont {{Starkenburg}}}, \bibinfo {author} {\bibfnamefont {M.}~\bibnamefont {{Ho}}}, \bibinfo {author} {\bibfnamefont {D.}~\bibnamefont {{Angl{\'e}s-Alc{\'a}zar}}}, \bibinfo {author} {\bibfnamefont {R.}~\bibnamefont {{Dav{\'e}}}}, \bibinfo {author} {\bibfnamefont {A.}~\bibnamefont {{Gabrielpillai}}}, \bibinfo {author} {\bibfnamefont {K.}~\bibnamefont {{Iyer}}}, \bibinfo {author} {\bibfnamefont {A.~E.}\ \bibnamefont {{Matthews}}}, \bibinfo {author} {\bibfnamefont {W.~J.}\ \bibnamefont {{Roper}}}, \bibinfo {author} {\bibfnamefont {R.}~\bibnamefont {{Somerville}}}, \bibinfo {author} {\bibfnamefont {L.}~\bibnamefont {{Sommovigo}}},\ and\ \bibinfo {author} {\bibfnamefont {F.}~\bibnamefont {{Villaescusa-Navarro}}},\ }\bibfield  {title} {\bibinfo {title} {{Learning the Universe: Cosmological and Astrophysical Parameter Inference with Galaxy Luminosity Functions and Colours}},\ }\href
  {https://doi.org/10.48550/arXiv.2411.13960} {\bibfield  {journal} {\bibinfo  {journal} {arXiv e-prints}\ ,\ \bibinfo {eid} {arXiv:2411.13960}} (\bibinfo {year} {2024})},\ \Eprint {https://arxiv.org/abs/2411.13960} {arXiv:2411.13960 [astro-ph.GA]} \BibitemShut {NoStop}%
\bibitem [{Note18()}]{Note18}%
  \BibitemOpen
  \bibinfo {note} {As a technical note, in Figure~\ref {fig:hydro_gmr_camels}, we used the older CAMELS SDSS band photometry, while in Figure~\ref {fig:hydro_gmr_camels_dust}, we use the more recent simulated photometry from Ref.~\cite {lovell_camels_photometry}}\BibitemShut {NoStop}%
\bibitem [{\citenamefont {{Maartens}}\ \emph {et~al.}(2021)\citenamefont {{Maartens}}, \citenamefont {{Fonseca}}, \citenamefont {{Camera}}, \citenamefont {{Jolicoeur}}, \citenamefont {{Viljoen}},\ and\ \citenamefont {{Clarkson}}}]{maartens_bev}%
  \BibitemOpen
  \bibfield  {author} {\bibinfo {author} {\bibfnamefont {R.}~\bibnamefont {{Maartens}}}, \bibinfo {author} {\bibfnamefont {J.}~\bibnamefont {{Fonseca}}}, \bibinfo {author} {\bibfnamefont {S.}~\bibnamefont {{Camera}}}, \bibinfo {author} {\bibfnamefont {S.}~\bibnamefont {{Jolicoeur}}}, \bibinfo {author} {\bibfnamefont {J.-A.}\ \bibnamefont {{Viljoen}}},\ and\ \bibinfo {author} {\bibfnamefont {C.}~\bibnamefont {{Clarkson}}},\ }\bibfield  {title} {\bibinfo {title} {{Magnification and evolution biases in large-scale structure surveys}},\ }\href {https://doi.org/10.1088/1475-7516/2021/12/009} {\bibfield  {journal} {\bibinfo  {journal} {\jcap}\ }\textbf {\bibinfo {volume} {2021}},\ \bibinfo {eid} {009} (\bibinfo {year} {2021})},\ \Eprint {https://arxiv.org/abs/2107.13401} {arXiv:2107.13401 [astro-ph.CO]} \BibitemShut {NoStop}%
\bibitem [{Note19()}]{Note19}%
  \BibitemOpen
  \bibinfo {note} {In the fictitious case where we observe all galaxies in the universe, the number of objects would not be altered by this magnification, only the observed flux.}\BibitemShut {Stop}%
\bibitem [{\citenamefont {{Yoo}}\ \emph {et~al.}(2009)\citenamefont {{Yoo}}, \citenamefont {{Fitzpatrick}},\ and\ \citenamefont {{Zaldarriaga}}}]{yoo_plus_09_gr}%
  \BibitemOpen
  \bibfield  {author} {\bibinfo {author} {\bibfnamefont {J.}~\bibnamefont {{Yoo}}}, \bibinfo {author} {\bibfnamefont {A.~L.}\ \bibnamefont {{Fitzpatrick}}},\ and\ \bibinfo {author} {\bibfnamefont {M.}~\bibnamefont {{Zaldarriaga}}},\ }\bibfield  {title} {\bibinfo {title} {{New perspective on galaxy clustering as a cosmological probe: General relativistic effects}},\ }\href {https://doi.org/10.1103/PhysRevD.80.083514} {\bibfield  {journal} {\bibinfo  {journal} {\prd}\ }\textbf {\bibinfo {volume} {80}},\ \bibinfo {eid} {083514} (\bibinfo {year} {2009})},\ \Eprint {https://arxiv.org/abs/0907.0707} {arXiv:0907.0707 [astro-ph.CO]} \BibitemShut {NoStop}%
\bibitem [{\citenamefont {{Bartelmann}}\ and\ \citenamefont {{Schneider}}(2001)}]{bartelmann_schneider_wl_review}%
  \BibitemOpen
  \bibfield  {author} {\bibinfo {author} {\bibfnamefont {M.}~\bibnamefont {{Bartelmann}}}\ and\ \bibinfo {author} {\bibfnamefont {P.}~\bibnamefont {{Schneider}}},\ }\bibfield  {title} {\bibinfo {title} {{Weak gravitational lensing}},\ }\href {https://doi.org/10.1016/S0370-1573(00)00082-X} {\bibfield  {journal} {\bibinfo  {journal} {\physrep}\ }\textbf {\bibinfo {volume} {340}},\ \bibinfo {pages} {291} (\bibinfo {year} {2001})},\ \Eprint {https://arxiv.org/abs/astro-ph/9912508} {arXiv:astro-ph/9912508 [astro-ph]} \BibitemShut {NoStop}%
\bibitem [{\citenamefont {{Reid}}\ \emph {et~al.}(2016)\citenamefont {{Reid}}, \citenamefont {{Ho}}, \citenamefont {{Padmanabhan}}, \citenamefont {{Percival}}, \citenamefont {{Tinker}}, \citenamefont {{Tojeiro}}, \citenamefont {{White}}, \citenamefont {{Eisenstein}}, \citenamefont {{Maraston}}, \citenamefont {{Ross}}, \citenamefont {{S{\'a}nchez}}, \citenamefont {{Schlegel}}, \citenamefont {{Sheldon}}, \citenamefont {{Strauss}}, \citenamefont {{Thomas}}, \citenamefont {{Wake}}, \citenamefont {{Beutler}}, \citenamefont {{Bizyaev}}, \citenamefont {{Bolton}}, \citenamefont {{Brownstein}}, \citenamefont {{Chuang}}, \citenamefont {{Dawson}}, \citenamefont {{Harding}}, \citenamefont {{Kitaura}}, \citenamefont {{Leauthaud}}, \citenamefont {{Masters}}, \citenamefont {{McBride}}, \citenamefont {{More}}, \citenamefont {{Olmstead}}, \citenamefont {{Oravetz}}, \citenamefont {{Nuza}}, \citenamefont {{Pan}}, \citenamefont {{Parejko}}, \citenamefont {{Pforr}}, \citenamefont {{Prada}}, \citenamefont
  {{Rodr{\'\i}guez-Torres}}, \citenamefont {{Salazar-Albornoz}}, \citenamefont {{Samushia}}, \citenamefont {{Schneider}}, \citenamefont {{Sc{\'o}ccola}}, \citenamefont {{Simmons}},\ and\ \citenamefont {{Vargas-Magana}}}]{reid_target_selection}%
  \BibitemOpen
  \bibfield  {author} {\bibinfo {author} {\bibfnamefont {B.}~\bibnamefont {{Reid}}}, \bibinfo {author} {\bibfnamefont {S.}~\bibnamefont {{Ho}}}, \bibinfo {author} {\bibfnamefont {N.}~\bibnamefont {{Padmanabhan}}}, \bibinfo {author} {\bibfnamefont {W.~J.}\ \bibnamefont {{Percival}}}, \bibinfo {author} {\bibfnamefont {J.}~\bibnamefont {{Tinker}}}, \bibinfo {author} {\bibfnamefont {R.}~\bibnamefont {{Tojeiro}}}, \bibinfo {author} {\bibfnamefont {M.}~\bibnamefont {{White}}}, \bibinfo {author} {\bibfnamefont {D.~J.}\ \bibnamefont {{Eisenstein}}}, \bibinfo {author} {\bibfnamefont {C.}~\bibnamefont {{Maraston}}}, \bibinfo {author} {\bibfnamefont {A.~J.}\ \bibnamefont {{Ross}}}, \bibinfo {author} {\bibfnamefont {A.~G.}\ \bibnamefont {{S{\'a}nchez}}}, \bibinfo {author} {\bibfnamefont {D.}~\bibnamefont {{Schlegel}}}, \bibinfo {author} {\bibfnamefont {E.}~\bibnamefont {{Sheldon}}}, \bibinfo {author} {\bibfnamefont {M.~A.}\ \bibnamefont {{Strauss}}}, \bibinfo {author} {\bibfnamefont {D.}~\bibnamefont {{Thomas}}},
  \bibinfo {author} {\bibfnamefont {D.}~\bibnamefont {{Wake}}}, \bibinfo {author} {\bibfnamefont {F.}~\bibnamefont {{Beutler}}}, \bibinfo {author} {\bibfnamefont {D.}~\bibnamefont {{Bizyaev}}}, \bibinfo {author} {\bibfnamefont {A.~S.}\ \bibnamefont {{Bolton}}}, \bibinfo {author} {\bibfnamefont {J.~R.}\ \bibnamefont {{Brownstein}}}, \bibinfo {author} {\bibfnamefont {C.-H.}\ \bibnamefont {{Chuang}}}, \bibinfo {author} {\bibfnamefont {K.}~\bibnamefont {{Dawson}}}, \bibinfo {author} {\bibfnamefont {P.}~\bibnamefont {{Harding}}}, \bibinfo {author} {\bibfnamefont {F.-S.}\ \bibnamefont {{Kitaura}}}, \bibinfo {author} {\bibfnamefont {A.}~\bibnamefont {{Leauthaud}}}, \bibinfo {author} {\bibfnamefont {K.}~\bibnamefont {{Masters}}}, \bibinfo {author} {\bibfnamefont {C.~K.}\ \bibnamefont {{McBride}}}, \bibinfo {author} {\bibfnamefont {S.}~\bibnamefont {{More}}}, \bibinfo {author} {\bibfnamefont {M.~D.}\ \bibnamefont {{Olmstead}}}, \bibinfo {author} {\bibfnamefont {D.}~\bibnamefont {{Oravetz}}}, \bibinfo {author}
  {\bibfnamefont {S.~E.}\ \bibnamefont {{Nuza}}}, \bibinfo {author} {\bibfnamefont {K.}~\bibnamefont {{Pan}}}, \bibinfo {author} {\bibfnamefont {J.}~\bibnamefont {{Parejko}}}, \bibinfo {author} {\bibfnamefont {J.}~\bibnamefont {{Pforr}}}, \bibinfo {author} {\bibfnamefont {F.}~\bibnamefont {{Prada}}}, \bibinfo {author} {\bibfnamefont {S.}~\bibnamefont {{Rodr{\'\i}guez-Torres}}}, \bibinfo {author} {\bibfnamefont {S.}~\bibnamefont {{Salazar-Albornoz}}}, \bibinfo {author} {\bibfnamefont {L.}~\bibnamefont {{Samushia}}}, \bibinfo {author} {\bibfnamefont {D.~P.}\ \bibnamefont {{Schneider}}}, \bibinfo {author} {\bibfnamefont {C.~G.}\ \bibnamefont {{Sc{\'o}ccola}}}, \bibinfo {author} {\bibfnamefont {A.}~\bibnamefont {{Simmons}}},\ and\ \bibinfo {author} {\bibfnamefont {M.}~\bibnamefont {{Vargas-Magana}}},\ }\bibfield  {title} {\bibinfo {title} {{SDSS-III Baryon Oscillation Spectroscopic Survey Data Release 12: galaxy target selection and large-scale structure catalogues}},\ }\href
  {https://doi.org/10.1093/mnras/stv2382} {\bibfield  {journal} {\bibinfo  {journal} {\mnras}\ }\textbf {\bibinfo {volume} {455}},\ \bibinfo {pages} {1553} (\bibinfo {year} {2016})},\ \Eprint {https://arxiv.org/abs/1509.06529} {arXiv:1509.06529 [astro-ph.CO]} \BibitemShut {NoStop}%
\bibitem [{\citenamefont {{Hogg}}\ \emph {et~al.}(2002)\citenamefont {{Hogg}}, \citenamefont {{Baldry}}, \citenamefont {{Blanton}},\ and\ \citenamefont {{Eisenstein}}}]{hogg_kcorr}%
  \BibitemOpen
  \bibfield  {author} {\bibinfo {author} {\bibfnamefont {D.~W.}\ \bibnamefont {{Hogg}}}, \bibinfo {author} {\bibfnamefont {I.~K.}\ \bibnamefont {{Baldry}}}, \bibinfo {author} {\bibfnamefont {M.~R.}\ \bibnamefont {{Blanton}}},\ and\ \bibinfo {author} {\bibfnamefont {D.~J.}\ \bibnamefont {{Eisenstein}}},\ }\bibfield  {title} {\bibinfo {title} {{The K correction}},\ }\href {https://doi.org/10.48550/arXiv.astro-ph/0210394} {\bibfield  {journal} {\bibinfo  {journal} {arXiv e-prints}\ ,\ \bibinfo {eid} {astro-ph/0210394}} (\bibinfo {year} {2002})},\ \Eprint {https://arxiv.org/abs/astro-ph/0210394} {arXiv:astro-ph/0210394 [astro-ph]} \BibitemShut {NoStop}%
\bibitem [{\citenamefont {{Guo}}\ \emph {et~al.}(2013)\citenamefont {{Guo}}, \citenamefont {{Zehavi}}, \citenamefont {{Zheng}}, \citenamefont {{Weinberg}}, \citenamefont {{Berlind}}, \citenamefont {{Blanton}}, \citenamefont {{Chen}}, \citenamefont {{Eisenstein}}, \citenamefont {{Ho}}, \citenamefont {{Kazin}}, \citenamefont {{Manera}}, \citenamefont {{Maraston}}, \citenamefont {{McBride}}, \citenamefont {{Nuza}}, \citenamefont {{Padmanabhan}}, \citenamefont {{Parejko}}, \citenamefont {{Percival}}, \citenamefont {{Ross}}, \citenamefont {{Ross}}, \citenamefont {{Samushia}}, \citenamefont {{S{\'a}nchez}}, \citenamefont {{Schlegel}}, \citenamefont {{Schneider}}, \citenamefont {{Skibba}}, \citenamefont {{Swanson}}, \citenamefont {{Tinker}}, \citenamefont {{Tojeiro}}, \citenamefont {{Wake}}, \citenamefont {{White}}, \citenamefont {{Bahcall}}, \citenamefont {{Bizyaev}}, \citenamefont {{Brewington}}, \citenamefont {{Bundy}}, \citenamefont {{da Costa}}, \citenamefont {{Ebelke}}, \citenamefont {{Malanushenko}},
  \citenamefont {{Malanushenko}}, \citenamefont {{Oravetz}}, \citenamefont {{Rossi}}, \citenamefont {{Simmons}}, \citenamefont {{Snedden}}, \citenamefont {{Streblyanska}},\ and\ \citenamefont {{Thomas}}}]{guo_13_sdss_kpluse}%
  \BibitemOpen
  \bibfield  {author} {\bibinfo {author} {\bibfnamefont {H.}~\bibnamefont {{Guo}}}, \bibinfo {author} {\bibfnamefont {I.}~\bibnamefont {{Zehavi}}}, \bibinfo {author} {\bibfnamefont {Z.}~\bibnamefont {{Zheng}}}, \bibinfo {author} {\bibfnamefont {D.~H.}\ \bibnamefont {{Weinberg}}}, \bibinfo {author} {\bibfnamefont {A.~A.}\ \bibnamefont {{Berlind}}}, \bibinfo {author} {\bibfnamefont {M.}~\bibnamefont {{Blanton}}}, \bibinfo {author} {\bibfnamefont {Y.}~\bibnamefont {{Chen}}}, \bibinfo {author} {\bibfnamefont {D.~J.}\ \bibnamefont {{Eisenstein}}}, \bibinfo {author} {\bibfnamefont {S.}~\bibnamefont {{Ho}}}, \bibinfo {author} {\bibfnamefont {E.}~\bibnamefont {{Kazin}}}, \bibinfo {author} {\bibfnamefont {M.}~\bibnamefont {{Manera}}}, \bibinfo {author} {\bibfnamefont {C.}~\bibnamefont {{Maraston}}}, \bibinfo {author} {\bibfnamefont {C.~K.}\ \bibnamefont {{McBride}}}, \bibinfo {author} {\bibfnamefont {S.~E.}\ \bibnamefont {{Nuza}}}, \bibinfo {author} {\bibfnamefont {N.}~\bibnamefont {{Padmanabhan}}}, \bibinfo {author}
  {\bibfnamefont {J.~K.}\ \bibnamefont {{Parejko}}}, \bibinfo {author} {\bibfnamefont {W.~J.}\ \bibnamefont {{Percival}}}, \bibinfo {author} {\bibfnamefont {A.~J.}\ \bibnamefont {{Ross}}}, \bibinfo {author} {\bibfnamefont {N.~P.}\ \bibnamefont {{Ross}}}, \bibinfo {author} {\bibfnamefont {L.}~\bibnamefont {{Samushia}}}, \bibinfo {author} {\bibfnamefont {A.~G.}\ \bibnamefont {{S{\'a}nchez}}}, \bibinfo {author} {\bibfnamefont {D.~J.}\ \bibnamefont {{Schlegel}}}, \bibinfo {author} {\bibfnamefont {D.~P.}\ \bibnamefont {{Schneider}}}, \bibinfo {author} {\bibfnamefont {R.~A.}\ \bibnamefont {{Skibba}}}, \bibinfo {author} {\bibfnamefont {M.~E.~C.}\ \bibnamefont {{Swanson}}}, \bibinfo {author} {\bibfnamefont {J.~L.}\ \bibnamefont {{Tinker}}}, \bibinfo {author} {\bibfnamefont {R.}~\bibnamefont {{Tojeiro}}}, \bibinfo {author} {\bibfnamefont {D.~A.}\ \bibnamefont {{Wake}}}, \bibinfo {author} {\bibfnamefont {M.}~\bibnamefont {{White}}}, \bibinfo {author} {\bibfnamefont {N.~A.}\ \bibnamefont {{Bahcall}}}, \bibinfo {author}
  {\bibfnamefont {D.}~\bibnamefont {{Bizyaev}}}, \bibinfo {author} {\bibfnamefont {H.}~\bibnamefont {{Brewington}}}, \bibinfo {author} {\bibfnamefont {K.}~\bibnamefont {{Bundy}}}, \bibinfo {author} {\bibfnamefont {L.~N.~A.}\ \bibnamefont {{da Costa}}}, \bibinfo {author} {\bibfnamefont {G.}~\bibnamefont {{Ebelke}}}, \bibinfo {author} {\bibfnamefont {E.}~\bibnamefont {{Malanushenko}}}, \bibinfo {author} {\bibfnamefont {V.}~\bibnamefont {{Malanushenko}}}, \bibinfo {author} {\bibfnamefont {D.}~\bibnamefont {{Oravetz}}}, \bibinfo {author} {\bibfnamefont {G.}~\bibnamefont {{Rossi}}}, \bibinfo {author} {\bibfnamefont {A.}~\bibnamefont {{Simmons}}}, \bibinfo {author} {\bibfnamefont {S.}~\bibnamefont {{Snedden}}}, \bibinfo {author} {\bibfnamefont {A.}~\bibnamefont {{Streblyanska}}},\ and\ \bibinfo {author} {\bibfnamefont {D.}~\bibnamefont {{Thomas}}},\ }\bibfield  {title} {\bibinfo {title} {{The Clustering of Galaxies in the SDSS-III Baryon Oscillation Spectroscopic Survey: Luminosity and Color Dependence and Redshift
  Evolution}},\ }\href {https://doi.org/10.1088/0004-637X/767/2/122} {\bibfield  {journal} {\bibinfo  {journal} {\apj}\ }\textbf {\bibinfo {volume} {767}},\ \bibinfo {eid} {122} (\bibinfo {year} {2013})},\ \Eprint {https://arxiv.org/abs/1212.1211} {arXiv:1212.1211 [astro-ph.CO]} \BibitemShut {NoStop}%
\bibitem [{\citenamefont {{Sobral-Blanco}}\ \emph {et~al.}(2024)\citenamefont {{Sobral-Blanco}}, \citenamefont {{Bonvin}}, \citenamefont {{Clarkson}},\ and\ \citenamefont {{Maartens}}}]{Sobral-Blanco:2024qlb}%
  \BibitemOpen
  \bibfield  {author} {\bibinfo {author} {\bibfnamefont {D.}~\bibnamefont {{Sobral-Blanco}}}, \bibinfo {author} {\bibfnamefont {C.}~\bibnamefont {{Bonvin}}}, \bibinfo {author} {\bibfnamefont {C.}~\bibnamefont {{Clarkson}}},\ and\ \bibinfo {author} {\bibfnamefont {R.}~\bibnamefont {{Maartens}}},\ }\bibfield  {title} {\bibinfo {title} {{Using relativistic effects in large-scale structure to constrain astrophysical properties of galaxy populations}},\ }\href {https://doi.org/10.48550/arXiv.2406.19908} {\bibfield  {journal} {\bibinfo  {journal} {arXiv e-prints}\ ,\ \bibinfo {eid} {arXiv:2406.19908}} (\bibinfo {year} {2024})},\ \Eprint {https://arxiv.org/abs/2406.19908} {arXiv:2406.19908 [astro-ph.CO]} \BibitemShut {NoStop}%
\bibitem [{\citenamefont {{Zhou}}\ \emph {et~al.}(2023)\citenamefont {{Zhou}}, \citenamefont {{Dey}}, \citenamefont {{Newman}}, \citenamefont {{Eisenstein}}, \citenamefont {{Dawson}}, \citenamefont {{Bailey}}, \citenamefont {{Berti}}, \citenamefont {{Guy}}, \citenamefont {{Lan}}, \citenamefont {{Zou}}, \citenamefont {{Aguilar}}, \citenamefont {{Ahlen}}, \citenamefont {{Alam}}, \citenamefont {{Brooks}}, \citenamefont {{de la Macorra}}, \citenamefont {{Dey}}, \citenamefont {{Dhungana}}, \citenamefont {{Fanning}}, \citenamefont {{Font-Ribera}}, \citenamefont {{Gontcho}}, \citenamefont {{Honscheid}}, \citenamefont {{Ishak}}, \citenamefont {{Kisner}}, \citenamefont {{Kov{\'a}cs}}, \citenamefont {{Kremin}}, \citenamefont {{Landriau}}, \citenamefont {{Levi}}, \citenamefont {{Magneville}}, \citenamefont {{Manera}}, \citenamefont {{Martini}}, \citenamefont {{Meisner}}, \citenamefont {{Miquel}}, \citenamefont {{Moustakas}}, \citenamefont {{Myers}}, \citenamefont {{Nie}}, \citenamefont {{Palanque-Delabrouille}},
  \citenamefont {{Percival}}, \citenamefont {{Poppett}}, \citenamefont {{Prada}}, \citenamefont {{Raichoor}}, \citenamefont {{Ross}}, \citenamefont {{Schlafly}}, \citenamefont {{Schlegel}}, \citenamefont {{Schubnell}}, \citenamefont {{Tarl{\'e}}}, \citenamefont {{Weaver}}, \citenamefont {{Wechsler}}, \citenamefont {{Y{\'e}che}},\ and\ \citenamefont {{Zhou}}}]{zhou_lrg_target}%
  \BibitemOpen
  \bibfield  {author} {\bibinfo {author} {\bibfnamefont {R.}~\bibnamefont {{Zhou}}}, \bibinfo {author} {\bibfnamefont {B.}~\bibnamefont {{Dey}}}, \bibinfo {author} {\bibfnamefont {J.~A.}\ \bibnamefont {{Newman}}}, \bibinfo {author} {\bibfnamefont {D.~J.}\ \bibnamefont {{Eisenstein}}}, \bibinfo {author} {\bibfnamefont {K.}~\bibnamefont {{Dawson}}}, \bibinfo {author} {\bibfnamefont {S.}~\bibnamefont {{Bailey}}}, \bibinfo {author} {\bibfnamefont {A.}~\bibnamefont {{Berti}}}, \bibinfo {author} {\bibfnamefont {J.}~\bibnamefont {{Guy}}}, \bibinfo {author} {\bibfnamefont {T.-W.}\ \bibnamefont {{Lan}}}, \bibinfo {author} {\bibfnamefont {H.}~\bibnamefont {{Zou}}}, \bibinfo {author} {\bibfnamefont {J.}~\bibnamefont {{Aguilar}}}, \bibinfo {author} {\bibfnamefont {S.}~\bibnamefont {{Ahlen}}}, \bibinfo {author} {\bibfnamefont {S.}~\bibnamefont {{Alam}}}, \bibinfo {author} {\bibfnamefont {D.}~\bibnamefont {{Brooks}}}, \bibinfo {author} {\bibfnamefont {A.}~\bibnamefont {{de la Macorra}}}, \bibinfo {author} {\bibfnamefont
  {A.}~\bibnamefont {{Dey}}}, \bibinfo {author} {\bibfnamefont {G.}~\bibnamefont {{Dhungana}}}, \bibinfo {author} {\bibfnamefont {K.}~\bibnamefont {{Fanning}}}, \bibinfo {author} {\bibfnamefont {A.}~\bibnamefont {{Font-Ribera}}}, \bibinfo {author} {\bibfnamefont {S.~G.~A.}\ \bibnamefont {{Gontcho}}}, \bibinfo {author} {\bibfnamefont {K.}~\bibnamefont {{Honscheid}}}, \bibinfo {author} {\bibfnamefont {M.}~\bibnamefont {{Ishak}}}, \bibinfo {author} {\bibfnamefont {T.}~\bibnamefont {{Kisner}}}, \bibinfo {author} {\bibfnamefont {A.}~\bibnamefont {{Kov{\'a}cs}}}, \bibinfo {author} {\bibfnamefont {A.}~\bibnamefont {{Kremin}}}, \bibinfo {author} {\bibfnamefont {M.}~\bibnamefont {{Landriau}}}, \bibinfo {author} {\bibfnamefont {M.~E.}\ \bibnamefont {{Levi}}}, \bibinfo {author} {\bibfnamefont {C.}~\bibnamefont {{Magneville}}}, \bibinfo {author} {\bibfnamefont {M.}~\bibnamefont {{Manera}}}, \bibinfo {author} {\bibfnamefont {P.}~\bibnamefont {{Martini}}}, \bibinfo {author} {\bibfnamefont {A.~M.}\ \bibnamefont
  {{Meisner}}}, \bibinfo {author} {\bibfnamefont {R.}~\bibnamefont {{Miquel}}}, \bibinfo {author} {\bibfnamefont {J.}~\bibnamefont {{Moustakas}}}, \bibinfo {author} {\bibfnamefont {A.~D.}\ \bibnamefont {{Myers}}}, \bibinfo {author} {\bibfnamefont {J.}~\bibnamefont {{Nie}}}, \bibinfo {author} {\bibfnamefont {N.}~\bibnamefont {{Palanque-Delabrouille}}}, \bibinfo {author} {\bibfnamefont {W.~J.}\ \bibnamefont {{Percival}}}, \bibinfo {author} {\bibfnamefont {C.}~\bibnamefont {{Poppett}}}, \bibinfo {author} {\bibfnamefont {F.}~\bibnamefont {{Prada}}}, \bibinfo {author} {\bibfnamefont {A.}~\bibnamefont {{Raichoor}}}, \bibinfo {author} {\bibfnamefont {A.~J.}\ \bibnamefont {{Ross}}}, \bibinfo {author} {\bibfnamefont {E.}~\bibnamefont {{Schlafly}}}, \bibinfo {author} {\bibfnamefont {D.}~\bibnamefont {{Schlegel}}}, \bibinfo {author} {\bibfnamefont {M.}~\bibnamefont {{Schubnell}}}, \bibinfo {author} {\bibfnamefont {G.}~\bibnamefont {{Tarl{\'e}}}}, \bibinfo {author} {\bibfnamefont {B.~A.}\ \bibnamefont {{Weaver}}},
  \bibinfo {author} {\bibfnamefont {R.~H.}\ \bibnamefont {{Wechsler}}}, \bibinfo {author} {\bibfnamefont {C.}~\bibnamefont {{Y{\'e}che}}},\ and\ \bibinfo {author} {\bibfnamefont {Z.}~\bibnamefont {{Zhou}}},\ }\bibfield  {title} {\bibinfo {title} {{Target Selection and Validation of DESI Luminous Red Galaxies}},\ }\href {https://doi.org/10.3847/1538-3881/aca5fb} {\bibfield  {journal} {\bibinfo  {journal} {\aj}\ }\textbf {\bibinfo {volume} {165}},\ \bibinfo {eid} {58} (\bibinfo {year} {2023})},\ \Eprint {https://arxiv.org/abs/2208.08515} {arXiv:2208.08515 [astro-ph.CO]} \BibitemShut {NoStop}%
\bibitem [{\citenamefont {{Dor{\'e}}}\ \emph {et~al.}(2014)\citenamefont {{Dor{\'e}}}, \citenamefont {{Bock}}, \citenamefont {{Ashby}}, \citenamefont {{Capak}}, \citenamefont {{Cooray}}, \citenamefont {{de Putter}}, \citenamefont {{Eifler}}, \citenamefont {{Flagey}}, \citenamefont {{Gong}}, \citenamefont {{Habib}}, \citenamefont {{Heitmann}}, \citenamefont {{Hirata}}, \citenamefont {{Jeong}}, \citenamefont {{Katti}}, \citenamefont {{Korngut}}, \citenamefont {{Krause}}, \citenamefont {{Lee}}, \citenamefont {{Masters}}, \citenamefont {{Mauskopf}}, \citenamefont {{Melnick}}, \citenamefont {{Mennesson}}, \citenamefont {{Nguyen}}, \citenamefont {{{\"O}berg}}, \citenamefont {{Pullen}}, \citenamefont {{Raccanelli}}, \citenamefont {{Smith}}, \citenamefont {{Song}}, \citenamefont {{Tolls}}, \citenamefont {{Unwin}}, \citenamefont {{Venumadhav}}, \citenamefont {{Viero}}, \citenamefont {{Werner}},\ and\ \citenamefont {{Zemcov}}}]{spherex14}%
  \BibitemOpen
  \bibfield  {author} {\bibinfo {author} {\bibfnamefont {O.}~\bibnamefont {{Dor{\'e}}}}, \bibinfo {author} {\bibfnamefont {J.}~\bibnamefont {{Bock}}}, \bibinfo {author} {\bibfnamefont {M.}~\bibnamefont {{Ashby}}}, \bibinfo {author} {\bibfnamefont {P.}~\bibnamefont {{Capak}}}, \bibinfo {author} {\bibfnamefont {A.}~\bibnamefont {{Cooray}}}, \bibinfo {author} {\bibfnamefont {R.}~\bibnamefont {{de Putter}}}, \bibinfo {author} {\bibfnamefont {T.}~\bibnamefont {{Eifler}}}, \bibinfo {author} {\bibfnamefont {N.}~\bibnamefont {{Flagey}}}, \bibinfo {author} {\bibfnamefont {Y.}~\bibnamefont {{Gong}}}, \bibinfo {author} {\bibfnamefont {S.}~\bibnamefont {{Habib}}}, \bibinfo {author} {\bibfnamefont {K.}~\bibnamefont {{Heitmann}}}, \bibinfo {author} {\bibfnamefont {C.}~\bibnamefont {{Hirata}}}, \bibinfo {author} {\bibfnamefont {W.-S.}\ \bibnamefont {{Jeong}}}, \bibinfo {author} {\bibfnamefont {R.}~\bibnamefont {{Katti}}}, \bibinfo {author} {\bibfnamefont {P.}~\bibnamefont {{Korngut}}}, \bibinfo {author} {\bibfnamefont
  {E.}~\bibnamefont {{Krause}}}, \bibinfo {author} {\bibfnamefont {D.-H.}\ \bibnamefont {{Lee}}}, \bibinfo {author} {\bibfnamefont {D.}~\bibnamefont {{Masters}}}, \bibinfo {author} {\bibfnamefont {P.}~\bibnamefont {{Mauskopf}}}, \bibinfo {author} {\bibfnamefont {G.}~\bibnamefont {{Melnick}}}, \bibinfo {author} {\bibfnamefont {B.}~\bibnamefont {{Mennesson}}}, \bibinfo {author} {\bibfnamefont {H.}~\bibnamefont {{Nguyen}}}, \bibinfo {author} {\bibfnamefont {K.}~\bibnamefont {{{\"O}berg}}}, \bibinfo {author} {\bibfnamefont {A.}~\bibnamefont {{Pullen}}}, \bibinfo {author} {\bibfnamefont {A.}~\bibnamefont {{Raccanelli}}}, \bibinfo {author} {\bibfnamefont {R.}~\bibnamefont {{Smith}}}, \bibinfo {author} {\bibfnamefont {Y.-S.}\ \bibnamefont {{Song}}}, \bibinfo {author} {\bibfnamefont {V.}~\bibnamefont {{Tolls}}}, \bibinfo {author} {\bibfnamefont {S.}~\bibnamefont {{Unwin}}}, \bibinfo {author} {\bibfnamefont {T.}~\bibnamefont {{Venumadhav}}}, \bibinfo {author} {\bibfnamefont {M.}~\bibnamefont {{Viero}}}, \bibinfo
  {author} {\bibfnamefont {M.}~\bibnamefont {{Werner}}},\ and\ \bibinfo {author} {\bibfnamefont {M.}~\bibnamefont {{Zemcov}}},\ }\bibfield  {title} {\bibinfo {title} {{Cosmology with the SPHEREX All-Sky Spectral Survey}},\ }\href {https://doi.org/10.48550/arXiv.1412.4872} {\bibfield  {journal} {\bibinfo  {journal} {arXiv e-prints}\ ,\ \bibinfo {eid} {arXiv:1412.4872}} (\bibinfo {year} {2014})},\ \Eprint {https://arxiv.org/abs/1412.4872} {arXiv:1412.4872 [astro-ph.CO]} \BibitemShut {NoStop}%
\bibitem [{\citenamefont {{Besuner}}\ \emph {et~al.}(2025)\citenamefont {{Besuner}}, \citenamefont {{Dey}}, \citenamefont {{Drlica-Wagner}}, \citenamefont {{Ebina}}, \citenamefont {{Fernandez Moroni}}, \citenamefont {{Ferraro}}, \citenamefont {{Forero-Romero}}, \citenamefont {{Honscheid}}, \citenamefont {{Jelinsky}}, \citenamefont {{Lang}}, \citenamefont {{Levi}}, \citenamefont {{Martini}}, \citenamefont {{Myers}}, \citenamefont {{Palanque-Delabrouille}}, \citenamefont {{Panda}}, \citenamefont {{Poppett}}, \citenamefont {{Sailer}}, \citenamefont {{Schlegel}}, \citenamefont {{Shafieloo}}, \citenamefont {{Silber}}, \citenamefont {{White}}, \citenamefont {{Abbott}}, \citenamefont {{Allen}}, \citenamefont {{Avila}}, \citenamefont {{Bailey}}, \citenamefont {{Bault}}, \citenamefont {{Bouri}}, \citenamefont {{Boutsia}}, \citenamefont {{Burtin}}, \citenamefont {{Chierchie}}, \citenamefont {{Coulton}}, \citenamefont {{Dawson}}, \citenamefont {{Dey}}, \citenamefont {{Dunlop}}, \citenamefont {{Eisenstein}},
  \citenamefont {{Emanuele}}, \citenamefont {{Escoffier}}, \citenamefont {{Estrada}}, \citenamefont {{Fagrelius}}, \citenamefont {{Fanning}}, \citenamefont {{Fanning}}, \citenamefont {{Font-Ribera}}, \citenamefont {{Frieman}}, \citenamefont {{Galal}}, \citenamefont {{Gluscevic}}, \citenamefont {{Gontcho}}, \citenamefont {{Green}}, \citenamefont {{Gutierrez}}, \citenamefont {{Guy}}, \citenamefont {{Hashemi}}, \citenamefont {{Heathcote}}, \citenamefont {{Holland}}, \citenamefont {{Hou}}, \citenamefont {{Huterer}}, \citenamefont {{Irigoyen Gimenez}}, \citenamefont {{Ivanov}}, \citenamefont {{Joyce}}, \citenamefont {{Jullo}}, \citenamefont {{Juneau}}, \citenamefont {{Juramy}}, \citenamefont {{Karcher}}, \citenamefont {{Kent}}, \citenamefont {{Kirkby}}, \citenamefont {{Kneib}}, \citenamefont {{Krause}}, \citenamefont {{Krolewski}}, \citenamefont {{Lahav}}, \citenamefont {{Lapi}}, \citenamefont {{Leauthaud}}, \citenamefont {{Lewandowski}}, \citenamefont {{Li}}, \citenamefont {{Lin}}, \citenamefont {{Loverde}},
  \citenamefont {{MacBride}}, \citenamefont {{Magneville}}, \citenamefont {{Marshall}}, \citenamefont {{McDonald}}, \citenamefont {{Miller}}, \citenamefont {{Moustakas}}, \citenamefont {{M{\"u}nchmeyer}}, \citenamefont {{Najita}}, \citenamefont {{Newman}}, \citenamefont {{Percival}}, \citenamefont {{Philcox}}, \citenamefont {{Pires}}, \citenamefont {{Raichoor}}, \citenamefont {{Roach}}, \citenamefont {{Rockosi}}, \citenamefont {{Rombach}}, \citenamefont {{Ross}}, \citenamefont {{Sanchez}}, \citenamefont {{Schmidt}}, \citenamefont {{Schubnell}}, \citenamefont {{Sebok}}, \citenamefont {{Seljak}}, \citenamefont {{Silverstein}}, \citenamefont {{Slepian}}, \citenamefont {{Stupak}}, \citenamefont {{Tarl{\'e}}}, \citenamefont {{Tyas}}, \citenamefont {{Vargas-Maga{\~n}a}}, \citenamefont {{Walker}}, \citenamefont {{Wenner}}, \citenamefont {{Y{\`e}che}}, \citenamefont {{Zhang}},\ and\ \citenamefont {{Zhou}}}]{spec_s5}%
  \BibitemOpen
  \bibfield  {author} {\bibinfo {author} {\bibfnamefont {R.}~\bibnamefont {{Besuner}}}, \bibinfo {author} {\bibfnamefont {A.}~\bibnamefont {{Dey}}}, \bibinfo {author} {\bibfnamefont {A.}~\bibnamefont {{Drlica-Wagner}}}, \bibinfo {author} {\bibfnamefont {H.}~\bibnamefont {{Ebina}}}, \bibinfo {author} {\bibfnamefont {G.}~\bibnamefont {{Fernandez Moroni}}}, \bibinfo {author} {\bibfnamefont {S.}~\bibnamefont {{Ferraro}}}, \bibinfo {author} {\bibfnamefont {J.}~\bibnamefont {{Forero-Romero}}}, \bibinfo {author} {\bibfnamefont {K.}~\bibnamefont {{Honscheid}}}, \bibinfo {author} {\bibfnamefont {P.}~\bibnamefont {{Jelinsky}}}, \bibinfo {author} {\bibfnamefont {D.}~\bibnamefont {{Lang}}}, \bibinfo {author} {\bibfnamefont {M.}~\bibnamefont {{Levi}}}, \bibinfo {author} {\bibfnamefont {P.}~\bibnamefont {{Martini}}}, \bibinfo {author} {\bibfnamefont {A.}~\bibnamefont {{Myers}}}, \bibinfo {author} {\bibfnamefont {N.}~\bibnamefont {{Palanque-Delabrouille}}}, \bibinfo {author} {\bibfnamefont {S.}~\bibnamefont {{Panda}}},
  \bibinfo {author} {\bibfnamefont {C.}~\bibnamefont {{Poppett}}}, \bibinfo {author} {\bibfnamefont {N.}~\bibnamefont {{Sailer}}}, \bibinfo {author} {\bibfnamefont {D.}~\bibnamefont {{Schlegel}}}, \bibinfo {author} {\bibfnamefont {A.}~\bibnamefont {{Shafieloo}}}, \bibinfo {author} {\bibfnamefont {J.}~\bibnamefont {{Silber}}}, \bibinfo {author} {\bibfnamefont {M.}~\bibnamefont {{White}}}, \bibinfo {author} {\bibfnamefont {T.}~\bibnamefont {{Abbott}}}, \bibinfo {author} {\bibfnamefont {L.}~\bibnamefont {{Allen}}}, \bibinfo {author} {\bibfnamefont {S.}~\bibnamefont {{Avila}}}, \bibinfo {author} {\bibfnamefont {S.}~\bibnamefont {{Bailey}}}, \bibinfo {author} {\bibfnamefont {A.}~\bibnamefont {{Bault}}}, \bibinfo {author} {\bibfnamefont {M.}~\bibnamefont {{Bouri}}}, \bibinfo {author} {\bibfnamefont {K.}~\bibnamefont {{Boutsia}}}, \bibinfo {author} {\bibfnamefont {E.}~\bibnamefont {{Burtin}}}, \bibinfo {author} {\bibfnamefont {F.}~\bibnamefont {{Chierchie}}}, \bibinfo {author} {\bibfnamefont {W.}~\bibnamefont
  {{Coulton}}}, \bibinfo {author} {\bibfnamefont {K.}~\bibnamefont {{Dawson}}}, \bibinfo {author} {\bibfnamefont {B.}~\bibnamefont {{Dey}}}, \bibinfo {author} {\bibfnamefont {P.}~\bibnamefont {{Dunlop}}}, \bibinfo {author} {\bibfnamefont {D.}~\bibnamefont {{Eisenstein}}}, \bibinfo {author} {\bibfnamefont {C.}~\bibnamefont {{Emanuele}}}, \bibinfo {author} {\bibfnamefont {S.}~\bibnamefont {{Escoffier}}}, \bibinfo {author} {\bibfnamefont {J.}~\bibnamefont {{Estrada}}}, \bibinfo {author} {\bibfnamefont {P.}~\bibnamefont {{Fagrelius}}}, \bibinfo {author} {\bibfnamefont {K.}~\bibnamefont {{Fanning}}}, \bibinfo {author} {\bibfnamefont {T.}~\bibnamefont {{Fanning}}}, \bibinfo {author} {\bibfnamefont {A.}~\bibnamefont {{Font-Ribera}}}, \bibinfo {author} {\bibfnamefont {J.}~\bibnamefont {{Frieman}}}, \bibinfo {author} {\bibfnamefont {M.}~\bibnamefont {{Galal}}}, \bibinfo {author} {\bibfnamefont {V.}~\bibnamefont {{Gluscevic}}}, \bibinfo {author} {\bibfnamefont {S.~G.~A.}\ \bibnamefont {{Gontcho}}}, \bibinfo {author}
  {\bibfnamefont {D.}~\bibnamefont {{Green}}}, \bibinfo {author} {\bibfnamefont {G.}~\bibnamefont {{Gutierrez}}}, \bibinfo {author} {\bibfnamefont {J.}~\bibnamefont {{Guy}}}, \bibinfo {author} {\bibfnamefont {K.}~\bibnamefont {{Hashemi}}}, \bibinfo {author} {\bibfnamefont {S.}~\bibnamefont {{Heathcote}}}, \bibinfo {author} {\bibfnamefont {S.}~\bibnamefont {{Holland}}}, \bibinfo {author} {\bibfnamefont {J.}~\bibnamefont {{Hou}}}, \bibinfo {author} {\bibfnamefont {D.}~\bibnamefont {{Huterer}}}, \bibinfo {author} {\bibfnamefont {B.}~\bibnamefont {{Irigoyen Gimenez}}}, \bibinfo {author} {\bibfnamefont {M.}~\bibnamefont {{Ivanov}}}, \bibinfo {author} {\bibfnamefont {R.}~\bibnamefont {{Joyce}}}, \bibinfo {author} {\bibfnamefont {E.}~\bibnamefont {{Jullo}}}, \bibinfo {author} {\bibfnamefont {S.}~\bibnamefont {{Juneau}}}, \bibinfo {author} {\bibfnamefont {C.}~\bibnamefont {{Juramy}}}, \bibinfo {author} {\bibfnamefont {A.}~\bibnamefont {{Karcher}}}, \bibinfo {author} {\bibfnamefont {S.}~\bibnamefont {{Kent}}},
  \bibinfo {author} {\bibfnamefont {D.}~\bibnamefont {{Kirkby}}}, \bibinfo {author} {\bibfnamefont {J.-P.}\ \bibnamefont {{Kneib}}}, \bibinfo {author} {\bibfnamefont {E.}~\bibnamefont {{Krause}}}, \bibinfo {author} {\bibfnamefont {A.}~\bibnamefont {{Krolewski}}}, \bibinfo {author} {\bibfnamefont {O.}~\bibnamefont {{Lahav}}}, \bibinfo {author} {\bibfnamefont {A.}~\bibnamefont {{Lapi}}}, \bibinfo {author} {\bibfnamefont {A.}~\bibnamefont {{Leauthaud}}}, \bibinfo {author} {\bibfnamefont {M.}~\bibnamefont {{Lewandowski}}}, \bibinfo {author} {\bibfnamefont {T.}~\bibnamefont {{Li}}}, \bibinfo {author} {\bibfnamefont {K.}~\bibnamefont {{Lin}}}, \bibinfo {author} {\bibfnamefont {M.}~\bibnamefont {{Loverde}}}, \bibinfo {author} {\bibfnamefont {S.}~\bibnamefont {{MacBride}}}, \bibinfo {author} {\bibfnamefont {C.}~\bibnamefont {{Magneville}}}, \bibinfo {author} {\bibfnamefont {J.}~\bibnamefont {{Marshall}}}, \bibinfo {author} {\bibfnamefont {P.}~\bibnamefont {{McDonald}}}, \bibinfo {author} {\bibfnamefont
  {T.}~\bibnamefont {{Miller}}}, \bibinfo {author} {\bibfnamefont {J.}~\bibnamefont {{Moustakas}}}, \bibinfo {author} {\bibfnamefont {M.}~\bibnamefont {{M{\"u}nchmeyer}}}, \bibinfo {author} {\bibfnamefont {J.}~\bibnamefont {{Najita}}}, \bibinfo {author} {\bibfnamefont {J.}~\bibnamefont {{Newman}}}, \bibinfo {author} {\bibfnamefont {W.}~\bibnamefont {{Percival}}}, \bibinfo {author} {\bibfnamefont {O.}~\bibnamefont {{Philcox}}}, \bibinfo {author} {\bibfnamefont {P.}~\bibnamefont {{Pires}}}, \bibinfo {author} {\bibfnamefont {A.}~\bibnamefont {{Raichoor}}}, \bibinfo {author} {\bibfnamefont {B.}~\bibnamefont {{Roach}}}, \bibinfo {author} {\bibfnamefont {C.}~\bibnamefont {{Rockosi}}}, \bibinfo {author} {\bibfnamefont {M.}~\bibnamefont {{Rombach}}}, \bibinfo {author} {\bibfnamefont {A.}~\bibnamefont {{Ross}}}, \bibinfo {author} {\bibfnamefont {E.}~\bibnamefont {{Sanchez}}}, \bibinfo {author} {\bibfnamefont {L.}~\bibnamefont {{Schmidt}}}, \bibinfo {author} {\bibfnamefont {M.}~\bibnamefont {{Schubnell}}}, \bibinfo
  {author} {\bibfnamefont {R.}~\bibnamefont {{Sebok}}}, \bibinfo {author} {\bibfnamefont {U.}~\bibnamefont {{Seljak}}}, \bibinfo {author} {\bibfnamefont {E.}~\bibnamefont {{Silverstein}}}, \bibinfo {author} {\bibfnamefont {Z.}~\bibnamefont {{Slepian}}}, \bibinfo {author} {\bibfnamefont {R.}~\bibnamefont {{Stupak}}}, \bibinfo {author} {\bibfnamefont {G.}~\bibnamefont {{Tarl{\'e}}}}, \bibinfo {author} {\bibfnamefont {L.}~\bibnamefont {{Tyas}}}, \bibinfo {author} {\bibfnamefont {M.}~\bibnamefont {{Vargas-Maga{\~n}a}}}, \bibinfo {author} {\bibfnamefont {A.}~\bibnamefont {{Walker}}}, \bibinfo {author} {\bibfnamefont {N.}~\bibnamefont {{Wenner}}}, \bibinfo {author} {\bibfnamefont {C.}~\bibnamefont {{Y{\`e}che}}}, \bibinfo {author} {\bibfnamefont {Y.}~\bibnamefont {{Zhang}}},\ and\ \bibinfo {author} {\bibfnamefont {R.}~\bibnamefont {{Zhou}}},\ }\bibfield  {title} {\bibinfo {title} {{The Spectroscopic Stage-5 Experiment}},\ }\href {https://doi.org/10.48550/arXiv.2503.07923} {\bibfield  {journal} {\bibinfo  {journal}
  {arXiv e-prints}\ ,\ \bibinfo {eid} {arXiv:2503.07923}} (\bibinfo {year} {2025})},\ \Eprint {https://arxiv.org/abs/2503.07923} {arXiv:2503.07923 [astro-ph.CO]} \BibitemShut {NoStop}%
\bibitem [{Note20()}]{Note20}%
  \BibitemOpen
  \bibinfo {note} {E.g., such priors may be obtained from simulations \cite {ivanov_sfpng_priors}}\BibitemShut {NoStop}%
\bibitem [{\citenamefont {{Voivodic}}\ and\ \citenamefont {{Barreira}}(2021)}]{voivodic_response_phi}%
  \BibitemOpen
  \bibfield  {author} {\bibinfo {author} {\bibfnamefont {R.}~\bibnamefont {{Voivodic}}}\ and\ \bibinfo {author} {\bibfnamefont {A.}~\bibnamefont {{Barreira}}},\ }\bibfield  {title} {\bibinfo {title} {{Responses of Halo Occupation Distributions: a new ingredient in the halo model \& the impact on galaxy bias}},\ }\href {https://doi.org/10.1088/1475-7516/2021/05/069} {\bibfield  {journal} {\bibinfo  {journal} {\jcap}\ }\textbf {\bibinfo {volume} {2021}},\ \bibinfo {eid} {069} (\bibinfo {year} {2021})},\ \Eprint {https://arxiv.org/abs/2012.04637} {arXiv:2012.04637 [astro-ph.CO]} \BibitemShut {NoStop}%
\bibitem [{Note21()}]{Note21}%
  \BibitemOpen
  \bibinfo {note} {For the case of strong assembly bias, the difference in $\Delta b_{\phi }$ is most important, but the sensitivity of multitracer in the sample variance limit involves the product of $b_{\phi }$ and $b_1$ across samples.}\BibitemShut {Stop}%
\bibitem [{Note22()}]{Note22}%
  \BibitemOpen
  \bibinfo {note} {For example, while argued not to be an issue for LRGs, for emission line galaxies (ELGs), it may be necessary to correct for the interaction between galaxy velocities and color selection \cite {doppler_bias}.}\BibitemShut {Stop}%
\bibitem [{\citenamefont {{Chaussidon}}\ \emph {et~al.}(2024)\citenamefont {{Chaussidon}}, \citenamefont {{Y{\`e}che}}, \citenamefont {{de Mattia}}, \citenamefont {{Payerne}}, \citenamefont {{McDonald}}, \citenamefont {{Ross}}, \citenamefont {{Ahlen}}, \citenamefont {{Bianchi}}, \citenamefont {{Brooks}}, \citenamefont {{Burtin}}, \citenamefont {{Claybaugh}}, \citenamefont {{de la Macorra}}, \citenamefont {{Doel}}, \citenamefont {{Ferraro}}, \citenamefont {{Font-Ribera}}, \citenamefont {{Forero-Romero}}, \citenamefont {{Gazta{\~n}aga}}, \citenamefont {{Gil-Mar{\'\i}n}}, \citenamefont {{Gontcho}}, \citenamefont {{Gutierrez}}, \citenamefont {{Guy}}, \citenamefont {{Honscheid}}, \citenamefont {{Howlett}}, \citenamefont {{Huterer}}, \citenamefont {{Kehoe}}, \citenamefont {{Kirkby}}, \citenamefont {{Kisner}}, \citenamefont {{Kremin}}, \citenamefont {{Le Guillou}}, \citenamefont {{Levi}}, \citenamefont {{Manera}}, \citenamefont {{Meisner}}, \citenamefont {{Miquel}}, \citenamefont {{Moustakas}}, \citenamefont
  {{Newman}}, \citenamefont {{Niz}}, \citenamefont {{Palanque-Delabrouille}}, \citenamefont {{Percival}}, \citenamefont {{Prada}}, \citenamefont {{P{\'e}rez-R{\`a}fols}}, \citenamefont {{Ravoux}}, \citenamefont {{Rossi}}, \citenamefont {{Sanchez}}, \citenamefont {{Schlegel}}, \citenamefont {{Schubnell}}, \citenamefont {{Seo}}, \citenamefont {{Sprayberry}}, \citenamefont {{Tarl{\'e}}}, \citenamefont {{Vargas-Maga{\~n}a}}, \citenamefont {{Weaver}}, \citenamefont {{Zhao}},\ and\ \citenamefont {{Zou}}}]{edmonnd_qso_lrg_desi_lpng}%
  \BibitemOpen
  \bibfield  {author} {\bibinfo {author} {\bibfnamefont {E.}~\bibnamefont {{Chaussidon}}}, \bibinfo {author} {\bibfnamefont {C.}~\bibnamefont {{Y{\`e}che}}}, \bibinfo {author} {\bibfnamefont {A.}~\bibnamefont {{de Mattia}}}, \bibinfo {author} {\bibfnamefont {C.}~\bibnamefont {{Payerne}}}, \bibinfo {author} {\bibfnamefont {P.}~\bibnamefont {{McDonald}}}, \bibinfo {author} {\bibfnamefont {A.~J.}\ \bibnamefont {{Ross}}}, \bibinfo {author} {\bibfnamefont {S.}~\bibnamefont {{Ahlen}}}, \bibinfo {author} {\bibfnamefont {D.}~\bibnamefont {{Bianchi}}}, \bibinfo {author} {\bibfnamefont {D.}~\bibnamefont {{Brooks}}}, \bibinfo {author} {\bibfnamefont {E.}~\bibnamefont {{Burtin}}}, \bibinfo {author} {\bibfnamefont {T.}~\bibnamefont {{Claybaugh}}}, \bibinfo {author} {\bibfnamefont {A.}~\bibnamefont {{de la Macorra}}}, \bibinfo {author} {\bibfnamefont {P.}~\bibnamefont {{Doel}}}, \bibinfo {author} {\bibfnamefont {S.}~\bibnamefont {{Ferraro}}}, \bibinfo {author} {\bibfnamefont {A.}~\bibnamefont {{Font-Ribera}}}, \bibinfo
  {author} {\bibfnamefont {J.~E.}\ \bibnamefont {{Forero-Romero}}}, \bibinfo {author} {\bibfnamefont {E.}~\bibnamefont {{Gazta{\~n}aga}}}, \bibinfo {author} {\bibfnamefont {H.}~\bibnamefont {{Gil-Mar{\'\i}n}}}, \bibinfo {author} {\bibfnamefont {S.~G.~A.}\ \bibnamefont {{Gontcho}}}, \bibinfo {author} {\bibfnamefont {G.}~\bibnamefont {{Gutierrez}}}, \bibinfo {author} {\bibfnamefont {J.}~\bibnamefont {{Guy}}}, \bibinfo {author} {\bibfnamefont {K.}~\bibnamefont {{Honscheid}}}, \bibinfo {author} {\bibfnamefont {C.}~\bibnamefont {{Howlett}}}, \bibinfo {author} {\bibfnamefont {D.}~\bibnamefont {{Huterer}}}, \bibinfo {author} {\bibfnamefont {R.}~\bibnamefont {{Kehoe}}}, \bibinfo {author} {\bibfnamefont {D.}~\bibnamefont {{Kirkby}}}, \bibinfo {author} {\bibfnamefont {T.}~\bibnamefont {{Kisner}}}, \bibinfo {author} {\bibfnamefont {A.}~\bibnamefont {{Kremin}}}, \bibinfo {author} {\bibfnamefont {L.}~\bibnamefont {{Le Guillou}}}, \bibinfo {author} {\bibfnamefont {M.~E.}\ \bibnamefont {{Levi}}}, \bibinfo {author}
  {\bibfnamefont {M.}~\bibnamefont {{Manera}}}, \bibinfo {author} {\bibfnamefont {A.}~\bibnamefont {{Meisner}}}, \bibinfo {author} {\bibfnamefont {R.}~\bibnamefont {{Miquel}}}, \bibinfo {author} {\bibfnamefont {J.}~\bibnamefont {{Moustakas}}}, \bibinfo {author} {\bibfnamefont {J.~A.}\ \bibnamefont {{Newman}}}, \bibinfo {author} {\bibfnamefont {G.}~\bibnamefont {{Niz}}}, \bibinfo {author} {\bibfnamefont {N.}~\bibnamefont {{Palanque-Delabrouille}}}, \bibinfo {author} {\bibfnamefont {W.~J.}\ \bibnamefont {{Percival}}}, \bibinfo {author} {\bibfnamefont {F.}~\bibnamefont {{Prada}}}, \bibinfo {author} {\bibfnamefont {I.}~\bibnamefont {{P{\'e}rez-R{\`a}fols}}}, \bibinfo {author} {\bibfnamefont {C.}~\bibnamefont {{Ravoux}}}, \bibinfo {author} {\bibfnamefont {G.}~\bibnamefont {{Rossi}}}, \bibinfo {author} {\bibfnamefont {E.}~\bibnamefont {{Sanchez}}}, \bibinfo {author} {\bibfnamefont {D.}~\bibnamefont {{Schlegel}}}, \bibinfo {author} {\bibfnamefont {M.}~\bibnamefont {{Schubnell}}}, \bibinfo {author} {\bibfnamefont
  {H.}~\bibnamefont {{Seo}}}, \bibinfo {author} {\bibfnamefont {D.}~\bibnamefont {{Sprayberry}}}, \bibinfo {author} {\bibfnamefont {G.}~\bibnamefont {{Tarl{\'e}}}}, \bibinfo {author} {\bibfnamefont {M.}~\bibnamefont {{Vargas-Maga{\~n}a}}}, \bibinfo {author} {\bibfnamefont {B.~A.}\ \bibnamefont {{Weaver}}}, \bibinfo {author} {\bibfnamefont {C.}~\bibnamefont {{Zhao}}},\ and\ \bibinfo {author} {\bibfnamefont {H.}~\bibnamefont {{Zou}}},\ }\bibfield  {title} {\bibinfo {title} {{Constraining primordial non-Gaussianity with DESI 2024 LRG and QSO samples}},\ }\href {https://doi.org/10.48550/arXiv.2411.17623} {\bibfield  {journal} {\bibinfo  {journal} {arXiv e-prints}\ ,\ \bibinfo {eid} {arXiv:2411.17623}} (\bibinfo {year} {2024})},\ \Eprint {https://arxiv.org/abs/2411.17623} {arXiv:2411.17623 [astro-ph.CO]} \BibitemShut {NoStop}%
\bibitem [{\citenamefont {{Cagliari}}\ \emph {et~al.}(2024)\citenamefont {{Cagliari}}, \citenamefont {{Castorina}}, \citenamefont {{Bonici}},\ and\ \citenamefont {{Bianchi}}}]{cagliari_qso_eboss_24}%
  \BibitemOpen
  \bibfield  {author} {\bibinfo {author} {\bibfnamefont {M.~S.}\ \bibnamefont {{Cagliari}}}, \bibinfo {author} {\bibfnamefont {E.}~\bibnamefont {{Castorina}}}, \bibinfo {author} {\bibfnamefont {M.}~\bibnamefont {{Bonici}}},\ and\ \bibinfo {author} {\bibfnamefont {D.}~\bibnamefont {{Bianchi}}},\ }\bibfield  {title} {\bibinfo {title} {{Optimal constraints on Primordial non-Gaussianity with the eBOSS DR16 quasars in Fourier space}},\ }\href {https://doi.org/10.1088/1475-7516/2024/08/036} {\bibfield  {journal} {\bibinfo  {journal} {\jcap}\ }\textbf {\bibinfo {volume} {2024}},\ \bibinfo {eid} {036} (\bibinfo {year} {2024})},\ \Eprint {https://arxiv.org/abs/2309.15814} {arXiv:2309.15814 [astro-ph.CO]} \BibitemShut {NoStop}%
\bibitem [{Note23()}]{Note23}%
  \BibitemOpen
  \bibinfo {note} {Quasars potentially also have a less-complicated sample definition, at least at the high-redshift end \cite {palanque-delabrouille_qso_lf_sdss,palanque-delabrouille_qso_lf_eboss, wang_quasar_bev}}\BibitemShut {NoStop}%
\bibitem [{\citenamefont {{Alam}}\ \emph {et~al.}(2015)\citenamefont {{Alam}}, \citenamefont {{Albareti}}, \citenamefont {{Allende Prieto}}, \citenamefont {{Anders}}, \citenamefont {{Anderson}}, \citenamefont {{Anderton}}, \citenamefont {{Andrews}}, \citenamefont {{Armengaud}}, \citenamefont {{Aubourg}}, \citenamefont {{Bailey}}, \citenamefont {{Basu}}, \citenamefont {{Bautista}}, \citenamefont {{Beaton}}, \citenamefont {{Beers}}, \citenamefont {{Bender}}, \citenamefont {{Berlind}}, \citenamefont {{Beutler}}, \citenamefont {{Bhardwaj}}, \citenamefont {{Bird}}, \citenamefont {{Bizyaev}}, \citenamefont {{Blake}}, \citenamefont {{Blanton}}, \citenamefont {{Blomqvist}}, \citenamefont {{Bochanski}}, \citenamefont {{Bolton}}, \citenamefont {{Bovy}}, \citenamefont {{Shelden Bradley}}, \citenamefont {{Brandt}}, \citenamefont {{Brauer}}, \citenamefont {{Brinkmann}}, \citenamefont {{Brown}}, \citenamefont {{Brownstein}}, \citenamefont {{Burden}}, \citenamefont {{Burtin}}, \citenamefont {{Busca}}, \citenamefont {{Cai}},
  \citenamefont {{Capozzi}}, \citenamefont {{Carnero Rosell}}, \citenamefont {{Carr}}, \citenamefont {{Carrera}}, \citenamefont {{Chambers}}, \citenamefont {{Chaplin}}, \citenamefont {{Chen}}, \citenamefont {{Chiappini}}, \citenamefont {{Chojnowski}}, \citenamefont {{Chuang}}, \citenamefont {{Clerc}}, \citenamefont {{Comparat}}, \citenamefont {{Covey}}, \citenamefont {{Croft}}, \citenamefont {{Cuesta}}, \citenamefont {{Cunha}}, \citenamefont {{da Costa}}, \citenamefont {{Da Rio}}, \citenamefont {{Davenport}}, \citenamefont {{Dawson}}, \citenamefont {{De Lee}}, \citenamefont {{Delubac}}, \citenamefont {{Deshpande}}, \citenamefont {{Dhital}}, \citenamefont {{Dutra-Ferreira}}, \citenamefont {{Dwelly}}, \citenamefont {{Ealet}}, \citenamefont {{Ebelke}}, \citenamefont {{Edmondson}}, \citenamefont {{Eisenstein}}, \citenamefont {{Ellsworth}}, \citenamefont {{Elsworth}}, \citenamefont {{Epstein}}, \citenamefont {{Eracleous}}, \citenamefont {{Escoffier}}, \citenamefont {{Esposito}}, \citenamefont {{Evans}},
  \citenamefont {{Fan}}, \citenamefont {{Fern{\'a}ndez-Alvar}}, \citenamefont {{Feuillet}}, \citenamefont {{Filiz Ak}}, \citenamefont {{Finley}}, \citenamefont {{Finoguenov}}, \citenamefont {{Flaherty}}, \citenamefont {{Fleming}}, \citenamefont {{Font-Ribera}}, \citenamefont {{Foster}}, \citenamefont {{Frinchaboy}}, \citenamefont {{Galbraith-Frew}}, \citenamefont {{Garc{\'\i}a}}, \citenamefont {{Garc{\'\i}a-Hern{\'a}ndez}}, \citenamefont {{Garc{\'\i}a P{\'e}rez}}, \citenamefont {{Gaulme}}, \citenamefont {{Ge}}, \citenamefont {{G{\'e}nova-Santos}}, \citenamefont {{Georgakakis}}, \citenamefont {{Ghezzi}}, \citenamefont {{Gillespie}}, \citenamefont {{Girardi}}, \citenamefont {{Goddard}}, \citenamefont {{Gontcho}}, \citenamefont {{Gonz{\'a}lez Hern{\'a}ndez}}, \citenamefont {{Grebel}}, \citenamefont {{Green}}, \citenamefont {{Grieb}}, \citenamefont {{Grieves}}, \citenamefont {{Gunn}}, \citenamefont {{Guo}}, \citenamefont {{Harding}}, \citenamefont {{Hasselquist}}, \citenamefont {{Hawley}}, \citenamefont
  {{Hayden}}, \citenamefont {{Hearty}}, \citenamefont {{Hekker}}, \citenamefont {{Ho}}, \citenamefont {{Hogg}}, \citenamefont {{Holley-Bockelmann}}, \citenamefont {{Holtzman}}, \citenamefont {{Honscheid}}, \citenamefont {{Huber}}, \citenamefont {{Huehnerhoff}}, \citenamefont {{Ivans}}, \citenamefont {{Jiang}}, \citenamefont {{Johnson}}, \citenamefont {{Kinemuchi}}, \citenamefont {{Kirkby}}, \citenamefont {{Kitaura}}, \citenamefont {{Klaene}}, \citenamefont {{Knapp}}, \citenamefont {{Kneib}}, \citenamefont {{Koenig}}, \citenamefont {{Lam}}, \citenamefont {{Lan}}, \citenamefont {{Lang}}, \citenamefont {{Laurent}}, \citenamefont {{Le Goff}}, \citenamefont {{Leauthaud}}, \citenamefont {{Lee}}, \citenamefont {{Lee}}, \citenamefont {{Licquia}}, \citenamefont {{Liu}}, \citenamefont {{Long}}, \citenamefont {{L{\'o}pez-Corredoira}}, \citenamefont {{Lorenzo-Oliveira}}, \citenamefont {{Lucatello}}, \citenamefont {{Lundgren}}, \citenamefont {{Lupton}}, \citenamefont {{Mack}}, \citenamefont {{Mahadevan}}, \citenamefont
  {{Maia}}, \citenamefont {{Majewski}}, \citenamefont {{Malanushenko}}, \citenamefont {{Malanushenko}}, \citenamefont {{Manchado}}, \citenamefont {{Manera}}, \citenamefont {{Mao}}, \citenamefont {{Maraston}}, \citenamefont {{Marchwinski}}, \citenamefont {{Margala}}, \citenamefont {{Martell}}, \citenamefont {{Martig}}, \citenamefont {{Masters}}, \citenamefont {{Mathur}}, \citenamefont {{McBride}}, \citenamefont {{McGehee}}, \citenamefont {{McGreer}}, \citenamefont {{McMahon}}, \citenamefont {{M{\'e}nard}}, \citenamefont {{Menzel}}, \citenamefont {{Merloni}}, \citenamefont {{M{\'e}sz{\'a}ros}}, \citenamefont {{Miller}}, \citenamefont {{Miralda-Escud{\'e}}}, \citenamefont {{Miyatake}}, \citenamefont {{Montero-Dorta}}, \citenamefont {{More}}, \citenamefont {{Morganson}}, \citenamefont {{Morice-Atkinson}}, \citenamefont {{Morrison}}, \citenamefont {{Mosser}}, \citenamefont {{Muna}}, \citenamefont {{Myers}}, \citenamefont {{Nandra}}, \citenamefont {{Newman}}, \citenamefont {{Neyrinck}}, \citenamefont {{Nguyen}},
  \citenamefont {{Nichol}}, \citenamefont {{Nidever}}, \citenamefont {{Noterdaeme}}, \citenamefont {{Nuza}}, \citenamefont {{O'Connell}}, \citenamefont {{O'Connell}}, \citenamefont {{O'Connell}}, \citenamefont {{Ogando}}, \citenamefont {{Olmstead}}, \citenamefont {{Oravetz}}, \citenamefont {{Oravetz}}, \citenamefont {{Osumi}}, \citenamefont {{Owen}}, \citenamefont {{Padgett}}, \citenamefont {{Padmanabhan}}, \citenamefont {{Paegert}}, \citenamefont {{Palanque-Delabrouille}},\ and\ \citenamefont {{Pan}}}]{alam_boss_data}%
  \BibitemOpen
  \bibfield  {author} {\bibinfo {author} {\bibfnamefont {S.}~\bibnamefont {{Alam}}}, \bibinfo {author} {\bibfnamefont {F.~D.}\ \bibnamefont {{Albareti}}}, \bibinfo {author} {\bibfnamefont {C.}~\bibnamefont {{Allende Prieto}}}, \bibinfo {author} {\bibfnamefont {F.}~\bibnamefont {{Anders}}}, \bibinfo {author} {\bibfnamefont {S.~F.}\ \bibnamefont {{Anderson}}}, \bibinfo {author} {\bibfnamefont {T.}~\bibnamefont {{Anderton}}}, \bibinfo {author} {\bibfnamefont {B.~H.}\ \bibnamefont {{Andrews}}}, \bibinfo {author} {\bibfnamefont {E.}~\bibnamefont {{Armengaud}}}, \bibinfo {author} {\bibfnamefont {{\'E}.}~\bibnamefont {{Aubourg}}}, \bibinfo {author} {\bibfnamefont {S.}~\bibnamefont {{Bailey}}}, \bibinfo {author} {\bibfnamefont {S.}~\bibnamefont {{Basu}}}, \bibinfo {author} {\bibfnamefont {J.~E.}\ \bibnamefont {{Bautista}}}, \bibinfo {author} {\bibfnamefont {R.~L.}\ \bibnamefont {{Beaton}}}, \bibinfo {author} {\bibfnamefont {T.~C.}\ \bibnamefont {{Beers}}}, \bibinfo {author} {\bibfnamefont {C.~F.}\ \bibnamefont
  {{Bender}}}, \bibinfo {author} {\bibfnamefont {A.~A.}\ \bibnamefont {{Berlind}}}, \bibinfo {author} {\bibfnamefont {F.}~\bibnamefont {{Beutler}}}, \bibinfo {author} {\bibfnamefont {V.}~\bibnamefont {{Bhardwaj}}}, \bibinfo {author} {\bibfnamefont {J.~C.}\ \bibnamefont {{Bird}}}, \bibinfo {author} {\bibfnamefont {D.}~\bibnamefont {{Bizyaev}}}, \bibinfo {author} {\bibfnamefont {C.~H.}\ \bibnamefont {{Blake}}}, \bibinfo {author} {\bibfnamefont {M.~R.}\ \bibnamefont {{Blanton}}}, \bibinfo {author} {\bibfnamefont {M.}~\bibnamefont {{Blomqvist}}}, \bibinfo {author} {\bibfnamefont {J.~J.}\ \bibnamefont {{Bochanski}}}, \bibinfo {author} {\bibfnamefont {A.~S.}\ \bibnamefont {{Bolton}}}, \bibinfo {author} {\bibfnamefont {J.}~\bibnamefont {{Bovy}}}, \bibinfo {author} {\bibfnamefont {A.}~\bibnamefont {{Shelden Bradley}}}, \bibinfo {author} {\bibfnamefont {W.~N.}\ \bibnamefont {{Brandt}}}, \bibinfo {author} {\bibfnamefont {D.~E.}\ \bibnamefont {{Brauer}}}, \bibinfo {author} {\bibfnamefont {J.}~\bibnamefont
  {{Brinkmann}}}, \bibinfo {author} {\bibfnamefont {P.~J.}\ \bibnamefont {{Brown}}}, \bibinfo {author} {\bibfnamefont {J.~R.}\ \bibnamefont {{Brownstein}}}, \bibinfo {author} {\bibfnamefont {A.}~\bibnamefont {{Burden}}}, \bibinfo {author} {\bibfnamefont {E.}~\bibnamefont {{Burtin}}}, \bibinfo {author} {\bibfnamefont {N.~G.}\ \bibnamefont {{Busca}}}, \bibinfo {author} {\bibfnamefont {Z.}~\bibnamefont {{Cai}}}, \bibinfo {author} {\bibfnamefont {D.}~\bibnamefont {{Capozzi}}}, \bibinfo {author} {\bibfnamefont {A.}~\bibnamefont {{Carnero Rosell}}}, \bibinfo {author} {\bibfnamefont {M.~A.}\ \bibnamefont {{Carr}}}, \bibinfo {author} {\bibfnamefont {R.}~\bibnamefont {{Carrera}}}, \bibinfo {author} {\bibfnamefont {K.~C.}\ \bibnamefont {{Chambers}}}, \bibinfo {author} {\bibfnamefont {W.~J.}\ \bibnamefont {{Chaplin}}}, \bibinfo {author} {\bibfnamefont {Y.-C.}\ \bibnamefont {{Chen}}}, \bibinfo {author} {\bibfnamefont {C.}~\bibnamefont {{Chiappini}}}, \bibinfo {author} {\bibfnamefont {S.~D.}\ \bibnamefont {{Chojnowski}}},
  \bibinfo {author} {\bibfnamefont {C.-H.}\ \bibnamefont {{Chuang}}}, \bibinfo {author} {\bibfnamefont {N.}~\bibnamefont {{Clerc}}}, \bibinfo {author} {\bibfnamefont {J.}~\bibnamefont {{Comparat}}}, \bibinfo {author} {\bibfnamefont {K.}~\bibnamefont {{Covey}}}, \bibinfo {author} {\bibfnamefont {R.~A.~C.}\ \bibnamefont {{Croft}}}, \bibinfo {author} {\bibfnamefont {A.~J.}\ \bibnamefont {{Cuesta}}}, \bibinfo {author} {\bibfnamefont {K.}~\bibnamefont {{Cunha}}}, \bibinfo {author} {\bibfnamefont {L.~N.}\ \bibnamefont {{da Costa}}}, \bibinfo {author} {\bibfnamefont {N.}~\bibnamefont {{Da Rio}}}, \bibinfo {author} {\bibfnamefont {J.~R.~A.}\ \bibnamefont {{Davenport}}}, \bibinfo {author} {\bibfnamefont {K.~S.}\ \bibnamefont {{Dawson}}}, \bibinfo {author} {\bibfnamefont {N.}~\bibnamefont {{De Lee}}}, \bibinfo {author} {\bibfnamefont {T.}~\bibnamefont {{Delubac}}}, \bibinfo {author} {\bibfnamefont {R.}~\bibnamefont {{Deshpande}}}, \bibinfo {author} {\bibfnamefont {S.}~\bibnamefont {{Dhital}}}, \bibinfo {author}
  {\bibfnamefont {L.}~\bibnamefont {{Dutra-Ferreira}}}, \bibinfo {author} {\bibfnamefont {T.}~\bibnamefont {{Dwelly}}}, \bibinfo {author} {\bibfnamefont {A.}~\bibnamefont {{Ealet}}}, \bibinfo {author} {\bibfnamefont {G.~L.}\ \bibnamefont {{Ebelke}}}, \bibinfo {author} {\bibfnamefont {E.~M.}\ \bibnamefont {{Edmondson}}}, \bibinfo {author} {\bibfnamefont {D.~J.}\ \bibnamefont {{Eisenstein}}}, \bibinfo {author} {\bibfnamefont {T.}~\bibnamefont {{Ellsworth}}}, \bibinfo {author} {\bibfnamefont {Y.}~\bibnamefont {{Elsworth}}}, \bibinfo {author} {\bibfnamefont {C.~R.}\ \bibnamefont {{Epstein}}}, \bibinfo {author} {\bibfnamefont {M.}~\bibnamefont {{Eracleous}}}, \bibinfo {author} {\bibfnamefont {S.}~\bibnamefont {{Escoffier}}}, \bibinfo {author} {\bibfnamefont {M.}~\bibnamefont {{Esposito}}}, \bibinfo {author} {\bibfnamefont {M.~L.}\ \bibnamefont {{Evans}}}, \bibinfo {author} {\bibfnamefont {X.}~\bibnamefont {{Fan}}}, \bibinfo {author} {\bibfnamefont {E.}~\bibnamefont {{Fern{\'a}ndez-Alvar}}}, \bibinfo {author}
  {\bibfnamefont {D.}~\bibnamefont {{Feuillet}}}, \bibinfo {author} {\bibfnamefont {N.}~\bibnamefont {{Filiz Ak}}}, \bibinfo {author} {\bibfnamefont {H.}~\bibnamefont {{Finley}}}, \bibinfo {author} {\bibfnamefont {A.}~\bibnamefont {{Finoguenov}}}, \bibinfo {author} {\bibfnamefont {K.}~\bibnamefont {{Flaherty}}}, \bibinfo {author} {\bibfnamefont {S.~W.}\ \bibnamefont {{Fleming}}}, \bibinfo {author} {\bibfnamefont {A.}~\bibnamefont {{Font-Ribera}}}, \bibinfo {author} {\bibfnamefont {J.}~\bibnamefont {{Foster}}}, \bibinfo {author} {\bibfnamefont {P.~M.}\ \bibnamefont {{Frinchaboy}}}, \bibinfo {author} {\bibfnamefont {J.~G.}\ \bibnamefont {{Galbraith-Frew}}}, \bibinfo {author} {\bibfnamefont {R.~A.}\ \bibnamefont {{Garc{\'\i}a}}}, \bibinfo {author} {\bibfnamefont {D.~A.}\ \bibnamefont {{Garc{\'\i}a-Hern{\'a}ndez}}}, \bibinfo {author} {\bibfnamefont {A.~E.}\ \bibnamefont {{Garc{\'\i}a P{\'e}rez}}}, \bibinfo {author} {\bibfnamefont {P.}~\bibnamefont {{Gaulme}}}, \bibinfo {author} {\bibfnamefont {J.}~\bibnamefont
  {{Ge}}}, \bibinfo {author} {\bibfnamefont {R.}~\bibnamefont {{G{\'e}nova-Santos}}}, \bibinfo {author} {\bibfnamefont {A.}~\bibnamefont {{Georgakakis}}}, \bibinfo {author} {\bibfnamefont {L.}~\bibnamefont {{Ghezzi}}}, \bibinfo {author} {\bibfnamefont {B.~A.}\ \bibnamefont {{Gillespie}}}, \bibinfo {author} {\bibfnamefont {L.}~\bibnamefont {{Girardi}}}, \bibinfo {author} {\bibfnamefont {D.}~\bibnamefont {{Goddard}}}, \bibinfo {author} {\bibfnamefont {S.~G.~A.}\ \bibnamefont {{Gontcho}}}, \bibinfo {author} {\bibfnamefont {J.~I.}\ \bibnamefont {{Gonz{\'a}lez Hern{\'a}ndez}}}, \bibinfo {author} {\bibfnamefont {E.~K.}\ \bibnamefont {{Grebel}}}, \bibinfo {author} {\bibfnamefont {P.~J.}\ \bibnamefont {{Green}}}, \bibinfo {author} {\bibfnamefont {J.~N.}\ \bibnamefont {{Grieb}}}, \bibinfo {author} {\bibfnamefont {N.}~\bibnamefont {{Grieves}}}, \bibinfo {author} {\bibfnamefont {J.~E.}\ \bibnamefont {{Gunn}}}, \bibinfo {author} {\bibfnamefont {H.}~\bibnamefont {{Guo}}}, \bibinfo {author} {\bibfnamefont {P.}~\bibnamefont
  {{Harding}}}, \bibinfo {author} {\bibfnamefont {S.}~\bibnamefont {{Hasselquist}}}, \bibinfo {author} {\bibfnamefont {S.~L.}\ \bibnamefont {{Hawley}}}, \bibinfo {author} {\bibfnamefont {M.}~\bibnamefont {{Hayden}}}, \bibinfo {author} {\bibfnamefont {F.~R.}\ \bibnamefont {{Hearty}}}, \bibinfo {author} {\bibfnamefont {S.}~\bibnamefont {{Hekker}}}, \bibinfo {author} {\bibfnamefont {S.}~\bibnamefont {{Ho}}}, \bibinfo {author} {\bibfnamefont {D.~W.}\ \bibnamefont {{Hogg}}}, \bibinfo {author} {\bibfnamefont {K.}~\bibnamefont {{Holley-Bockelmann}}}, \bibinfo {author} {\bibfnamefont {J.~A.}\ \bibnamefont {{Holtzman}}}, \bibinfo {author} {\bibfnamefont {K.}~\bibnamefont {{Honscheid}}}, \bibinfo {author} {\bibfnamefont {D.}~\bibnamefont {{Huber}}}, \bibinfo {author} {\bibfnamefont {J.}~\bibnamefont {{Huehnerhoff}}}, \bibinfo {author} {\bibfnamefont {I.~I.}\ \bibnamefont {{Ivans}}}, \bibinfo {author} {\bibfnamefont {L.}~\bibnamefont {{Jiang}}}, \bibinfo {author} {\bibfnamefont {J.~A.}\ \bibnamefont {{Johnson}}},
  \bibinfo {author} {\bibfnamefont {K.}~\bibnamefont {{Kinemuchi}}}, \bibinfo {author} {\bibfnamefont {D.}~\bibnamefont {{Kirkby}}}, \bibinfo {author} {\bibfnamefont {F.}~\bibnamefont {{Kitaura}}}, \bibinfo {author} {\bibfnamefont {M.~A.}\ \bibnamefont {{Klaene}}}, \bibinfo {author} {\bibfnamefont {G.~R.}\ \bibnamefont {{Knapp}}}, \bibinfo {author} {\bibfnamefont {J.-P.}\ \bibnamefont {{Kneib}}}, \bibinfo {author} {\bibfnamefont {X.~P.}\ \bibnamefont {{Koenig}}}, \bibinfo {author} {\bibfnamefont {C.~R.}\ \bibnamefont {{Lam}}}, \bibinfo {author} {\bibfnamefont {T.-W.}\ \bibnamefont {{Lan}}}, \bibinfo {author} {\bibfnamefont {D.}~\bibnamefont {{Lang}}}, \bibinfo {author} {\bibfnamefont {P.}~\bibnamefont {{Laurent}}}, \bibinfo {author} {\bibfnamefont {J.-M.}\ \bibnamefont {{Le Goff}}}, \bibinfo {author} {\bibfnamefont {A.}~\bibnamefont {{Leauthaud}}}, \bibinfo {author} {\bibfnamefont {K.-G.}\ \bibnamefont {{Lee}}}, \bibinfo {author} {\bibfnamefont {Y.~S.}\ \bibnamefont {{Lee}}}, \bibinfo {author} {\bibfnamefont
  {T.~C.}\ \bibnamefont {{Licquia}}}, \bibinfo {author} {\bibfnamefont {J.}~\bibnamefont {{Liu}}}, \bibinfo {author} {\bibfnamefont {D.~C.}\ \bibnamefont {{Long}}}, \bibinfo {author} {\bibfnamefont {M.}~\bibnamefont {{L{\'o}pez-Corredoira}}}, \bibinfo {author} {\bibfnamefont {D.}~\bibnamefont {{Lorenzo-Oliveira}}}, \bibinfo {author} {\bibfnamefont {S.}~\bibnamefont {{Lucatello}}}, \bibinfo {author} {\bibfnamefont {B.}~\bibnamefont {{Lundgren}}}, \bibinfo {author} {\bibfnamefont {R.~H.}\ \bibnamefont {{Lupton}}}, \bibinfo {author} {\bibfnamefont {C.~E.}\ \bibnamefont {{Mack}}, \bibfnamefont {III}}, \bibinfo {author} {\bibfnamefont {S.}~\bibnamefont {{Mahadevan}}}, \bibinfo {author} {\bibfnamefont {M.~A.~G.}\ \bibnamefont {{Maia}}}, \bibinfo {author} {\bibfnamefont {S.~R.}\ \bibnamefont {{Majewski}}}, \bibinfo {author} {\bibfnamefont {E.}~\bibnamefont {{Malanushenko}}}, \bibinfo {author} {\bibfnamefont {V.}~\bibnamefont {{Malanushenko}}}, \bibinfo {author} {\bibfnamefont {A.}~\bibnamefont {{Manchado}}},
  \bibinfo {author} {\bibfnamefont {M.}~\bibnamefont {{Manera}}}, \bibinfo {author} {\bibfnamefont {Q.}~\bibnamefont {{Mao}}}, \bibinfo {author} {\bibfnamefont {C.}~\bibnamefont {{Maraston}}}, \bibinfo {author} {\bibfnamefont {R.~C.}\ \bibnamefont {{Marchwinski}}}, \bibinfo {author} {\bibfnamefont {D.}~\bibnamefont {{Margala}}}, \bibinfo {author} {\bibfnamefont {S.~L.}\ \bibnamefont {{Martell}}}, \bibinfo {author} {\bibfnamefont {M.}~\bibnamefont {{Martig}}}, \bibinfo {author} {\bibfnamefont {K.~L.}\ \bibnamefont {{Masters}}}, \bibinfo {author} {\bibfnamefont {S.}~\bibnamefont {{Mathur}}}, \bibinfo {author} {\bibfnamefont {C.~K.}\ \bibnamefont {{McBride}}}, \bibinfo {author} {\bibfnamefont {P.~M.}\ \bibnamefont {{McGehee}}}, \bibinfo {author} {\bibfnamefont {I.~D.}\ \bibnamefont {{McGreer}}}, \bibinfo {author} {\bibfnamefont {R.~G.}\ \bibnamefont {{McMahon}}}, \bibinfo {author} {\bibfnamefont {B.}~\bibnamefont {{M{\'e}nard}}}, \bibinfo {author} {\bibfnamefont {M.-L.}\ \bibnamefont {{Menzel}}}, \bibinfo
  {author} {\bibfnamefont {A.}~\bibnamefont {{Merloni}}}, \bibinfo {author} {\bibfnamefont {S.}~\bibnamefont {{M{\'e}sz{\'a}ros}}}, \bibinfo {author} {\bibfnamefont {A.~A.}\ \bibnamefont {{Miller}}}, \bibinfo {author} {\bibfnamefont {J.}~\bibnamefont {{Miralda-Escud{\'e}}}}, \bibinfo {author} {\bibfnamefont {H.}~\bibnamefont {{Miyatake}}}, \bibinfo {author} {\bibfnamefont {A.~D.}\ \bibnamefont {{Montero-Dorta}}}, \bibinfo {author} {\bibfnamefont {S.}~\bibnamefont {{More}}}, \bibinfo {author} {\bibfnamefont {E.}~\bibnamefont {{Morganson}}}, \bibinfo {author} {\bibfnamefont {X.}~\bibnamefont {{Morice-Atkinson}}}, \bibinfo {author} {\bibfnamefont {H.~L.}\ \bibnamefont {{Morrison}}}, \bibinfo {author} {\bibfnamefont {B.}~\bibnamefont {{Mosser}}}, \bibinfo {author} {\bibfnamefont {D.}~\bibnamefont {{Muna}}}, \bibinfo {author} {\bibfnamefont {A.~D.}\ \bibnamefont {{Myers}}}, \bibinfo {author} {\bibfnamefont {K.}~\bibnamefont {{Nandra}}}, \bibinfo {author} {\bibfnamefont {J.~A.}\ \bibnamefont {{Newman}}}, \bibinfo
  {author} {\bibfnamefont {M.}~\bibnamefont {{Neyrinck}}}, \bibinfo {author} {\bibfnamefont {D.~C.}\ \bibnamefont {{Nguyen}}}, \bibinfo {author} {\bibfnamefont {R.~C.}\ \bibnamefont {{Nichol}}}, \bibinfo {author} {\bibfnamefont {D.~L.}\ \bibnamefont {{Nidever}}}, \bibinfo {author} {\bibfnamefont {P.}~\bibnamefont {{Noterdaeme}}}, \bibinfo {author} {\bibfnamefont {S.~E.}\ \bibnamefont {{Nuza}}}, \bibinfo {author} {\bibfnamefont {J.~E.}\ \bibnamefont {{O'Connell}}}, \bibinfo {author} {\bibfnamefont {R.~W.}\ \bibnamefont {{O'Connell}}}, \bibinfo {author} {\bibfnamefont {R.}~\bibnamefont {{O'Connell}}}, \bibinfo {author} {\bibfnamefont {R.~L.~C.}\ \bibnamefont {{Ogando}}}, \bibinfo {author} {\bibfnamefont {M.~D.}\ \bibnamefont {{Olmstead}}}, \bibinfo {author} {\bibfnamefont {A.~E.}\ \bibnamefont {{Oravetz}}}, \bibinfo {author} {\bibfnamefont {D.~J.}\ \bibnamefont {{Oravetz}}}, \bibinfo {author} {\bibfnamefont {K.}~\bibnamefont {{Osumi}}}, \bibinfo {author} {\bibfnamefont {R.}~\bibnamefont {{Owen}}}, \bibinfo
  {author} {\bibfnamefont {D.~L.}\ \bibnamefont {{Padgett}}}, \bibinfo {author} {\bibfnamefont {N.}~\bibnamefont {{Padmanabhan}}}, \bibinfo {author} {\bibfnamefont {M.}~\bibnamefont {{Paegert}}}, \bibinfo {author} {\bibfnamefont {N.}~\bibnamefont {{Palanque-Delabrouille}}},\ and\ \bibinfo {author} {\bibfnamefont {K.}~\bibnamefont {{Pan}}},\ }\bibfield  {title} {\bibinfo {title} {{The Eleventh and Twelfth Data Releases of the Sloan Digital Sky Survey: Final Data from SDSS-III}},\ }\href {https://doi.org/10.1088/0067-0049/219/1/12} {\bibfield  {journal} {\bibinfo  {journal} {\apjs}\ }\textbf {\bibinfo {volume} {219}},\ \bibinfo {eid} {12} (\bibinfo {year} {2015})},\ \Eprint {https://arxiv.org/abs/1501.00963} {arXiv:1501.00963 [astro-ph.IM]} \BibitemShut {NoStop}%
\bibitem [{\citenamefont {{Maraston}}\ \emph {et~al.}(2009)\citenamefont {{Maraston}}, \citenamefont {{Str{\"o}mb{\"a}ck}}, \citenamefont {{Thomas}}, \citenamefont {{Wake}},\ and\ \citenamefont {{Nichol}}}]{maraston_passive_lrg_sps_model}%
  \BibitemOpen
  \bibfield  {author} {\bibinfo {author} {\bibfnamefont {C.}~\bibnamefont {{Maraston}}}, \bibinfo {author} {\bibfnamefont {G.}~\bibnamefont {{Str{\"o}mb{\"a}ck}}}, \bibinfo {author} {\bibfnamefont {D.}~\bibnamefont {{Thomas}}}, \bibinfo {author} {\bibfnamefont {D.~A.}\ \bibnamefont {{Wake}}},\ and\ \bibinfo {author} {\bibfnamefont {R.~C.}\ \bibnamefont {{Nichol}}},\ }\bibfield  {title} {\bibinfo {title} {{Modelling the colour evolution of luminous red galaxies - improvements with empirical stellar spectra}},\ }\href {https://doi.org/10.1111/j.1745-3933.2009.00621.x} {\bibfield  {journal} {\bibinfo  {journal} {\mnras}\ }\textbf {\bibinfo {volume} {394}},\ \bibinfo {pages} {L107} (\bibinfo {year} {2009})},\ \Eprint {https://arxiv.org/abs/0809.1867} {arXiv:0809.1867 [astro-ph]} \BibitemShut {NoStop}%
\bibitem [{\citenamefont {{Maraston}}\ \emph {et~al.}(2013)\citenamefont {{Maraston}}, \citenamefont {{Pforr}}, \citenamefont {{Henriques}}, \citenamefont {{Thomas}}, \citenamefont {{Wake}}, \citenamefont {{Brownstein}}, \citenamefont {{Capozzi}}, \citenamefont {{Tinker}}, \citenamefont {{Bundy}}, \citenamefont {{Skibba}}, \citenamefont {{Beifiori}}, \citenamefont {{Nichol}}, \citenamefont {{Edmondson}}, \citenamefont {{Schneider}}, \citenamefont {{Chen}}, \citenamefont {{Masters}}, \citenamefont {{Steele}}, \citenamefont {{Bolton}}, \citenamefont {{York}}, \citenamefont {{Weaver}}, \citenamefont {{Higgs}}, \citenamefont {{Bizyaev}}, \citenamefont {{Brewington}}, \citenamefont {{Malanushenko}}, \citenamefont {{Malanushenko}}, \citenamefont {{Snedden}}, \citenamefont {{Oravetz}}, \citenamefont {{Pan}}, \citenamefont {{Shelden}},\ and\ \citenamefont {{Simmons}}}]{maraston_cmass_stellar_mass}%
  \BibitemOpen
  \bibfield  {author} {\bibinfo {author} {\bibfnamefont {C.}~\bibnamefont {{Maraston}}}, \bibinfo {author} {\bibfnamefont {J.}~\bibnamefont {{Pforr}}}, \bibinfo {author} {\bibfnamefont {B.~M.}\ \bibnamefont {{Henriques}}}, \bibinfo {author} {\bibfnamefont {D.}~\bibnamefont {{Thomas}}}, \bibinfo {author} {\bibfnamefont {D.}~\bibnamefont {{Wake}}}, \bibinfo {author} {\bibfnamefont {J.~R.}\ \bibnamefont {{Brownstein}}}, \bibinfo {author} {\bibfnamefont {D.}~\bibnamefont {{Capozzi}}}, \bibinfo {author} {\bibfnamefont {J.}~\bibnamefont {{Tinker}}}, \bibinfo {author} {\bibfnamefont {K.}~\bibnamefont {{Bundy}}}, \bibinfo {author} {\bibfnamefont {R.~A.}\ \bibnamefont {{Skibba}}}, \bibinfo {author} {\bibfnamefont {A.}~\bibnamefont {{Beifiori}}}, \bibinfo {author} {\bibfnamefont {R.~C.}\ \bibnamefont {{Nichol}}}, \bibinfo {author} {\bibfnamefont {E.}~\bibnamefont {{Edmondson}}}, \bibinfo {author} {\bibfnamefont {D.~P.}\ \bibnamefont {{Schneider}}}, \bibinfo {author} {\bibfnamefont {Y.}~\bibnamefont {{Chen}}}, \bibinfo
  {author} {\bibfnamefont {K.~L.}\ \bibnamefont {{Masters}}}, \bibinfo {author} {\bibfnamefont {O.}~\bibnamefont {{Steele}}}, \bibinfo {author} {\bibfnamefont {A.~S.}\ \bibnamefont {{Bolton}}}, \bibinfo {author} {\bibfnamefont {D.~G.}\ \bibnamefont {{York}}}, \bibinfo {author} {\bibfnamefont {B.~A.}\ \bibnamefont {{Weaver}}}, \bibinfo {author} {\bibfnamefont {T.}~\bibnamefont {{Higgs}}}, \bibinfo {author} {\bibfnamefont {D.}~\bibnamefont {{Bizyaev}}}, \bibinfo {author} {\bibfnamefont {H.}~\bibnamefont {{Brewington}}}, \bibinfo {author} {\bibfnamefont {E.}~\bibnamefont {{Malanushenko}}}, \bibinfo {author} {\bibfnamefont {V.}~\bibnamefont {{Malanushenko}}}, \bibinfo {author} {\bibfnamefont {S.}~\bibnamefont {{Snedden}}}, \bibinfo {author} {\bibfnamefont {D.}~\bibnamefont {{Oravetz}}}, \bibinfo {author} {\bibfnamefont {K.}~\bibnamefont {{Pan}}}, \bibinfo {author} {\bibfnamefont {A.}~\bibnamefont {{Shelden}}},\ and\ \bibinfo {author} {\bibfnamefont {A.}~\bibnamefont {{Simmons}}},\ }\bibfield  {title} {\bibinfo
  {title} {{Stellar masses of SDSS-III/BOSS galaxies at z {\ensuremath{\sim}} 0.5 and constraints to galaxy formation models}},\ }\href {https://doi.org/10.1093/mnras/stt1424} {\bibfield  {journal} {\bibinfo  {journal} {\mnras}\ }\textbf {\bibinfo {volume} {435}},\ \bibinfo {pages} {2764} (\bibinfo {year} {2013})},\ \Eprint {https://arxiv.org/abs/1207.6114} {arXiv:1207.6114 [astro-ph.CO]} \BibitemShut {NoStop}%
\bibitem [{\citenamefont {{Wenzl}}\ \emph {et~al.}(2024)\citenamefont {{Wenzl}}, \citenamefont {{Chen}},\ and\ \citenamefont {{Bean}}}]{wenzl_mag_bias_boss}%
  \BibitemOpen
  \bibfield  {author} {\bibinfo {author} {\bibfnamefont {L.}~\bibnamefont {{Wenzl}}}, \bibinfo {author} {\bibfnamefont {S.-F.}\ \bibnamefont {{Chen}}},\ and\ \bibinfo {author} {\bibfnamefont {R.}~\bibnamefont {{Bean}}},\ }\bibfield  {title} {\bibinfo {title} {{Magnification bias estimators for realistic surveys: an application to the BOSS survey}},\ }\href {https://doi.org/10.1093/mnras/stad3314} {\bibfield  {journal} {\bibinfo  {journal} {\mnras}\ }\textbf {\bibinfo {volume} {527}},\ \bibinfo {pages} {1760} (\bibinfo {year} {2024})},\ \Eprint {https://arxiv.org/abs/2308.05892} {arXiv:2308.05892 [astro-ph.CO]} \BibitemShut {NoStop}%
\bibitem [{\citenamefont {{Friedman-Shaw}}\ \emph {et~al.}(2024)\citenamefont {{Friedman-Shaw}}, \citenamefont {{Krolewski}}, \citenamefont {{Foglieni}},\ and\ \citenamefont {{Afshordi}}}]{doppler_bias}%
  \BibitemOpen
  \bibfield  {author} {\bibinfo {author} {\bibfnamefont {B.}~\bibnamefont {{Friedman-Shaw}}}, \bibinfo {author} {\bibfnamefont {A.}~\bibnamefont {{Krolewski}}}, \bibinfo {author} {\bibfnamefont {M.}~\bibnamefont {{Foglieni}}},\ and\ \bibinfo {author} {\bibfnamefont {N.}~\bibnamefont {{Afshordi}}},\ }\bibfield  {title} {\bibinfo {title} {{Doppler bias: impact of peculiar velocities on color selection and the large scale structure of galaxy surveys}},\ }\href {https://doi.org/10.48550/arXiv.2410.04705} {\bibfield  {journal} {\bibinfo  {journal} {arXiv e-prints}\ ,\ \bibinfo {eid} {arXiv:2410.04705}} (\bibinfo {year} {2024})},\ \Eprint {https://arxiv.org/abs/2410.04705} {arXiv:2410.04705 [astro-ph.CO]} \BibitemShut {NoStop}%
\bibitem [{\citenamefont {{Ivanov}}\ \emph {et~al.}(2024)\citenamefont {{Ivanov}}, \citenamefont {{Cuesta-Lazaro}}, \citenamefont {{Mishra-Sharma}}, \citenamefont {{Obuljen}},\ and\ \citenamefont {{Toomey}}}]{ivanov_sfpng_priors}%
  \BibitemOpen
  \bibfield  {author} {\bibinfo {author} {\bibfnamefont {M.~M.}\ \bibnamefont {{Ivanov}}}, \bibinfo {author} {\bibfnamefont {C.}~\bibnamefont {{Cuesta-Lazaro}}}, \bibinfo {author} {\bibfnamefont {S.}~\bibnamefont {{Mishra-Sharma}}}, \bibinfo {author} {\bibfnamefont {A.}~\bibnamefont {{Obuljen}}},\ and\ \bibinfo {author} {\bibfnamefont {M.~W.}\ \bibnamefont {{Toomey}}},\ }\bibfield  {title} {\bibinfo {title} {{Full-shape analysis with simulation-based priors: Constraints on single field inflation from BOSS}},\ }\href {https://doi.org/10.1103/PhysRevD.110.063538} {\bibfield  {journal} {\bibinfo  {journal} {\prd}\ }\textbf {\bibinfo {volume} {110}},\ \bibinfo {eid} {063538} (\bibinfo {year} {2024})},\ \Eprint {https://arxiv.org/abs/2402.13310} {arXiv:2402.13310 [astro-ph.CO]} \BibitemShut {NoStop}%
\bibitem [{\citenamefont {{Palanque-Delabrouille}}\ \emph {et~al.}(2013)\citenamefont {{Palanque-Delabrouille}}, \citenamefont {{Magneville}}, \citenamefont {{Y{\`e}che}}, \citenamefont {{Eftekharzadeh}}, \citenamefont {{Myers}}, \citenamefont {{Petitjean}}, \citenamefont {{P{\^a}ris}}, \citenamefont {{Aubourg}}, \citenamefont {{McGreer}}, \citenamefont {{Fan}}, \citenamefont {{Dey}}, \citenamefont {{Schlegel}}, \citenamefont {{Bailey}}, \citenamefont {{Bizayev}}, \citenamefont {{Bolton}}, \citenamefont {{Dawson}}, \citenamefont {{Ebelke}}, \citenamefont {{Ge}}, \citenamefont {{Malanushenko}}, \citenamefont {{Malanushenko}}, \citenamefont {{Oravetz}}, \citenamefont {{Pan}}, \citenamefont {{Ross}}, \citenamefont {{Schneider}}, \citenamefont {{Sheldon}}, \citenamefont {{Simmons}}, \citenamefont {{Tinker}}, \citenamefont {{White}},\ and\ \citenamefont {{Willmer}}}]{palanque-delabrouille_qso_lf_sdss}%
  \BibitemOpen
  \bibfield  {author} {\bibinfo {author} {\bibfnamefont {N.}~\bibnamefont {{Palanque-Delabrouille}}}, \bibinfo {author} {\bibfnamefont {C.}~\bibnamefont {{Magneville}}}, \bibinfo {author} {\bibfnamefont {C.}~\bibnamefont {{Y{\`e}che}}}, \bibinfo {author} {\bibfnamefont {S.}~\bibnamefont {{Eftekharzadeh}}}, \bibinfo {author} {\bibfnamefont {A.~D.}\ \bibnamefont {{Myers}}}, \bibinfo {author} {\bibfnamefont {P.}~\bibnamefont {{Petitjean}}}, \bibinfo {author} {\bibfnamefont {I.}~\bibnamefont {{P{\^a}ris}}}, \bibinfo {author} {\bibfnamefont {E.}~\bibnamefont {{Aubourg}}}, \bibinfo {author} {\bibfnamefont {I.}~\bibnamefont {{McGreer}}}, \bibinfo {author} {\bibfnamefont {X.}~\bibnamefont {{Fan}}}, \bibinfo {author} {\bibfnamefont {A.}~\bibnamefont {{Dey}}}, \bibinfo {author} {\bibfnamefont {D.}~\bibnamefont {{Schlegel}}}, \bibinfo {author} {\bibfnamefont {S.}~\bibnamefont {{Bailey}}}, \bibinfo {author} {\bibfnamefont {D.}~\bibnamefont {{Bizayev}}}, \bibinfo {author} {\bibfnamefont {A.}~\bibnamefont {{Bolton}}},
  \bibinfo {author} {\bibfnamefont {K.}~\bibnamefont {{Dawson}}}, \bibinfo {author} {\bibfnamefont {G.}~\bibnamefont {{Ebelke}}}, \bibinfo {author} {\bibfnamefont {J.}~\bibnamefont {{Ge}}}, \bibinfo {author} {\bibfnamefont {E.}~\bibnamefont {{Malanushenko}}}, \bibinfo {author} {\bibfnamefont {V.}~\bibnamefont {{Malanushenko}}}, \bibinfo {author} {\bibfnamefont {D.}~\bibnamefont {{Oravetz}}}, \bibinfo {author} {\bibfnamefont {K.}~\bibnamefont {{Pan}}}, \bibinfo {author} {\bibfnamefont {N.~P.}\ \bibnamefont {{Ross}}}, \bibinfo {author} {\bibfnamefont {D.~P.}\ \bibnamefont {{Schneider}}}, \bibinfo {author} {\bibfnamefont {E.}~\bibnamefont {{Sheldon}}}, \bibinfo {author} {\bibfnamefont {A.}~\bibnamefont {{Simmons}}}, \bibinfo {author} {\bibfnamefont {J.}~\bibnamefont {{Tinker}}}, \bibinfo {author} {\bibfnamefont {M.}~\bibnamefont {{White}}},\ and\ \bibinfo {author} {\bibfnamefont {C.}~\bibnamefont {{Willmer}}},\ }\bibfield  {title} {\bibinfo {title} {{Luminosity function from dedicated SDSS-III and MMT data of
  quasars in 0.7 < z < 4.0 selected with a new approach}},\ }\href {https://doi.org/10.1051/0004-6361/201220379} {\bibfield  {journal} {\bibinfo  {journal} {\aap}\ }\textbf {\bibinfo {volume} {551}},\ \bibinfo {eid} {A29} (\bibinfo {year} {2013})},\ \Eprint {https://arxiv.org/abs/1209.3968} {arXiv:1209.3968 [astro-ph.CO]} \BibitemShut {NoStop}%
\bibitem [{\citenamefont {{Palanque-Delabrouille}}\ \emph {et~al.}(2016)\citenamefont {{Palanque-Delabrouille}}, \citenamefont {{Magneville}}, \citenamefont {{Y{\`e}che}}, \citenamefont {{P{\^a}ris}}, \citenamefont {{Petitjean}}, \citenamefont {{Burtin}}, \citenamefont {{Dawson}}, \citenamefont {{McGreer}}, \citenamefont {{Myers}}, \citenamefont {{Rossi}}, \citenamefont {{Schlegel}}, \citenamefont {{Schneider}}, \citenamefont {{Streblyanska}},\ and\ \citenamefont {{Tinker}}}]{palanque-delabrouille_qso_lf_eboss}%
  \BibitemOpen
  \bibfield  {author} {\bibinfo {author} {\bibfnamefont {N.}~\bibnamefont {{Palanque-Delabrouille}}}, \bibinfo {author} {\bibfnamefont {C.}~\bibnamefont {{Magneville}}}, \bibinfo {author} {\bibfnamefont {C.}~\bibnamefont {{Y{\`e}che}}}, \bibinfo {author} {\bibfnamefont {I.}~\bibnamefont {{P{\^a}ris}}}, \bibinfo {author} {\bibfnamefont {P.}~\bibnamefont {{Petitjean}}}, \bibinfo {author} {\bibfnamefont {E.}~\bibnamefont {{Burtin}}}, \bibinfo {author} {\bibfnamefont {K.}~\bibnamefont {{Dawson}}}, \bibinfo {author} {\bibfnamefont {I.}~\bibnamefont {{McGreer}}}, \bibinfo {author} {\bibfnamefont {A.~D.}\ \bibnamefont {{Myers}}}, \bibinfo {author} {\bibfnamefont {G.}~\bibnamefont {{Rossi}}}, \bibinfo {author} {\bibfnamefont {D.}~\bibnamefont {{Schlegel}}}, \bibinfo {author} {\bibfnamefont {D.}~\bibnamefont {{Schneider}}}, \bibinfo {author} {\bibfnamefont {A.}~\bibnamefont {{Streblyanska}}},\ and\ \bibinfo {author} {\bibfnamefont {J.}~\bibnamefont {{Tinker}}},\ }\bibfield  {title} {\bibinfo {title} {{The extended
  Baryon Oscillation Spectroscopic Survey: Variability selection and quasar luminosity function}},\ }\href {https://doi.org/10.1051/0004-6361/201527392} {\bibfield  {journal} {\bibinfo  {journal} {\aap}\ }\textbf {\bibinfo {volume} {587}},\ \bibinfo {eid} {A41} (\bibinfo {year} {2016})},\ \Eprint {https://arxiv.org/abs/1509.05607} {arXiv:1509.05607 [astro-ph.CO]} \BibitemShut {NoStop}%
\bibitem [{\citenamefont {{Wang}}\ \emph {et~al.}(2020)\citenamefont {{Wang}}, \citenamefont {{Beutler}},\ and\ \citenamefont {{Bacon}}}]{wang_quasar_bev}%
  \BibitemOpen
  \bibfield  {author} {\bibinfo {author} {\bibfnamefont {M.~S.}\ \bibnamefont {{Wang}}}, \bibinfo {author} {\bibfnamefont {F.}~\bibnamefont {{Beutler}}},\ and\ \bibinfo {author} {\bibfnamefont {D.}~\bibnamefont {{Bacon}}},\ }\bibfield  {title} {\bibinfo {title} {{Impact of relativistic effects on the primordial non-Gaussianity signature in the large-scale clustering of quasars}},\ }\href {https://doi.org/10.1093/mnras/staa2998} {\bibfield  {journal} {\bibinfo  {journal} {\mnras}\ }\textbf {\bibinfo {volume} {499}},\ \bibinfo {pages} {2598} (\bibinfo {year} {2020})},\ \Eprint {https://arxiv.org/abs/2007.01802} {arXiv:2007.01802 [astro-ph.CO]} \BibitemShut {NoStop}%
\end{thebibliography}%

\end{document}